\documentstyle[12pt]{report}
\begin{document}
\title{Tumbling and Technicolor Theory\thanks{Doctoral thesis}}
\author{
Noriaki Kitazawa\thanks{JSPS fellow}\\
\\
Department of Physics, Nagoya University\\
Nagoya, 464-01, Japan
}
\date{January, 1994\\ \quad \\ DPNU-94-03}
\maketitle
\begin{abstract}

The extended technicolor theory
 is a candidate of the physics beyond the standard model.
To explain the mass hierarchy of the quarks and leptons,
 the extended technicolor gauge symmetry
 must hierarchically break to the technicolor gauge symmetry.
Tumbling gauge theory is considered
 as a candidate of the dynamics of such hierarchical symmetry breaking,
 since the sequential self-breaking of the gauge symmetry (``tumbling'')
 can be expected in that theory.

It is well known that
 the extended technicolor theory induces
 too strong flavor-changing neutral current interactions
 to be consistent with the experiments.
This problem can be solved if the technicolor dynamics is the special one
 with very large anomalous dimension of the composite operator
 ${\bar T}T$ composed by the technifermion field $T$.
Two types of the models with large anomalous dimension have been proposed.
One is the gauge theory with slowly running coupling,
 another is the gauge theory with strong four fermion interaction.
It is expected that
 the large anomalous dimension is realized in the tumbling gauge theory.

In this thesis we systematically estimate
 the strength of the effective four fermion interactions
 induced in the tumbling gauge theory
 by using the effective action
 within the one-gauge-boson-exchange approximation.
It is shown that the couplings of the effective four fermion interactions
 cannot be large enough to realize sufficiently large anomalous dimension.
This result is important
 for the model building of the extended technicolor theory.
The decoupling of the gauge bosons and fermions
 which get their masses in the process of tumbling is also discussed.
It is non-perturbatively shown that
 their effect on the low energy dynamics is negligibly small.

Before the estimations,
 we give the explanation of the fundamental techniques
 to treat the dynamical symmetry breaking.
The scenario and the phenomenology of the extended technicolor theory
 is also reviewed.
The restriction on the extended technicolor theory
 from the precision experiments is discussed.
It is emphasized that
 the non-oblique correction is important as well as the oblique correction,
 since the top quark is heavy.
\end{abstract}

\chapter{Introduction}

The standard model of the elementary particle physics
 almost perfectly explains many present experiments.
Especially,
 the agreement with the recent precision experiments in LEP is significant.
The unified description of electromagnetic interaction
 and weak interaction\cite{SM} is now established
 \footnote{More rigorously,
 the final establishment will be done by LEPII experiment.}.
These interactions are described
 as the gauge interaction of the symmetry
 $SU(2)_L \times U(1)_Y$ (electroweak symmetry)
 which contains four massless gauge bosons.
This gauge symmetry
 is spontaneously broken to the electromagnetic gauge symmetry $U(1)_{em}$,
 and three gauge bosons (weak gauge bosons) get their mass
 by the Higgs mechanism\cite{Higgs}.
The mass of these gauge bosons
 explains why the weak interaction is short range.
The electroweak symmetry forbids the mass of quarks and leptons.
Quarks and leptons
 cannot be massive till the electroweak symmetry breaking is achieved.
Therefore, the mechanism of the mass generation
 must tightly be related with the mechanism
 of the electroweak symmetry breaking.

In the standard model,
 electroweak symmetry is spontaneously broken
 by the vacuum expectation value of elementary Higgs field.
The magnitude of vacuum expectation value
 decides the masses of weak gauge bosons.
The masses of fermions (quarks and leptons) are generated
 through the Yukawa interactions between the fermion fields and Higgs field.
The mass mixing among the quarks with the same electric charge
 is also generated.
This mixing
 is the origin of Cabibbo-Kobayashi-Maskawa (CKM) mixing angles
 and CP violation in the weak interaction\cite{Kobayashi-Maskawa}.

This mechanism of electroweak symmetry breaking and fermion mass generation
 contains many arbitrary parameters.
The masses of fermions, CKM mixing angles, and CP violating phase
 cannot be predicted in the framework of the standard model.
This suggests that
 there are some unknown physics which explain the origin of fermion masses
 as well as the origin of CKM mixing and CP violation in the weak interaction.
We must make some efforts to find this new physics,
 though it is a difficult problem.

The approaches to solve this problem may be classified into two categories.
One is the solution in the grand unified theories (GUTs)\cite{GUTs}.
The Yukawa couplings as well as the gauge couplings
 must be unified at some very high energy scale in this theories.
The variety of the Yukawa couplings at the low energy
 is due to the complex particle contents.
The interactions originated by the physics at the Plank scale
 may also be important\cite{Dimopoulos-Hall-Raby}.
This type of solutions
 {\it kinematically} explain the variety of fermion masses.

Another solution {\it dynamically} explains the fermion masses.
The extended technicolor theory\cite{ETC} is contained in this category.
The theory is based on the technicolor theory\cite{TC}
 in which the electroweak symmetry is dynamically broken
 by the technifermion condensates
 due to the strong coupling technicolor interaction.
In the technicolor theory
 the interaction which correspond to the Yukawa interactions
 are the effective four-fermion interactions
 between the ordinary fermions and technifermions.
In the extended technicolor theory,
 the effective four-fermion interactions are dynamically generated.
The variety of the fermion masses
 is originated in the dynamics of extended technicolor interaction.
The top condensation model\cite{Top}
 and composite model of quarks and leptons
 are also contained in this category.

To our regret, both approaches are not completely successful.
While the approach with GUTs
 can describe the fermion masses, CKM mixing angles, and CP violation,
 the potential of vacuum is very complicated one,
 and many fine-tuned parameters should be contained in it.
We must consider the physics at GUT scale and above
 to answer the question why the potential is so complicated.

The dynamical approach also is not successful.
One of the reason is
 that the non-perturbative treatment of the strong coupling gauge theory
 is quite difficult.
Since our knowledge of the dynamical spontaneous symmetry breaking
 is very limited, we must put some postulates to make models.
Other reason is that
 the model tends to become very complicated one
 to explain the complicated fermion mass spectrum.
Many people are disgusted by complicated model with many postulations.

There is no definite reason to select which approach is better,
 though recent precision experiments at LEP
 favor the GUT approach (with supersymmetry) but not dynamical approach.
It is interesting that
 the dynamical approach predicts new rich physics
 (new particles, new interactions) just above the present accessible energy.
It is now the time to extensively study this approach,
 since many intended experiments can be accessible to this new physics.

It is very important
 to estimate the non-perturbative behavior (dynamics) of the gauge theory
 for the dynamical approach.
In this thesis we treat an interesting gauge theory
 which can be applied as the dynamics of the extended technicolor theory.
It is called tumbling gauge theory\cite{tumbling}
 in which many dynamical gauge symmetry breakings
 with different energy scales are expected.
Natural existence of such many different mass scales
 has been expected as the source of many different fermion masses.
We formulate this theory
 by using the effective action in one-gauge-boson-exchange approximation,
 and estimate the strength (energy scale)
 of the effective four fermion interactions
 due to the massive gauge boson exchange.
The strength is directly related with the fermion masses.

The strength is also important
 to solve the flavor-changing neutral current (FCNC) problem.
In the naive model,
 extended technicolor theory generates large FCNC interactions.
The extended technicolor gauge bosons and the pseudo-Nambu-Goldstone bosons
 cause too large $K^0$-${\bar K}^0$ mixing, for example.
We may avoid this problem
 by devising the gauge group and the representation of fermions,
 but the model should become very complicated one.
This problem can simply be solved
 if the technicolor dynamics is the one with large $\gamma_m$,
 where $\gamma_m$ is the anomalous dimension of composite operator ${\bar T}T$
 composed by the technifermion field $T$\cite{large}.

Two types of large $\gamma_m$ dynamics have been proposed.
One is the gauge theory with slowly running coupling
 (so called walking gauge theory)\cite{walking}
 and another is the gauge theory with strong four-fermion interaction
 \cite{strong-4F}.
Since many effective four-fermion interactions
 are induced in tumbling gauge theory,
 the second type of large $\gamma_m$ dynamics is likely be realized.
But we show that
 the effective four-fermion interaction
 is too weak to realize sufficiently large $\gamma_m$.
This result is important
 for the model building of the extended technicolor theory.

We see that
 the massive gauge bosons can be neglected,
 when we consider the low energy dynamics.
The critical gauge coupling of the fermion pair condensation
 and the anomalous dimension $\gamma_m$ do not change,
 whether the massive gauge bosons are considered or not.
The massive fermions which get their masses in the process of tumbling
 can also be neglected, when we consider the low energy dynamics.
The critical gauge coupling
 of the pair condensation channel with the massive fermions
 is too large to be formed the condensate.
Although such decoupling of the heavy particles
 has already been proved based on the the perturbation theory,
 we non-perturbatively show that in the tumbling gauge theory
 by using our formalism in the one-gauge-boson exchange approximation.

This thesis is developed as follows.
In the next chapter,
 we review the fundamental properties of realization of symmetry in nature.
The fundamentals
 about spontaneous symmetry breaking and Higgs mechanism is reviewed.
In chapter 3,
 the fundamental properties of dynamical symmetry breaking is described.
Various techniques of the non-perturbative treatment of dynamics are reviewed.
In chapter 4 and 5,
 the technicolor models and its phenomenology are reviewed, respectively.
The relation with present precision experiments is also considered.
In chapter 6, our formalism of tumbling gauge theory is developed.
In chapter 7,
 the strength of effective four-fermion interaction is estimated,
 and the possibility of the large anomalous dimension dynamics is discussed.
The non-perturbative decoupling is also discussed.
Chapter 8 is devoted to the conclusion.
\chapter{Realization of Symmetry}

The symmetry is an important concept in the elementary particle physics.
Many hadrons composed by the quarks
 have been classified and understood by virtue of the symmetry in QCD.
We know that
 some decay processes of hadrons are forbidden by the conservation law,
 and the conservation law is directly connected with the symmetry.
To find the symmetry in nature has been a very important task
 to understand the particle spectrum and interactions.

The symmetry
 is described as the invariance of Lagrangian
 under the transformation
 \footnote{More accurately,
           we should consider the action instead of the Lagrangian.
           But here, we consider the Lagrangian for simplicity.}.
For example, the Lagrangian of complex-valued scalar field
\begin{equation}
 {\cal L} = (\partial_\mu \phi)^{\dag} (\partial^\mu \phi)
          - m^2 \phi^{\dag} \phi - \lambda (\phi^{\dag} \phi)^2
\label{example-global}
\end{equation}
 has a global symmetry,
 since this is invariant under the global transformation
\begin{equation}
 \phi \ \longrightarrow \ \phi' = e^{i \theta} \phi,
\label{example-global-transformation}
\end{equation}
 where $\theta$ is the parameter of transformation.
We can extend this symmetry to the local one
 by introducing gauge field $A_\mu$.
The Lagrangian
\begin{equation}
 {\cal L} = (D_\mu \phi)^{\dag} (D^\mu \phi)
          - m^2 \phi^{\dag} \phi - \lambda (\phi^{\dag} \phi)^2
          -{ 1 \over 4 } F_{\mu\nu} F^{\mu\nu}
\label{example-local}
\end{equation}
 is invariant under the local transformation
\begin{eqnarray}
 \phi &\longrightarrow& \phi' = e^{i \theta(x)} \phi,
\nonumber\\
 A_\mu &\longrightarrow& A_\mu' = A_\mu + \partial_\mu \theta(x),
\label{example-local-transformation}
\end{eqnarray}
 where
\begin{eqnarray}
 D^\mu \phi &=& \partial^\mu \phi - ig A^\mu \phi,
\\
 F^{\mu\nu} &=& \partial^\mu A^\nu - \partial^\nu A^\mu,
\end{eqnarray}
 and $g$ is the coupling constant of the gauge interaction.

The current and charge of the symmetry are defined as follows.
Consider the general fields $\Phi_i$
 and the (infinitesimal) transformation
\begin{equation}
 \Phi_i \ \longrightarrow \ \Phi'_i = \Phi_i + \theta^a (\delta \Phi)^a_i,
\end{equation}
 where index $i$ denotes the flavor of the fields
 and $a$ is the index of the transformation.
The current is defined as
\begin{equation}
 J_\mu^a(x) \equiv
            - {{\partial {\cal L}} \over {\partial {(\partial_\mu \Phi_i)}}}
              (\delta \Phi)^a_i,
\end{equation}
 and the corresponding charge is defined as
\begin{equation}
 Q^a \equiv - \int J_0^a(x) d^3x
\end{equation}
 by using this current.

We can show that
 the divergence of the currents are the variation of the Lagrangian
 by using Eular-Lagrange equation, namely
\begin{equation}
 \partial^\mu J_\mu^a = - (\delta {\cal L})^a,
\end{equation}
 where the Lagrangian is transformed as
\begin{equation}
 {\cal L} \rightarrow {\cal L} + (\delta {\cal L})^a.
\end{equation}
Therefore,
 if the Lagrangian is invariant under the transformation
 \footnote{For the invariance of the action,
           the variation of the Lagrangian do not always have to vanish.
           The form of the total divergence,
           $(\delta {\cal L})^a = \partial^\mu X^a_\mu$, is also acceptable.},
 the divergence of the current vanishes.
The current corresponding to the symmetry transformation
 is called Noether current.
The vanishing of the divergence of the current
 suggests the conservation of corresponding charge,
 because the charge is independent of time,
\begin{eqnarray}
 {d \over {dt}} Q^a &=& - \int {d \over {dt}} J_0^a d^3x
\nonumber\\
                    &=&  \int \nabla \cdot \mbox{\boldmath j}^a d^3x = 0.
\end{eqnarray}
Here we assume as the boundary condition that
 the fields and its derivatives vanish at infinity.

The charges generate the transformation
 when we consider the fields as operators in canonical quantization,
\begin{equation}
 \left[ Q^a, \Phi_i \right] = {1 \over i} (\delta \Phi)^a_i.
\end{equation}
The charge of the state is defined
 as the eigenvalue of the charge operator
 with respect to the corresponding state vector.
The transformations are classified by the algebra of the charge operators.
If the charge operators, for instance, satisfy the algebra
\begin{equation}
 \left[ Q^a, Q^b \right] = i \epsilon^{abc} Q^c,
\end{equation}
 the transformation forms the group $SU(2)$,
 where $a$, $b$, and $c$ run over $1$, $2$, and $3$,
 and $\epsilon^{abc}$ is the Levi-Civita symbol.
Therefore,
 the symmetry can be classified by the group theory.
The charges form the generators of the group.

The symmetry of the system (Lagrangian) can be realized in nature
 in two ways: Wigner phase or Nambu-Goldstone phase.
The generators of the symmetry annihilate the vacuum state in the Wigner phase,
 but not in Nambu-Goldstone phase. Namely,
\begin{eqnarray}
 \ Q \vert 0 \rangle = 0 \  & \qquad\quad & \mbox{Wigner phase},
\label{Wigner}
\\
 "Q \vert 0 \rangle \neq 0" & \qquad\quad & \mbox{Nambu-Goldstone phase},
\label{Nambu-Goldstone}
\end{eqnarray}
 where the vacuum state is defined
 by the annihilation operators of the asymptotic fields of the theory.
The double quotations in eq.(\ref{Nambu-Goldstone}) means that
 the formula is just a symbolic one.

Wigner phase is the naive realization of the symmetry.
In this phase the charge of the state is conserved in the physical processes.
Since the Hamiltonian ${\cal H}$ commutes with the charge,
\begin{eqnarray}
 0 &=& \langle \alpha \vert \left[ Q, {\cal H} \right] \vert \beta \rangle
\nonumber\\
   &=& \left( Q_\alpha - Q_\beta \right)
       \langle \alpha \vert {\cal H} \vert \beta \rangle,
\end{eqnarray}
 where $\vert \alpha \rangle$ and $\vert \beta \rangle$ are the eigenstates
 of the charge $Q$ with eigenvalues $Q_\alpha$ and $Q_\beta$, respectively.
This means that the transition amplitude do not vanish
 only if the charges of initial and finial state is equal to each order.

But it is not the case in the Nambu-Goldstone phase.
If there exists at least an operator $\Phi$
 which satisfies the condition
\begin{eqnarray}
 && \left[ Q, \Phi \right] = \delta \Phi \neq 0,
\nonumber\\
 && \langle 0 \vert \delta \Phi \vert 0 \rangle \neq 0.
\end{eqnarray}
 the symmetry is realized as Nambu-Goldstone phase.
The generator $Q$ does not annihilate the vacuum state,
 and then it is not the well-defined one.
Since $Q$ commutes with the energy-momentum operator,
 the state $Q \vert 0 \rangle$ must proportional
 to the vacuum state $\vert 0 \rangle$
 \footnote{Here, we postulate that the vacuum state
           is an unique zero energy state.
           But the result is not changed,
           if the other independent zero energy states exist.}.
Therefore,
\begin{eqnarray}
 c &=& \langle 0 \vert Q \vert 0 \rangle
\nonumber\\
   &=& - \int \langle 0 \vert J_0(x) \vert 0 \rangle d^3x
\nonumber\\
   &=& - \int \langle 0 \vert J_0(0) \vert 0 \rangle d^3x = 0,
\end{eqnarray}
 where $c$ is a constant.
In the last equality, we use the fact that
 there is no constant Lorentz vector in the theory.
If the generator $Q$ is well-defined one, then the constant $c$ must be zero.
Namely,
 if the generator $Q$ does not annihilate the vacuum state,
 then the generator is not well-defined one.
In this phase the symmetry is spontaneously broken,
 and the charge of the state as the eigenvalue of the generators
 cannot be defined.
The generator $Q$ is called the broken generator.

The symmetry is realized
 in the different way from the charge conservation
 in the Wigner phase.
The massless Nambu-Goldstone bosons
 corresponding to the broken generators exist in particle spectrum.
Nambu-Goldstone theorem says that if the conditions
\begin{enumerate}
 \item Existence of the translational invariance and Lorentz covariance.
 \item Existence of the conserved current $j_\mu$.
 \item Existence of the operator $\Phi$ which satisfies
       $\langle 0 \vert \left[ Q, \Phi \right] \vert 0 \rangle
        = {1 \over i} \langle 0 \vert \delta \Phi \vert 0 \rangle \neq 0$
       (finite).
\end{enumerate}
 are satisfied,
 the massless Nambu-Goldstone boson exists and it couples to the current.

The proof is the following.
Because of the current conservation,
\begin{equation}
 \langle 0 \vert \left[ Q, \Phi(0) \right] \vert 0 \rangle
 = {1 \over i} \langle 0 \vert \delta \Phi(0) \vert 0 \rangle
\label{Condition-3}
\end{equation}
 is independent of $x_0$, where $Q$ is defined as
\begin{equation}
 Q = - \int j_0(x) d^3x.
\end{equation}
The left-hand side of eq.(\ref{Condition-3}) can be expressed as
\begin{eqnarray}
 \langle 0 \vert \left[ Q, \Phi(0) \right] \vert 0 \rangle
 &=& - \int d^3x \sum_n
       \left\{ \langle 0 \vert j_0(x) \vert n \rangle
               \langle n \vert \Phi(0) \vert 0 \rangle
             - \langle 0 \vert \Phi(0) \vert n \rangle
               \langle n \vert j_0(x) \vert 0 \rangle \right\}
\nonumber\\
 &=& - \int d^3x \sum_n
       \Big\{
               \langle 0 \vert j_0(0) \vert n \rangle
               \langle n \vert \Phi(0) \vert 0 \rangle
               e^{-ip_nx}
\nonumber\\
 & & \qquad\qquad\qquad\qquad\qquad
             - \langle 0 \vert \Phi(0) \vert n \rangle
               \langle n \vert j_0(0) \vert 0 \rangle
               e^{ip_nx}
       \Big\}
\nonumber\\
 &=& - \sum_n (2\pi)^3 \delta^3(p_n)
       \Big\{
               \langle 0 \vert j_0(0) \vert n \rangle
               \langle n \vert \Phi(0) \vert 0 \rangle
               e^{-i E_n x^0}
\nonumber\\
 & & \qquad\qquad\qquad\qquad\qquad
             - \langle 0 \vert \Phi(0) \vert n \rangle
               \langle n \vert j_0(0) \vert 0 \rangle
               e^{i E_n x^0}
       \Big\}
\nonumber\\
 &=& - \sum_{n_1}
              \int {{d^3p_{n_1}} \over {2 E_{n_1}}} \delta^3(p_{n_1})
       \Big\{
               \langle 0 \vert j_0(0) \vert n_1(p_{n_1}) \rangle
               \langle n_1(p_{n_1}) \vert \Phi(0) \vert 0 \rangle
               e^{-i E_{n_1} x^0}
\nonumber\\
 & & \qquad\qquad\qquad\qquad\qquad
             - \langle 0 \vert \Phi(0) \vert n_1(p_{n_1}) \rangle
               \langle n_1(p_{n_1}) \vert j_0(0) \vert 0 \rangle
               e^{i E_{n_1} x^0}
       \Big\}
\nonumber\\
 & & - \sum_{n'} (2\pi)^3 \delta^3(p_n')
       \Big\{
               \langle 0 \vert j_0(0) \vert n' \rangle
               \langle n' \vert \Phi(0) \vert 0 \rangle
               e^{-i E_{n'} x^0}
\nonumber\\
 & & \qquad\qquad\qquad\qquad\qquad
             - \langle 0 \vert \Phi(0) \vert n' \rangle
               \langle n' \vert j_0(0) \vert 0 \rangle
               e^{i E_{n'} x^0}
       \Big\},
\end{eqnarray}
 where we use the translational invariance.
In the last equality,
 the sum of states is divided into two parts: one particle states and others.
In addition,
 we recover the invariant measure of the integration
 in the part of the one-particle states.
Because of the delta function and the independence of $x^0$,
 there must exist a state of $E=p^i=0$ except for the vacuum state.
{}From the spectral condition of the field theory,
 this means the existence of massless one particle state.
The state $\vert {\rm NG}(p) \rangle$, Nambu-Goldstone boson state,
 must satisfy the condition
\begin{eqnarray}
 \langle 0 \vert j_\mu(0) \vert {\rm NG}(p) \rangle &\propto& p_\mu \neq 0,
\\
 \langle {\rm NG}(p) \vert \Phi(0) \vert 0 \rangle &\neq& 0,
\end{eqnarray}
 where we use the Lorentz covariance
 and the finiteness of the vacuum expectation value.
This equation means that
 the Nambu-Goldstone boson couples to the current $j_\mu$.
It is apparent that
 the number of Nambu-Goldstone bosons
 equals to the number of independent broken generators.

Consider an example of the Lagrangian
 with spontaneous symmetry breaking.
The Lagrangian is the same as eq.(\ref{example-global})
 but the sign of the mass term is different.
\begin{equation}
 {\cal L} = (\partial_\mu \phi)^{\dag} (\partial^\mu \phi)
          + m^2 \phi^{\dag} \phi - \lambda (\phi^{\dag} \phi)^2,
\end{equation}
This is invariant under the transformation
 of eq.(\ref{example-global-transformation}).
We introduce the two real fields as $\phi=(\sigma + i \pi)/\sqrt2$.
The transformation properties of the fields $\sigma$ and $\pi$ are
\begin{eqnarray}
 \left[ Q, \sigma \right] &=& i \pi,
\nonumber\\
 \left[ Q, \pi    \right] &=& -i \sigma.
\end{eqnarray}
The potential of the fields $\sigma$ and $\pi$ is
\begin{equation}
 V(\sigma, \pi) = - {1 \over 2} m^2 ( \sigma^2 + \pi^2 )
                  + {\lambda \over 4} ( \sigma^2 + \pi^2 )^2,
\end{equation}
 and the minimum point is at $\sigma=\sqrt{m^2/\lambda}$, $\pi=0$
 (We can always set $\pi=0$ by using the transformation
 of eq.(\ref{example-global-transformation})).
This value of $\sigma$ is translated as the vacuum expectation value
\begin{equation}
 \langle \sigma \rangle \equiv v = \sqrt{{{m^2} \over \lambda}}.
\end{equation}
The physical field $\sigma_{phys}$ which has positive squared mass
 is given by the shift $\sigma_{phys} = \sigma - v$.
The mass of $\pi$ vanishes by this shift.
The current of symmetry is written as
\begin{equation}
 j_\mu = - v \partial_\mu \pi
         + \left( \pi \partial_\mu \sigma_{phys}
                - \sigma_{phys} \partial_\mu \pi \right)
\end{equation}
 by this physical field,
 and we can see that $\pi$ couples to the current as
\begin{eqnarray}
 \langle 0 \vert j_\mu(0) \vert \pi(p) \rangle = i p_\mu v \neq 0.
\end{eqnarray}
 in the lowest order of the perturbation
 (Consider the fields as the asymptotic ones.).
Therefore, $\pi$ is the field of Nambu-Goldstone boson.

If the local symmetry is spontaneously broken,
 the corresponding gauge boson becomes massive.
This is called Higgs mechanism.
Consider the Lagrangian of eq.(\ref{example-local})
 but with negative squared mass.
\begin{equation}
 {\cal L} = (D_\mu \phi)^{\dag} (D^\mu \phi)
          + m^2 \phi^{\dag} \phi - \lambda (\phi^{\dag} \phi)^2
          -{ 1 \over 4 } F_{\mu\nu} F^{\mu\nu}.
\end{equation}
The Lagrangian is invariant
 under the transformation of eq.(\ref{example-local-transformation}),
 but this symmetry is spontaneously broken
 by the vacuum expectation value of $\sigma$
 just the same as the above example.
By the shifting $\sigma_{phys} = \sigma - v$,
 the mass term of gauge boson $A_\mu$ is generated
 from the first term of the Lagrangian.
The mass is controlled
 by the gauge coupling and the vacuum expectation value of $\sigma$:
 $m_A^2 = g^2 v^2$.

The complete calculation for the gauge boson mass
 in the second order of the gauge coupling is as follows.
We rewrite the Lagrangian as
\begin{eqnarray}
 {\cal L} &=& {\cal L}_0 + g A^\mu j_\mu + g^2 \phi^{\dag} \phi A_\mu A^\mu,
\nonumber\\
 {\cal L}_0 &=& (\partial_\mu \phi)^{\dag} (\partial^\mu \phi)
                + m^2 \phi^{\dag} \phi - \lambda (\phi^{\dag} \phi)^2
                - {1 \over 4} F_{\mu\nu} F^{\mu\nu}.
\end{eqnarray}
The generating functional is
\begin{eqnarray}
 Z &=& \int {\cal D}A {\cal D}\phi {\cal D}\phi^{\dag}
       \exp i \int d^4x
             \left\{
              {\cal L}_0 + g A^\mu j_\mu + g^2 \phi^{\dag} \phi A_\mu A^\mu
             \right\}
\nonumber\\
   &\simeq& \int {\cal D}A {\cal D}\phi {\cal D}\phi^{\dag}
         \left[ 1 + i g^2 \int_x
                         \langle {\rm T} \phi^{\dag}(x) \phi(x) \rangle_\phi
                         A_\mu(x) A^\mu(x) \right.
\nonumber\\
   & & \qquad\qquad
         \left.   - {1 \over 2} g^2 \int_{x,y}
                         A^\mu(x)
                         \langle {\rm T} j_\mu(x) j_\nu(y) \rangle_\phi
                         A^\nu(y)
         \right]
    \exp i \int d^4x {\cal L}_0
\nonumber\\
   &\simeq& \int {\cal D}A {\cal D}\phi {\cal D}\phi^{\dag}
       \exp
        \left[ i \int d^4x {\cal L}_0
               + i g^2 \int_x
                       \langle {\rm T} \phi^{\dag}(x) \phi(x) \rangle_{c,\phi}
                       A_\mu(x) A^\mu(x) \right.
\nonumber\\
   & & \qquad\qquad
        \left. - {1 \over 2} g^2 \int_{x,y}
                       A^\mu(x)
                       \langle {\rm T} j_\mu(x) j_\nu(y) \rangle_{c,\phi}
                       A^\nu(y)
        \right],
\end{eqnarray}
 where $\langle \dots \rangle_{c, \phi}$ means the connected Green function
 containing only the loop effect due to the field $\phi$.
We used the relation
\begin{equation}
 \langle T {\cal O}(x) {\cal O}'(y) \rangle
 ={
   { \int {\cal D}A {\cal D}\phi {\cal D}\phi^{\dag}
          {\cal O}(x) {\cal O}'(y) e^{i \int d^4x {\cal L}_0}}
   \over
   { \int {\cal D}A {\cal D}\phi {\cal D}\phi^{\dag}
          e^{i \int d^4x {\cal L}_0}}
  }
\end{equation}
 for any operators ${\cal O}(x)$ and ${\cal O}'(y)$.
Therefore,
 the effective action for the gauge field in the second order of $g$ is
\begin{eqnarray}
 iS_{eff} &=& i \int_{x,y} {1 \over 2} A^\mu(x) \delta^4(x-y)
                \big( g_{\mu\nu} \Box
                       - (1-{1 \over \alpha}) \partial_\mu \partial_\nu
                \big) A^\nu(y)
\nonumber\\
          &+& i g^2 \int_{x}
                \langle {\rm T} \phi^{\dag}(x) \phi(x) \rangle_{c,\phi}
                A^\mu(x) A_\mu(x)
\nonumber\\
          &-& {1 \over 2} g^2 \int_{x,y}
                A^\mu(x)
                \langle {\rm T} j_\mu(x) j_\nu(y) \rangle_{c,\phi}
                A^\nu(y)
\nonumber\\
          &+& \dots,
\end{eqnarray}
 where $\alpha$ is a gauge parameter.
Therefore, the propagator of gauge boson is
\begin{equation}
 D_{\mu\nu}(k)
  = {1 \over {k^2 - g^2 v^2}}
    \left\{ g_{\mu\nu} - (1-\alpha) {{k_\mu k_\nu} \over {k^2}}
            - \alpha {{g^2 v^2} \over {k^2}} {{k_\mu k_\nu} \over {k^2}}
    \right\},
\end{equation}
 where we use
\begin{equation}
 \langle {\rm T} \phi^{\dag}(x) \phi(x) \rangle_{c,\phi}
  = {{v^2} \over 2} + \mbox{loop diagrams}
\label{vacuum-expectation}
\end{equation}
 and
\begin{equation}
 \mbox{F.T.} \langle {\rm T} j_\mu(x) j_\nu(y) \rangle_{c,\phi}
   =  i v^2 {{k_\mu k_\nu} \over {k^2}} + \mbox{loop diagrams},
\label{current-correlation}
\end{equation}
 and neglect the loop effects.
The propagator has a pole at $k^2 = m_A^2 = g^2 v^2$.
It is important that
 both the direct contribution of vacuum expectation value
 (eq.(\ref{vacuum-expectation}))
 and the current-current correlation (eq.(\ref{current-correlation}))
 contribute to the propagator.
In the next chapter
 we treat the spontaneous symmetry breaking
 without the vacuum expectation value of {\it elementary} field.
In that case,
 the contribution of current-current correlation is essential.
We can see that
 if the massless particles couple with the currents
 (massless pole $1/k^2$ in eq.(\ref{current-correlation})),
 the gauge bosons which couple to the current become massive
 (Schwinger mechanism).
\chapter{Dynamical Symmetry Breaking}

The symmetry of Lagrangian is spontaneously broken
 by the non-zero vacuum expectation value of the operator
 which is not invariant under the transformation of the symmetry.
The operator is not always be the elementary field,
 but also the operator composed by the elementary fields
 (composite operator).
The spontaneous symmetry breaking
 which is triggered by the vacuum expectation value of the composite operator
 is called dynamical symmetry breaking.

It is believed that
 the dynamical symmetry breaking is realized
 in the theory of strong interaction (quantum chromodynamics, QCD).
The Lagrangian of the theory is
\begin{equation}
 {\cal L}_{QCD}
          = {\bar \psi}_i i \not\!D \psi^i
          - {\bar \psi}_i m_i \psi^i
          - {1 \over 2} {\rm tr} \left( F_{\mu\nu} F^{\mu\nu} \right),
\end{equation}
 where
\begin{eqnarray}
 D_\mu \psi^i &=& \partial_\mu \psi^i - i g G_\mu \psi^i,
\\
 F_{\mu\nu} &=& \partial_\mu G_\nu - \partial_\nu G_\mu
                - i g [ G_\mu, G_\nu ],
\\
 G_\mu &=& G^a_\mu {{\lambda^a} \over 2}.
\end{eqnarray}
Here,
 $G^a_\mu$ denote the gluon fields
 ($a=1,2, \cdots, 8$, and $\lambda^a$ are the Gell-Mann matrixes.)
 which couple with the quark fields $\psi^i$ ($i=1,2, \cdots, N_f$)
 with coupling constant $g$.
If the quark masses $m_i$ vanish,
 the global chiral symmetry $SU(N_f)_L \times SU(N_f)_R$ exists.
The transformation of the symmetry is
\begin{eqnarray}
 \psi_L^i &\longrightarrow& U_L{}^i{}_j \psi_L^j
 \qquad \mbox{$U_L \in SU(N_f)_L$},
\\
 \psi_R^i &\longrightarrow& U_R{}^i{}_j \psi_R^j
 \qquad \mbox{$U_R \in SU(N_f)_R$}.
\end{eqnarray}
As long as
 we consider the light
 (compared with the typical scale of QCD dynamics) quarks,
 u-quark and d-quark (and s-quark),
 this is a good approximate symmetry.
It is believed that
 this chiral symmetry is dynamically broken to the symmetry $SU(N_f)_V$
 by the quark pair condensation
 $\langle {\bar \psi}_i \psi^i \rangle \neq 0$
 due to the QCD dynamics.
The composite operator ${\bar \psi}_i \psi^i$
 is not invariant under the chiral symmetry
 but invariant under the $SU(N_f)_V$ transformation
\begin{equation}
 \psi^i \longrightarrow U^i{}_j \psi^j
 \qquad \mbox{$U \in SU(N_f)_V$}.
\end{equation}

If we consider u and d quarks (u,d, and s quarks) ($N_f=2$ $(3)$)
 and neglect their masses (chiral limit),
 there should be three (eight) pseudo-scalar massless Nambu-Goldstone bosons.
The pseudo-scalar mesons, $\pi^0$ and $\pi^\pm$
 ($\pi^0$, $\pi^\pm$, $K^0$, ${\bar K}^0$, $K^\pm$, and ``$\eta_8$''),
 can be identified as the Nambu-Goldstone bosons.
The masses of these mesons are understood
 as the effect of the explicit chiral symmetry breaking
 due to the quark masses.
The effect can be perturbatively included,
 if the masses of the quarks are small enough
 (chiral perturbation)\cite{Gasser-Leutwyler}.
The interactions between these mesons at low energy
 (low energy $\pi$-$\pi$ scattering, for example)
 are described as the interactions of the Nambu-Goldstone bosons
 which are determined only by the symmetry (low energy theorems).
The correction due to the explicit breaking is perturbatively considered.
The fact that the picture of the chiral perturbation
 can satisfactory describe the low energy phenomenology of QCD
 is the reason why we believe the existence of the pair condensation in QCD.

In next section,
 we try to calculate the value of the pair condensation in QCD
 by using Schwinger-Dyson equation\cite{Schwinger-Dyson}
 and operator product expansion\cite{OPE}.
The Schwinger-Dyson equation (self-consistent equation for quark propagator)
 is solved in ladder approximation,
 and the non-trivial solution is converted into the pair condensation
 $\langle {\bar \psi}_i \psi^i \rangle$
 by using the operator product expansion.
In section 2,
 we calculate the decay constant of Nambu-Goldstone boson
 which corresponds to the vacuum expectation value
 of the elementary scalar field in the previous chapter.
The Pagels-Stokar formula\cite{Pagels-Stokar},
 which gives decay constant as a functional of the quark propagator,
 is introduced there.
In section 3,
 we discuss the Cornwall-Jackiw-Tomboulis effective action\cite{CJT}
 by which the general treatment of spontaneous symmetry breaking is possible.

\section{Schwinger-Dyson equation and quark condensation}

We consider the Schwinger-Dyson equation\cite{Schwinger-Dyson}
 for the quark and gluon propagators.
We want to solve the equation, and get the full quark propagator.
The equation is
\begin{eqnarray}
\lefteqn{
 i S^{-1}(x-y)^i{}_j \delta^\alpha_\beta
  = \left( i \not\!\partial - m_i \right)
    \delta^i_j \delta^\alpha_\beta \delta^4 (x-y)
}
\label{quark}
\\
&&
 + g^2 \int d^4x_1 d^4x_2 \gamma^\mu
                   \left( T^a \right)^\alpha{}_\gamma
                   S(x-x_1)^i{}_k
                   \Gamma^\nu (x_2,x_1,y)^k{}_j
                   \left( T^a \right)^\gamma{}_\beta
                   D_{\mu\nu}(x-x_2),
\nonumber
\end{eqnarray}
\begin{eqnarray}
 D^{-1}_{\mu\nu}(x-y) \delta^{ab}
  &=& - i \left\{ \Box g_{\mu\nu}
                - \left( 1-{1 \over \alpha} \right) \partial_\mu \partial_\nu
          \right\}
    \delta^{ab} \delta^4 (x-y)
\\
&&
 + i g^2 \int d^4x_1 d^4x_2
    {\rm tr} \left[
             \gamma_\mu T^a S(x-x_1) \Gamma_\nu (y,x_1,x_2) T^b S(x_2-x)
             \right],
\nonumber
\label{gluon}
\end{eqnarray}
 where $S$ and $D$ are the full propagators (bare) of quarks and gluons
\begin{eqnarray}
 S(x-y)^i{}_j \delta^\alpha_\beta
  &\equiv& \left.
           - {{\delta^2 \ln Z}
              \over
              {\delta i {\bar \eta}^{i\alpha}(x) \delta i \eta_{j\beta}(y)}}
           \right|_{J, \ \eta, \ {\bar \eta} = 0},
\\
 D_{\mu\nu}(x-y) \delta^{ab}
  &\equiv& \left.
            {{\delta^2 \ln Z}
            \over
            {\delta i J^{\mu a}(x) \delta i J^{\nu b}(y)}}
           \right|_{J, \ \eta, \ {\bar \eta} = 0},
\end{eqnarray}
(The left-derivative of the Grassmann variable is understood.)
 and $\Gamma$ is a full gluon-quark-quark vertex (bare)
\begin{eqnarray}
\lefteqn{
 \left.
 {{\delta^3 \ln Z}
  \over
  {\delta i J^{\mu a}(x)
   \delta i {\bar \eta}^{i\alpha} \delta i \eta_{j\beta}}}
 \right|_{J, \ \eta, \ {\bar \eta} = 0}
}
\\
&&
 \equiv -
   \int d^4x' d^4y' d^4z'
    D_{\mu\nu}(x-z')
    S(x-x')^i{}_k{}
    g \Gamma^\nu (z',x',y')^k{}_l \left( T^a \right)^\beta{}_\alpha
    S(y'-y)^l{}_j.
\nonumber
\end{eqnarray}
The matrixes $T^a$ are the generators of $SU(3)_c$ ($T^a = \lambda^a/2$).
The indexes $i$, $j$, $k$ denote flavor,
 $\alpha$, $\beta$, $\gamma$ denote the color of quarks,
 and $a$, $b$ denote the color of gluons.
We are assuming that the gauge symmetry $SU(3)_c$ is not broken.
The propagators of the quarks and gluons
 are set to diagonal in the color space.
The functional $Z$ is the generating functional of the Green functions
\begin{equation}
 Z[J, \eta, {\bar \eta}]
   = \int {\cal D}(G, \psi, {\bar \psi})
     \exp i S[G, \psi, {\bar \psi}, J, \eta, {\bar \eta}],
\end{equation}
\begin{equation}
 S[G, \psi, {\bar \psi}, J, \eta, {\bar \eta}]
  = \int d^4x \left[ {\cal L}_{QCD}
                + J_\mu G^\mu + {\bar \eta} \psi + {\bar \psi} \eta
              \right],
\end{equation}
 where $J_\mu$, $\eta$, and ${\bar \eta}$ are the source fields.

We can derive eq.(\ref{quark}) as follows.
The equation
\begin{equation}
 \int {\cal D}(G, \psi, {\bar \psi})
  {{\delta S[G, \psi, {\bar \psi}, J, \eta, {\bar \eta}]}
   \over
   {\delta {\bar \psi}^{i\alpha}(x)}}
  \exp i S[G, \psi, {\bar \psi}, J, \eta, {\bar \eta}] = 0
\end{equation}
 is satisfied as the field equation of quarks.
This equation means that
 the expectation value of the derivative of the action by quarks
 should vanish.
It is rewritten as
\begin{equation}
 \left[ \eta(x)^{i\alpha}
      + \left( \left( i \not\!\partial_x - m_i \right) \delta^\alpha_\gamma
             + g \gamma^\mu \left( T^a \right)^\alpha{}_\gamma
               { \delta \over {\delta i J^{a\mu} (x)}}
        \right) { \delta \over {\delta i {\bar \eta}^{i\gamma} (x)}}
 \right] Z[j, \eta, {\bar \eta}] = 0.
\end{equation}
Differentiating with respect to $\eta^{j\beta}(y)$ and dividing by $Z$ give
\begin{eqnarray}
\lefteqn{
 \delta^4 (x-y) \delta^i_j \delta^\alpha_\beta
}
\\
&&
 + \left( i \not\!\partial_x - m_i \right) \delta^\alpha_\gamma i {1 \over Z}
   {{\delta^2 Z}
    \over
    { \delta {\bar \eta}^{i\gamma}(x) \delta \eta_{j\beta}(y)}}
 + g \gamma^\mu \left( T^a \right)^\alpha{}_\gamma
                {1 \over Z} { \delta \over {\delta J^{a\mu} (x)}}
   {{\delta^2 Z}
    \over
    { \delta {\bar \eta}^{i\gamma}(x) \delta \eta_{j\beta}(y)}}
 = 0,
\nonumber
\end{eqnarray}
\begin{eqnarray}
\lefteqn{
 \delta^4 (x-y) \delta^i_j \delta^\alpha_\beta
}
\nonumber\\
&&
 + \left( i \not\!\partial_x - m_i \right) \delta^\alpha_\gamma
   i \left(
       {{\delta^2 \ln Z}
        \over
        { \delta {\bar \eta}^{i\gamma}(x) \delta \eta_{j\beta}(y)}}
     + {{\delta \ln Z} \over {\delta {\bar \eta}^{i\gamma}(x)}}
       {{\delta \ln Z} \over {\delta \eta_{j\beta}(y)}}
     \right)
\nonumber\\
&&
 + g \gamma^\mu \left( T^a \right)^\alpha{}_\gamma
   \biggl(
     {{\delta^3 \ln Z}
      \over
      { \delta J^{a\mu}(x)
        \delta {\bar \eta}^{i\gamma}(x) \delta \eta_{j\beta}(y)}}
   + { \delta \over {\delta J^{a\mu} (x)}}
     \left(
      {{\delta \ln Z} \over {\delta {\bar \eta}^{i\gamma}(x)}}
      {{\delta \ln Z} \over {\delta \eta_{j\beta}(y)}}
     \right)
\nonumber\\
&&
   + { {\delta \ln Z} \over {\delta J^{a\mu} (x)}}
     \left(
       {{\delta^2 \ln Z}
        \over
        { \delta {\bar \eta}^{i\gamma}(x) \delta \eta_{j\beta}(y)}}
     + {{\delta \ln Z} \over {\delta {\bar \eta}^{i\gamma}(x)}}
       {{\delta \ln Z} \over {\delta \eta_{j\beta}(y)}}
     \right)
   \biggr)
  = 0.
\end{eqnarray}
By setting $J, \ {\bar \eta}, \ \eta = 0$, and assuming
\begin{eqnarray}
 \left. {{\delta \ln Z} \over {\delta J^{a\mu}(x)}}
 \right|_{J, \ \eta, \ {\bar \eta} = 0}
 &=& i \left. \langle G^a_\mu \rangle
       \right|_{J, \ \eta, \ {\bar \eta} = 0}
  = 0,
\\
 \left. {{\delta \ln Z} \over {\delta {\bar \eta}^{i\alpha}(x)}}
 \right|_{J, \ \eta, \ {\bar \eta} = 0}
 &=& i \left. \langle \psi_{i\alpha}(x) \rangle
       \right|_{J, \ \eta, \ {\bar \eta} = 0}
 = 0,
\\
 \left. {{\delta \ln Z} \over {\delta \eta_{i\alpha}(x)}}
 \right|_{J, \ \eta, \ {\bar \eta} = 0}
 &=& - i
     \left. \langle {\bar \psi}^{i\alpha}(x) \rangle
     \right|_{J, \ \eta, \ {\bar \eta} = 0}
 = 0,
\end{eqnarray}
 we obtain
\begin{eqnarray}
\lefteqn{
 \delta^4(x-y) \delta^i_j \delta^\alpha_\beta
}
\nonumber\\
&&
 + \left( i \not\!\partial_x - m_i \right) \delta^\alpha{}_\gamma
   \left.
   i {{\delta^2 \ln Z}
      \over
      { \delta {\bar \eta}^{i\gamma}(x) \delta \eta_{j\beta}(y)}}
   \right|_{J, \ \eta, \ {\bar \eta} = 0}
\nonumber\\
&&
 + g \gamma^\mu \left( T^a \right)^\alpha{}_\gamma
     \left.
     {{\delta^3 \ln Z}
      \over
      { \delta J^{a\mu}(x)
        \delta {\bar \eta}^{i\gamma}(x) \delta \eta_{j\beta}(y)}}
     \right|_{J, \ \eta, \ {\bar \eta} = 0}
  = 0.
\nonumber
\end{eqnarray}
{}From the definition of $S$ and $\Gamma$ and
\begin{equation}
 \int d^4z S(x-z)^\alpha{}_\gamma{}^i{}_k S^{-1}(z-y)^\gamma{}_\beta{}^k{}_j
  = \delta^\alpha_\beta \delta^i_j \delta^4(x-y),
\end{equation}
 we get the Schwinger-Dyson equation for quark propagator eq.(\ref{quark}).
The Schwinger-Dyson equation for gluon propagator, eq.(\ref{gluon}),
 can also be obtained in the same way.

The eq.(\ref{quark}) and eq.(\ref{gluon}) can be written as follows
 in momentum space.
\begin{equation}
 S^{-1}(k)
  = \left( m - \not\!k \right)
  - g^2 C_2 \int {{d^4q} \over {(2\pi)^4 i}}
        \gamma^\mu S(q) \Gamma^\nu(k-q,q,-k) D_{\mu\nu}(k-q)
\label{SD-quark}
\end{equation}
\begin{eqnarray}
 D^{-1}_{\mu\nu}(k)
  &=& \left\{
       k^2 g_{\mu\nu} - \left( 1-{1 \over \alpha} \right) k_\mu k_\nu
      \right\}
\nonumber\\
  &-& i g^2 C_2 \int {{d^4q} \over {(2\pi)^4 i}}
                     {\rm tr} \left[
                              \gamma_\mu S(q) \Gamma_\nu(-k,q,k-q) S(q-k)
                              \right],
\label{SD-gluon}
\end{eqnarray}
 where $C_2 = {\rm tr} \left( T^a T^a \right) / N_c$ ($N_c=3$),
 $m = \mbox{\rm diag} (m_1, m_2, \cdots , m_{N_f})$,
\begin{eqnarray}
 S(k) &=& \left( \Sigma(k) - \not\!k \right)^{-1},
\\
 S^{-1}(k) &=& \Sigma(k) - \not\!k,
\\
 S(x-y) &=& \int {{d^4q} \over {(2\pi)^4 i}} S(q) e^{-iq(x-y)},
\\
 S^{-1}(x-y) &=& \int {{d^4q} \over {(2\pi)^4}} i S^{-1}(q) e^{-iq(x-y)},
\end{eqnarray}
\begin{eqnarray}
 D_{\mu\nu}(k)
 &=& {1 \over {k^2 (1 + \Pi(k^2))}}
     \left( g_{\mu\nu} - (1-\alpha) {{k_\mu k_\nu} \over {k^2}} \right),
\\
 D^{-1}_{\mu\nu}(k)
 &=& \left( 1 + \Pi(k^2) \right)
     \left( k^2 g_{\mu\nu} - (1-{1 \over \alpha}) k_\mu k_\nu \right),
\\
 D_{\mu\nu}(x-y)
 &=& \int {{d^4q} \over {(2\pi)^4 i}} D_{\mu\nu}(q) e^{-iq(x-y)},
\\
 D^{-1}_{\mu\nu}(x-y)
 &=& \int {{d^4q} \over {(2\pi)^4}} i D^{-1}_{\mu\nu}(q) e^{-iq(x-y)},
\end{eqnarray}
 and
\begin{eqnarray}
\lefteqn{
 \Gamma^\mu(y,x_1,x_2)
}
\\
&&
  = \int {{d^4q} \over {(2\pi)^4}}
         {{d^4p_1} \over {(2\pi)^4}} {{d^4p_2} \over {(2\pi)^4}}
         (2\pi)^4 i \delta^4(q+p_1+p_2)
         \Gamma^\mu(q,p_1,p_2) e^{-iqy} e^{-ip_1x_1} e^{-ip_2x_2}.
\nonumber
\end{eqnarray}
All the quantities in eqs.(\ref{SD-quark}) and (\ref{SD-gluon}) are bare ones.
The Schwinger-Dyson equation for renormalized quantities is
\begin{eqnarray}
\lefteqn{
 S_{R}^{-1}(k)
  = Z_2 \left( {{m_R + \delta m} \over {Z_2}} - \not\!k \right)
}
\nonumber\\
&&
  - Z_1 g_R^2 C_2 \int {{d^4q} \over {(2\pi)^4 i}}
        \gamma^\mu S_R(q) \Gamma_R^\nu(k-q,q,-k) D_{R\mu\nu}(k-q),
\end{eqnarray}
\begin{eqnarray}
\lefteqn{
 D^{-1}_{R\mu\nu}(k)
  = Z_3 \left( k^2 g_{\mu\nu} - (1-{1 \over \alpha}) k_\mu k_\nu \right)
}
\nonumber\\
&&
  - i Z_1 g_R^2 C_2 \int {{d^4q} \over {(2\pi)^4 i}}
                   {\rm tr} \left[
                            \gamma_\mu S_R(q) \Gamma_{R\nu}(-k,q,k-q)
                            S_R(q-k)
                            \right].
\end{eqnarray}
Z-factors and $\delta m_i$ should be fixed by the renormalization conditions,
 for example,
\begin{equation}
 \left. {1 \over 4} {\rm tr} S_R^{-1}(k) \right|_{k^2=\mu^2} = \mu,
\end{equation}
\begin{equation}
 \left.
 {\partial \over {\partial k^2}}
 {1 \over 4} {\rm tr} \not\!k S_R^{-1}(k)
 \right|_{k^2=\mu^2} = - 1.
\end{equation}

Now we solve the equation.
We consider the case that the chiral symmetry is exact, $m_i=0$.
The flavor dependence of the propagator will disappear,
 and the global $SU(N_f)_V$ symmetry will not be broken
 \footnote{We simply assume this.}.
Take the full vertex function (bare) as
\begin{equation}
 \Gamma^\mu(q,p_1,p_2)^i{}_j = \gamma^\mu \delta^i_j,
\end{equation}
 and take the gluon propagator in eq.(\ref{SD-quark}) as free one.
Then the equation becomes
\begin{equation}
 S^{-1}(k)
  = - \not\!k - g^2 C_2 \int {{d^4q} \over {(2\pi)^4 i}}
                \gamma^\mu S(q) \gamma^\nu D^{\mbox{\rm free}}_{\mu\nu}(k-q)
\label{SD-ladder}
\end{equation}
\begin{equation}
 D^{\mbox{\rm free}}_{\mu\nu}(k)
  = {1 \over {k^2}}
     \left\{
      g_{\mu\nu} - \left( 1-\alpha \right) {{k_\mu k_\nu} \over {k^2}}
     \right\},
\end{equation}
 where we set $S(q)^i_j = S(q) \delta^i_j$.
This {\it approximation} is called ladder approximation.
The equation of the quark propagator
 is independent of the equation of the gluon propagator.
Though there is no reason for supporting this {\it approximation},
 it is taken only for the analytic treatment of the equation.
Since the result is gauge dependent ($\alpha$ dependent),
 we can not do the quantitative discussion.
We believe that
 this {\it approximation} does not destroy the qualitative feature.

The quark propagator can be generally written as
\begin{equation}
 S^{-1}(k) = B(k^2) - A(k^2) \not\!k.
\label{quark-propagator}
\end{equation}
We get two equation for $A(k^2)$ and $B(k^2)$ from eq.(\ref{SD-ladder}).
By taking trace, we get
\begin{equation}
 B(k^2) = - {1 \over 4} g^2 C_2 {\rm tr} \int {{d^4q} \over {(2\pi)^4 i}}
          \gamma^\mu S(q) \gamma^\nu D^{\mbox{\rm free}}_{\mu\nu}(k-q),
\end{equation}
 and by taking trace after multiplying $-\not\!k / k^2$, we get
\begin{equation}
 A(k^2) = 1 + {1 \over 4} g^2 C_2 {\rm tr} {{\not\!k} \over {k^2}}
              \int {{d^4q} \over {(2\pi)^4 i}}
              \gamma^\mu S(q) \gamma^\nu D^{\mbox{\rm free}}_{\mu\nu}(k-q).
\end{equation}
In the Euclidean space
 \footnote{It is a non-trivial problem
            whether we can go to the Euclidean space or not.
           Here, we simply assume that we can do that.},
 they become
\begin{equation}
 B(-k^2) = {{3+\alpha} \over 3} \lambda \int dq^2
             {{q^2 B(-q^2)} \over {B(-q^2)^2 + A(-q^2)^2 q^2}}
             \left\{ {1 \over {k^2}} \theta(k^2-q^2)
                   + {1 \over {q^2}} \theta(q^2-k^2) \right\},
\end{equation}
\begin{equation}
 A(-k^2) = 1 + {\alpha \over 3} \lambda {1 \over {k^2}} \int dq^2
             {{q^2 A(-q^2)} \over {B(-q^2)^2 + A(-q^2)^2 q^2}}
             \left\{ {1 \over {k^2}} \theta(k^2-q^2)
                   + {1 \over {q^2}} \theta(q^2-k^2) \right\},
\end{equation}
 where
\begin{equation}
 \lambda = {3 \over {4\pi}} C_2 {{g^2} \over {4\pi}}.
\end{equation}
If we take Landau gauge $\alpha=0$
 \footnote{When we consider QED,
            to take Landau gauge makes the ladder approximation
            consistent with the Ward-Takahashi identity.},
 $A(-k^2)=1$ and
\begin{equation}
 B(-k^2) =  \lambda \int dq^2
             {{q^2 B(-q^2)} \over {B(-q^2)^2 + q^2}}
             \left\{ {1 \over {k^2}} \theta(k^2-q^2)
                   + {1 \over {q^2}} \theta(q^2-k^2) \right\}.
\label{SD-non-linear}
\end{equation}

This non-linear equation can not be solved analytically.
We use the bifurcation theory\cite{Atkinson}
 to estimate the existence of the non-trivial solution.
We expand the non-linear equation around the trivial solution as
\begin{equation}
 \Sigma(x) = \lambda \int y \Sigma(y)
             \left\{ {1 \over y} \theta(y-x)
                   + {1 \over x} \theta(x-y) \right\},
\end{equation}
 where $\Sigma(x)=B(-k^2)$ with $x=k^2$ and $y=q^2$.
$\Sigma(x)$ is called the mass function.
Bifurcation theory says that
 if this expanded equation
 has only one non-trivial linearly independent solution
 when $\lambda=\lambda_c$,
 the full equation has only one non-trivial solution (up to sign)
 when $\lambda > \lambda_c$ or $\lambda < \lambda_c$.
The linearized equation
 can be transformed to the differential equation with boundary conditions as
\begin{equation}
 x^2 {{d^2 \Sigma(x)} \over {dx^2}}
  + 2x {{d\Sigma(x)} \over {dx}} + \lambda \Sigma(x) = 0,
\label{differential-eq}
\end{equation}
\begin{eqnarray}
 \lim_{x \rightarrow 0} x^2 \Sigma'(x) &=& 0,
\label{IRBC-bif}
\\
 \lim_{x \rightarrow \infty} \left( x \Sigma(x) \right)' &=& 0.
\label{UVBC-bif}
\end{eqnarray}
This can be easily solved.
The general two linearly independent non-trivial solutions are
\begin{equation}
 \Sigma(x) = \left\{
  \begin{array}{l}
   x^{-(1-\gamma) / 2}\\
   x^{-(1+\gamma) / 2},
  \end{array}
 \right.
\label{solution-bifurcation}
\end{equation}
 where $\gamma=\sqrt{1-4\lambda}$.
These solutions satisfy the boundary conditions if $\lambda \neq 0$.
When $\lambda=\lambda_c \equiv 1/4$,
 the two solutions are not linearly independent.
Therefore, we know that
 the non-linear equation has only one non-trivial solution (up to sign)
 when $\lambda > \lambda_c$ or $\lambda < \lambda_c$.
$\lambda_c$ is called the critical coupling.

The non-linear equation is expected to have the non-trivial solution
 when $\lambda > \lambda_c$\cite{MNFK}.
By replacing the $B(-q^2)=\Sigma(y)$
 in the denominator of eq.(\ref{SD-non-linear})
 into the constant $m$ ($m=\Sigma(m)$), we get
\begin{equation}
 \Sigma(x) =  \lambda \int dy
             {{y \Sigma(y)} \over {m^2 + y}}
             \left\{ {1 \over x} \theta(x-y)
                   + {1 \over y} \theta(y-x) \right\}.
\end{equation}
This equation
 can be transformed to the differential equation and boundary conditions as
\begin{equation}
 \left( x \Sigma(x) \right)''
  + {{\lambda \Sigma(x)} \over {x+m^2}} = 0,
\end{equation}
\begin{eqnarray}
&& \lim_{x \rightarrow 0} x^2 \Sigma'(x) = 0,
\label{IRBC}
\\
&& \lim_{x \rightarrow \Lambda^2} \left( x \Sigma(x) \right)' = 0,
\label{UVBC}
\end{eqnarray}
 where $\Lambda$ is the ultraviolet cut off.
By changing the valuable $x$ to $y=-x/m^2$, we get
\begin{equation}
 y(1-y) {{d^2 \Sigma} \over {dy^2}}
  + 2(1-y) {{d\Sigma} \over {dy}} - \lambda \Sigma = 0.
\end{equation}
This equation is called hypergeometric differential equation
 which can be solved analytically.
The general solution is
\begin{eqnarray}
\lefteqn{
 \Sigma(x)
  = a_1 F({1 \over 2} + \gamma, {1 \over 2} - \gamma, 2 ; -{x \over {m^2}})
}
\nonumber\\
&& \qquad
  - a_2 \left( - {x \over {m^2}} \right)^{-1}
    F(- {1 \over 2} + \gamma, - {1 \over 2} - \gamma, 0 ; -{x \over {m^2}})
\end{eqnarray}
 for $x / m^2 < 1$, and
\begin{eqnarray}
\lefteqn{
 \Sigma(x)
  = b_1 \left( {{m^2} \over x} \right)^{{1+\gamma} \over 2}
    F({{1+\gamma} \over 2},-{{1-\gamma} \over 2}, 1+\gamma; -{{m^2} \over x})
}
\nonumber\\
&& \qquad
  + b_2 \left( {{m^2} \over x} \right)^{{1-\gamma} \over 2}
    F({{1-\gamma} \over 2},-{{1+\gamma} \over 2}, 1-\gamma; -{{m^2} \over x})
\end{eqnarray}
 for $m^2 / x < 1$, where $a_1$, $a_2$, $b_1$, and $b_2$ are the constants.
The function $F$ is hypergeometric series
\begin{equation}
 F(\alpha,\beta,\gamma;z)
  \equiv {{\Gamma(\gamma)} \over {\Gamma(\alpha) \Gamma(\beta)}}
         \sum_{n=0}^\infty
         {{\Gamma(\alpha+n) \Gamma(\beta+n)} \over {\Gamma(\gamma+n)}}
         {{z^n} \over {n!}}.
\end{equation}
{}From the infrared boundary condition, eq.(\ref{IRBC}),
 $a_2$ must be zero.
The value $\Sigma(x=0)=a_1$
 is finite unlike the solution when $m=0$ (eq.(\ref{solution-bifurcation})).
The ultraviolet boundary condition eq.(\ref{UVBC})
 is not satisfied unless $\lambda > \lambda_c$.
The non-trivial solution which satisfy the boundary condition is
\begin{equation}
 \Sigma(x)
  = a F({1 \over 2} + i\gamma', {1 \over 2} - i\gamma', 2 ; -{x \over {m^2}})
\end{equation}
 for $x / m^2 < 1$, and
\begin{equation}
 \Sigma(x)
  = {1 \over 2} b \left( {{m^2} \over x} \right)^{{1+i\gamma'} \over 2}
    F({{1+i\gamma'} \over 2},-{{1-i\gamma'} \over 2}, 1+i\gamma';
                                                       -{{m^2} \over x})
    + h.c.
\end{equation}
 for $m^2 / x < 1$,
 where $a$ and $b$ are real and complex constants, respectively,
 and $\gamma' = \sqrt{4\lambda-1} > 0$.
We are assuming that $\Sigma(x)$ is the real function.
The asymptotic form for $x \rightarrow \Lambda^2$ is
\begin{equation}
 \Sigma(x)
  = b \left( {{m^2} \over x} \right)^{1/2}
      \cos \left( {{\gamma'} \over 2} \ln ({{m^2} \over x}) \right).
\label{mass-function}
\end{equation}
$\Sigma(x)$ dumps as $1 / \sqrt{x}$.

Next,
 we connect the solution of the Schwinger-Dyson equation
 to the pair condensate of quark $\langle {\bar \psi}_i \psi^i \rangle$.
We use the technique of operator product expansion\cite{OPE}.
The quark propagator is expanded as
\begin{eqnarray}
 \int d^4x e^{ipx} \langle 0 | T \psi_L^i(x) {\bar \psi}_R^j(0) | 0 \rangle
 &=& U^{ij}_{\rm unit}(p,g,\mu) {{1-\gamma_5} \over 2}
\nonumber\\
 &+& U^{ij}_{{\bar \psi} \psi}(p,g,\mu) {{1-\gamma_5} \over 2}
     \langle 0 | T {\bar \psi}_R(0) \psi_L(0) | 0 \rangle
\nonumber\\
 &+& \mbox{(higher dimensional operators)},
\label{OPE}
\end{eqnarray}
 where $\mu$ is the renormalization point.
The first term in right-hand side
 is the effect of the explicit chiral symmetry breaking
 by which the chirality is flipped
 without the vacuum expectation value of operators.
Since we are neglecting the explicit breaking now,
\begin{equation}
 \lim_{p \rightarrow \infty}
 \int d^4x e^{ipx} \langle 0 | T \psi_L^i(x) {\bar \psi}_R^j(0) | 0 \rangle
 = \lim_{p \rightarrow \infty}
   U^{ij}_{{\bar \psi} \psi}(p,g,\mu) {{1-\gamma_5} \over 2}
   \langle 0 | {\bar \psi}_R(0) \psi_L(0) | 0 \rangle.
\label{OPE-limit}
\end{equation}
We can calculate the asymptotic form of the Wilson coefficient function
 $U^{ij}_{{\bar \psi} \psi}$ by the perturbation theory.
If the operator expansion is good, the relation
\begin{eqnarray}
\lefteqn{
 \lim_{p \rightarrow \infty} \int d^4x d^4y d^4z
 e^{ipx} e^{iqy} e^{ikz}
 \langle 0 | T \psi_L^i(x)_\alpha {\bar \psi}_R^j(0)_\delta
               \psi_R^k(y)_\gamma {\bar \psi}_L^l(z)_\beta | 0 \rangle
}
\\
&&
 = \lim_{p \rightarrow \infty}
   U^{ij}_{{\bar \psi} \psi}(p,g,\mu)
   \left( {{1-\gamma_5} \over 2} \right)_{\alpha\delta}
   \int d^4y d^4z e^{iqy} e^{ikz}
   \langle 0 | T \left( {\bar \psi}_R(0) \psi_L(0) \right)
               \psi_R^k(y)_\gamma {\bar \psi}_L^l(z)_\beta | 0 \rangle
\nonumber
\end{eqnarray}
 should be satisfied.
The Green functions can be calculated in tree level as (fig.(\ref{4-point}))
\begin{figure}[t]
\caption{Tree diagram for four-point Green function.}
\label{4-point}
\end{figure}
\begin{eqnarray}
\lefteqn{
 \lim_{p \rightarrow \infty} \int d^4x d^4y d^4z e^{ipx} e^{iqy} e^{ikz}
 \langle 0 | T \psi_L^i(x)_\alpha {\bar \psi}_R^j(0)_\delta
               \psi_R^k(y)_\gamma {\bar \psi}_L^l(z)_\beta | 0 \rangle
}
\nonumber\\
&&
 = - (2\pi)^4 \delta^4 (k+p)
     \left( {{1-\gamma_5} \over 2} {i \over {\not\!p}} \right)_{\alpha\beta}
     \left( {{1+\gamma_5} \over 2} {i \over {\not\!q}} \right)_{\gamma\delta}
     \delta^{il} \delta^{kj}
\\
&&
 - \left( {{1-\gamma_5} \over 2} {i \over {\not\!p}} \gamma_\mu
          {i \over {-\not\!k}} \right)_{\alpha\beta}
   g^2 (T^a)^{il} (T^a)^{kj} D^{\mu\nu}(p+k)
   \left( {{1+\gamma_5} \over 2} {i \over {\not\!q}} \gamma_\nu
          {i \over {\not\!q+\not\!p+\not\!k}} \right)_{\gamma\delta}
\nonumber
\end{eqnarray}
and (fig.(\ref{2-point}))
\begin{figure}[t]
\caption{Tree diagram for two point Green function with composite operator.}
\label{2-point}
\end{figure}
\begin{eqnarray}
\lefteqn{
 \int d^4y d^4z e^{iqy} e^{ikz}
 \langle 0 | T \left( {\bar \psi}_R(0) \psi_L(0) \right)
               \psi_R^k(y)_\gamma {\bar \psi}_L^l(z)_\beta | 0 \rangle
}
\nonumber\\
&& \qquad
 = \left( \left( {{1+\gamma_5} \over 2} {i \over {\not\!q}} \right)
          \left( {{1-\gamma_5} \over 2} {i \over {-\not\!k}} \right)
   \right)_{\gamma\beta} \delta^{kl},
\end{eqnarray}
 where $D^{\mu\nu}$ is the gluon free propagator
\begin{equation}
 D^{\mu\nu}
  = {1 \over i} {1 \over {q^2}}
    \left\{
     g^{\mu\nu} - \left( 1-\alpha \right) {{q^\mu q^\nu} \over {q^2}}
    \right\}.
\end{equation}
Therefore, when $k=-q$,
\begin{eqnarray}
\lefteqn{
 \lim_{p \rightarrow \infty}
  \left( {{1-\gamma_5} \over 2} {i \over {\not\!p}} \gamma_\mu
         {i \over {\not\!q}} \right)_{\alpha\beta}
  g^2 (T^a)^{il} (T^a)^{kj} D^{\mu\nu}(p-q)
  \left( {{1+\gamma_5} \over 2} {i \over {\not\!q}} \gamma_\nu
         {i \over {\not\!p}} \right)_{\gamma\delta}
}
\nonumber\\
& &
 = - \lim_{p \rightarrow \infty}
     U^{ij}_{{\bar \psi} \psi}(p,g,\mu)
     \left( {{1-\gamma_5} \over 2} \right)_{\alpha\delta}
     \left( {{1+\gamma_5} \over 2}
            {i \over {\not\!q}} {i \over {\not\!q}} \right)_{\gamma\beta}
     \delta^{kl}.
\end{eqnarray}
By taking trace for the indexes $\gamma\beta$, $\alpha\delta$, and $kl$,
 we get
\begin{equation}
 \lim_{p \rightarrow \infty} U^{ij}_{{\bar \psi} \psi}(p,g,\mu)
 = {1 \over i} {{3+\alpha} \over {2N_c}} g^2 C_2 \delta^{ij} {1 \over {p^4}},
\end{equation}
 where $N_c$ is the color degrees of freedom of quarks ($N_c=3$).

This result is improved by using the renormalization group equation.
The improvement makes it to depend on the renormalization point $\mu$.
The Wilson coefficient function
 satisfies the following renormalization group equation.
\begin{equation}
 U^{ij}_{{\bar \psi} \psi}(\kappa p,g,\mu)
 = \kappa^{d_U} U^{ij}_{{\bar \psi} \psi}(p,g(\kappa),\mu)
   \exp \int_1^\kappa \left[ \gamma_{{\bar \psi} \psi}(g(\kappa'))
                           - \gamma_{\psi {\bar \psi}}(g(\kappa')) \right]
                      {{d\kappa'} \over {\kappa'}},
\end{equation}
 where $d_U=-4$ is the dimension of the Wilson coefficient function,
 $\gamma_{{\bar \psi} \psi}$ and $\gamma_{\psi {\bar \psi}}$
 are the anomalous dimensions of the operator
 ${\bar \psi} \psi$ and $T \psi {\bar \psi}$, respectively,
 and $g(\kappa)$ is the running coupling.
This equation means the fact
 that the renormalization point dependence of the both side of eq.(\ref{OPE})
 should be the same.
We multiply the function $f(p^2/\mu^2)$
 to the Wilson coefficient given by the perturbative calculation
 to include the renormalization point dependence.
By substituting it
 into the renormalization group equation in $p \rightarrow 0$ limit, we get
\begin{eqnarray}
\lefteqn{
 \lim_{k \rightarrow \infty} U^{ij}_{{\bar \psi} \psi}(k,g,\mu)
}
\\
&&
 = \left( {{k^2} \over {p^2}} \right)^{d_U/2}
   {1 \over i} {{3+\alpha} \over {2N_c}}
   g^2((k^2 / p^2)^{1/2}) C_2 \delta^{ij}
   {1 \over {p^4}} f(p^2 / \mu^2)
   \exp \int_1^{(k^2 / p^2)^{1/2}}
        \gamma(g(\kappa')) {{d\kappa'} \over {\kappa'}},
\nonumber
\end{eqnarray}
 where $k=\kappa p$ and
\begin{equation}
 \gamma(g(\kappa')) = \gamma_{{\bar \psi} \psi}(g(\kappa'))
                    - \gamma_{\psi {\bar \psi}}(g(\kappa')).
\end{equation}
The function $f(p^2/\mu^2)$,
 is specified by assuming that
 the right-hand side is only depend on $k$ but not $p$ explicitly.
Namely,
\begin{equation}
 f(p^2 / \mu^2) g^2((k^2 / p^2)^{1/2})
 \exp \int_1^{(k^2 / p^2)^{1/2}}
      \gamma(g(\kappa')) {{d\kappa'} \over {\kappa'}} = \mbox{constant}.
\end{equation}
If the running of the coupling is very slow,
 we can replace it by a constant $g^*$.
Then
\begin{equation}
 f(p^2 / \mu^2) g^{*2}
 \left( {{k^2} \over {p^2}} \right)^{{{\gamma^*} \over 2}}
 = \mbox{constant},
\end{equation}
 where $\gamma^* = \gamma(g^*)$.
Therefore,
\begin{equation}
 f(p^2 / \mu^2)
 = c \left( {{p^2} \over {\mu^2}} \right)^{\gamma^* / 2}
\end{equation}
 with a constant $c$.
In this case,
 we can explicitly write the asymptotic form
 of the Wilson coefficient function as
\begin{equation}
 \lim_{k \rightarrow \infty} U^{ij}_{{\bar \psi} \psi}(k,g,\mu)
 = c {1 \over i} {{3+\alpha} \over {2N_c}}
   g^{*2} C_2 \delta^{ij}
   {1 \over {k^4}} \left( {{k^2} \over {\mu^2}} \right)^{\gamma^* / 2}.
\label{Wilson-coefficient}
\end{equation}
Treating the running coupling as a constant
 is a convenience {\it approximation}
 to consider the effect of anomalous dimension.

We now return to the pair condensate $\langle {\bar \psi}_i \psi^i \rangle$.
{}From the eq.(\ref{OPE-limit}), we get
\begin{equation}
 \langle {\bar \psi}_R \psi_L \rangle {{1-\gamma_5} \over 2}
 = \lim_{p \rightarrow \infty}
   {{ \int d^4x e^{ipx} \langle T \psi_L^i(x) {\bar \psi}^i_R(0) \rangle }
    \over
    {U^{ii}_{{\bar \psi} \psi}(p,g,\mu)}}.
\end{equation}
By substituting the result of eq.(\ref{Wilson-coefficient}), we obtain
\begin{equation}
 \langle {\bar \psi} \psi \rangle
 = - \lim_{p \rightarrow \infty} {{4N_c} \over {3 c g^{*2} C_2}}
     \left( {{\mu^2} \over {p^2}} \right)^{\gamma^* / 2}
     p^2 \Sigma(-p^2),
\end{equation}
 where we use the solution of Schwinger-Dyson equation in Landau gauge
 ($A(k^2)=1$ and $B(k^2)=\Sigma(-k^2)$ in eq.(\ref{quark-propagator})).
If $\Sigma=0$, the pair condensate vanishes.
But if $\Sigma$ is non-trivial and has the asymptotic behavior
\begin{equation}
 \lim_{p \rightarrow \infty} \Sigma(-p^2)
 = \mbox{constant} \times
   {1 \over {p^2}} \left( {{p^2} \over {\mu^2}} \right)^{\gamma^* / 2},
\label{asymptotic-form}
\end{equation}
 then the condensate takes finite value.

We have tried to calculate the value of the pair condensate in QCD.
Since the mass function $\Sigma$ and the anomalous dimension
 can not perfectly be estimated,
 we can not {\it proof} the existence of the pair condensate in QCD.
If we believe the existence of the pair condensate,
 the anomalous dimension can be obtained
 form eq.(\ref{asymptotic-form}) and eq.(\ref{mass-function}).
Since eq.(\ref{mass-function}) suggests that
 the asymptotic behavior of mass function is $1/\sqrt{p^2}$ (up to log),
 we know that the anomalous dimension must be unity, $\gamma^* = 1$,
 for the finite condensate.
This value of $\gamma^*$
 is large compared with the naive expectation from the perturbative QCD.

The Schwinger-Dyson equation
 which is considered in this section is called QED-like,
 since the non-abelian effect of gauge interaction
 (non-abelian vertex) is not included.
The form of the equation
 is just the same in QED, except for the Casimir coefficient $C_2$.
The effect can be approximately included
 by replacing the coupling constant by running coupling
 (Higashijima approximation \cite{Higashijima}).
The approximation results the small anomalous dimension $\gamma^* \sim 0$.
We can expect that
 the slower running coupling means the larger anomalous dimension $\gamma^*$.

\section{Decay constant of Nambu-Goldstone bosons}

The Nambu-Goldstone theorem says that
 the Nambu-Goldstone bosons couple with the broken currents as
\begin{equation}
 \langle 0 | J_\mu^a(0) | (NG)^b(q) \rangle = i q_\mu I^{ab}(q^2).
\end{equation}
The value $I^{ab}(q^2=0)=f^{ab}$ is generally referred as the decay constant,
 because the pions as the pseudo-Nambu-Goldstone bosons
 couple with the weak currents and decay to the leptons.

In the following we consider QCD
 in which the chiral symmetry is spontaneously broken by the quark condensate.
The currents which couple with the Nambu-Goldstone bosons
 are the axial currents ${\bar q} \gamma_\mu \gamma_5 T^a q$.
The above amplitude can be written as (fig.(\ref{Coupling}))
\begin{equation}
 - \mbox{\rm tr} \int {{d^4k} \over {(4\pi)^4 i}}
                 S(k) \gamma_\mu \gamma_5 T^a S(q+k) P^b(q+k,k)
 = i q_\mu I^{ab}(q^2),
\end{equation}
\begin{figure}[t]
\caption{Coupling between the Nambu-Goldstone bosons and currents.}
\label{Coupling}
\end{figure}
 where $S$ is the full quark propagator,
 and $P^b(q+k,k)$ is the Bethe-Salpeter amplitude.
Bethe-Salpeter amplitude
 describes the coupling between the quarks and the Nambu-Goldstone boson.
By differentiating both side with respect to $q^\nu$
 and taking limit $q \rightarrow 0$, we obtain
\begin{eqnarray}
 i g_{\mu\nu} f^{ab} =
 &-& \mbox{\rm tr} \int {{d^4k} \over {(4\pi)^4 i}}
                   S(k) \gamma_\mu \gamma_5 T^a
                   \left(
                    {\partial \over {\partial q^\nu}} S(q+k)
                   \right)_{q \rightarrow 0}
                   P^b(k,k)
\nonumber\\
 &-& \mbox{\rm tr} \int {{d^4k} \over {(4\pi)^4 i}}
                   S(k) \gamma_\mu \gamma_5 T^a S(k)
                   \left(
                    {\partial \over {\partial q^\nu}} P^b(q+k,k)
                   \right)_{q \rightarrow 0}.
\label{pre-Pagels-Stokar-1}
\end{eqnarray}

To discuss the ladder approximation of this formula,
 we consider the Schwinger-Dyson equation for the axial vertex
 (fig.(\ref{SD-vertex}))
\begin{eqnarray}
 \Gamma^a_{5\mu}(p,p+q)
  = \gamma_\mu \gamma_5 T^a
  - \mbox{\rm tr} \int {{d^4k} \over {(4\pi)^4 i}}
    \Gamma^a_{5\mu}(k,k+q) S(k+q) K(k+q,p,p+q) S(k),
\nonumber\\
\end{eqnarray}
\begin{figure}[t]
\caption{Schwinger-Dyson equation for the axial vertex.}
\label{SD-vertex}
\end{figure}
 where $K$ is the set of the horizontally one-particle irreducible diagrams.
The axial vertex can be decomposed in two parts as (fig.(\ref{Axial}))
\begin{equation}
 \Gamma^a_{5\mu}(p,p+q)
   = i q_\mu I^{ab}(q^2) {1 \over {-q^2}} P^b(p,p+q)
   + {\tilde \Gamma}^a_{5\mu}(p,p+q).
\label{decomposition}
\end{equation}
\begin{figure}[t]
\caption{Decomposition of the axial vertex.}
\label{Axial}
\end{figure}
The first term is the $1/q^2$ pole contribution of Nambu-Goldstone boson,
 and the second term is the another contribution which is regular at $q=0$.
The functions
 ${\tilde \Gamma}^a_{5\mu}(p,p+q)$, $P^a(p,p+q)$, and $K(k+q,p,p+q)$
 are regular at $q=0$.
By substituting this to the Schwinger-Dyson equation, we get
\begin{eqnarray}
\lefteqn{
 i q_\mu I^{ab}(q^2) {1 \over {-q^2}}
  \biggl\{ P^b(p,p+q)
        + \mbox{\rm tr} \int {{d^4k} \over {(4\pi)^4 i}}
          P^b(k,k+q) S(k+q) K(k+q,p,p+q) S(k)
  \biggr\}
}
\nonumber\\
&&
 = \gamma_\mu \gamma_5 T^a - {\tilde \Gamma}^a_{5\mu}(p,p+q)
 - \mbox{\rm tr} \int {{d^4k} \over {(4\pi)^4 i}}
   {\tilde \Gamma}^a_{5\mu}(k,k+q) S(k+q) K(k+q,p,p+q) S(k).
\nonumber\\
\end{eqnarray}
It is expected that
 each of the functions $P^a(p,p+q)$ and ${\tilde \Gamma}^a_{5\mu}(p,p+q)$
 satisfies the independent equation in the limit $q \rightarrow 0$,
 since they appear as the different order of $q$ in eq.(\ref{decomposition}).
Therefore, we get in $q \rightarrow 0$ limit
\begin{equation}
 i q_\mu I^{ab}(q^2) {1 \over {-q^2}}
  \biggl\{ P^b(p,p+q)
        + \mbox{\rm tr} \int {{d^4k} \over {(4\pi)^4 i}}
          P^b(k,k+q) S(k+q) K(k+q,p,p+q) S(k)
  \biggr\} = O(q),
\end{equation}
\begin{equation}
 \gamma_\mu \gamma_5 T^a - {\tilde \Gamma}^a_{5\mu}(p,p+q)
 - \mbox{\rm tr} \int {{d^4k} \over {(4\pi)^4 i}}
   {\tilde \Gamma}^a_{5\mu}(k,k+q) S(k+q) K(k+q,p,p+q) S(k) = O(q).
\end{equation}
Then, we obtain
\begin{equation}
 P^b(p,p+q) =
  - \mbox{\rm tr} \int {{d^4k} \over {(4\pi)^4 i}}
              P^b(k,k+q) S(k+q) K(k+q,p,p+q) S(k) + O(q^2),
\label{BS}
\end{equation}
\begin{equation}
 {\tilde \Gamma}^a_{5\mu}(p,p+q)
 = \gamma_\mu \gamma_5 T^a
 - \mbox{\rm tr} \int {{d^4k} \over {(4\pi)^4 i}}
   {\tilde \Gamma}^a_{5\mu}(k,k+q) S(k+q) K(k+q,p,p+q) S(k)
 + O(q).
\label{SD-vertex-regular}
\end{equation}
The first equation
 is called Bethe-Salpeter equation for the Nambu-Goldstone boson vertex.

The vertex $\gamma_\mu \gamma_5 T^a$ can be eliminated
 from eq.(\ref{pre-Pagels-Stokar-1}) by using eq.(\ref{SD-vertex-regular}).
\begin{eqnarray}
\lefteqn{
 i g_{\mu\nu} f^{ab} =
 - \mbox{\rm tr} \int_k
                   S(k) {\tilde \Gamma}^a_{5\mu}(k,k)
                   \left(
                    {\partial \over {\partial q^\nu}} S(q+k)
                   \right)_{q \rightarrow 0}
                   P^b(k,k)
}
\nonumber\\
&&
 - \mbox{\rm tr} \int_{k, k'}
   S(k) \mbox{\rm tr}
   \left\{
    {\tilde \Gamma}^a_{5\mu}(k',k') S(k') K(k',k,k) S(k')
   \right\}
   \left(
    {\partial \over {\partial q^\nu}} S(q+k)
   \right)_{q \rightarrow 0}
   P^b(k,k)
\nonumber\\
&&
 - \mbox{\rm tr} \int_k
                 S(k) {\tilde \Gamma}^a_{5\mu}(k,k) S(k)
                 \left(
                  {\partial \over {\partial q^\nu}} P^b(q+k,k)
                 \right)_{q \rightarrow 0}
\nonumber\\
&&
 - \mbox{\rm tr} \int_{k, k'}
   S(k) \mbox{\rm tr}
   \left\{
    {\tilde \Gamma}^a_{5\mu}(k',k') S(k') K(k',k,k) S(k')
   \right\} S(k)
   \left(
    {\partial \over {\partial q^\nu}} P^b(q+k,k)
   \right)_{q \rightarrow 0},
\nonumber\\
\end{eqnarray}
 where $\int_k$ means $\int d^4k / (4\pi)^4 i$.
Moreover, the derivative of $P^b$ can be eliminated by using eq.(\ref{BS})
 (fig.(\ref{pre-Pagels-Stokar-fig})).
\begin{eqnarray}
\lefteqn{
 i g_{\mu\nu} f^{ab} =
 - \mbox{\rm tr} \int_k
                 S(k) {\tilde \Gamma}^a_{5\mu}(k,k)
                 \left(
                  {\partial \over {\partial q^\nu}} S(q+k)
                 \right)_{q \rightarrow 0}
                 P^b(k,k)
}
\nonumber\\
&&
 + \mbox{\rm tr} \int_{k, k'}
   S(k) \mbox{\rm tr}
   \left\{
    {\tilde \Gamma}^a_{5\mu}(k',k') S(k')
    \left(
     {\partial \over {\partial q^\nu}}
     K(q+k',p,p+q)
    \right)_{q \rightarrow 0} S(k')
    \right\}
    S(k) P^b(k,k).
\nonumber\\
\label{pre-Pagels-Stokar-2}
\end{eqnarray}
\begin{figure}[t]
\caption{
 Graphic representation of eq.(3.90).
 Slash denotes the derivative with respect to $q$.
}
\label{pre-Pagels-Stokar-fig}
\end{figure}

Now we consider the ladder approximation of this formula.
In ladder approximation with Landau gauge,
 the renormalization of the axial vertex is trivial.
We can show that by using the chiral Ward-Takahashi identity
 \footnote{Here, we neglect the contribution of the chiral anomaly.}
\begin{equation}
 k^\mu \Gamma^a_{5\mu}(p,p+k)
  = S^{-1}(p) \gamma_5 T^a + \gamma_5 T^a S^{-1}(p+k)
\label{Chiral-WT}
\end{equation}
 and the fact that
 the renormalization of the quark field is trivial ($Z_2=1$)
 as we have already shown in previous section.
Therefore, we should set $K=0$ in eq.(\ref{SD-vertex-regular})
 and get ${\tilde \Gamma}^a_{5\mu}(k,k) = \gamma_\mu \gamma_5 T^a$.
The eq.(\ref{pre-Pagels-Stokar-2}) becomes
\begin{equation}
 i g_{\mu\nu} f^{ab} =
 - \mbox{\rm tr} \int {{d^4k} \over {(4\pi)^4 i}}
                 S(k) \gamma_\mu \gamma_5 T^a
                 \left(
                  {\partial \over {\partial q^\nu}} S(q+k)
                 \right)_{q \rightarrow 0}
                 P^b(k,k).
\label{pre-Pagels-Stokar-3}
\end{equation}
This equation also can be get
 by simply neglecting the second term in eq.(\ref{pre-Pagels-Stokar-1}).
Neglecting the derivative of Bethe-Salpeter amplitude
 corresponds to neglecting the effect of the axial vector resonance in it
 \footnote{
  The mixing between the axial vector resonance and Nambu-Goldstone boson
  is proportional to $q$}.

The Bethe-Salpeter amplitude at zero momentum transfer, $P(k,k)$,
 can be expressed by the mass function $\Sigma(k)$.
By taking the limit $k \rightarrow 0$ in eq.(\ref{Chiral-WT}),
 we extract the contribution of Nambu-Goldstone boson pole
 in the axial vertex function, and obtain
\begin{equation}
 -i f^{ab} P^b(k,k) = S^{-1}(k) \gamma_5 T^a + \gamma_5 T^a S^{-1}(k).
\end{equation}
Since we are taking ladder approximation with Landau gauge,
\begin{equation}
 f^{ab} P^b(k,k) = 2 i \Sigma (-k^2) \gamma_5 T^a.
\end{equation}
By substituting this equation into eq.(\ref{pre-Pagels-Stokar-3}), we get
\begin{equation}
 (f f^T)^{ab}
  = {{N_c} \over {4\pi^2}} \delta^{ab} \int dk^2
    {{k^2 \Sigma(k^2)} \over {\left( \Sigma^2(k^2) + k^2 \right)^2}}
    \left\{
     \Sigma(k^2) - {1 \over 2} k^2 {{d \Sigma(k^2)} \over {dk^2}}
    \right\}.
\end{equation}
This equation is expressed in the Euclidean momentum space.
If the $SU(N_f)_V$ symmetry is not broken,
 the decay constant matrix becomes diagonal.
In that case, we have
\begin{equation}
 f^2 = {{N_c} \over {4\pi^2}} \int dk^2
       {{k^2 \Sigma(k^2)} \over {\left( \Sigma^2(k^2) + k^2 \right)^2}}
       \left\{
        \Sigma(k^2) - {1 \over 2} k^2 {{d \Sigma(k^2)} \over {dk^2}}
       \right\}.
\label{PS-formula}
\end{equation}
This formula is called Pagels-Stokar formula\cite{Pagels-Stokar}.
We can obtain the value of the decay constant,
 if the ladder Schwinger-Dyson equation for quark propagator is solved.

\section{Cornwall-Jackiw-Tomboulis effective action}

In this section
 we introduce the Cornwall-Jackiw-Tomboulis (CJT) effective action\cite{CJT}.
The Schwinger-Dyson equation for quark propagator
 is systematically derived from this effective action.
It is also useful when we consider the more complicated system
 like that we will estimate in chapter 6 and 7.

The generating functional for the Green functions in QCD is
\begin{eqnarray}
\lefteqn{
 Z[K,J,{\bar J}]
 = {1 \over Z}
   \int {\cal D}\psi {\cal D}{\bar \psi} {\cal D}G \exp i \int d^4x
   \biggl\{
    {\bar \psi}(x) \left( i \not\!\partial - m + g \not\!G(x) \right) \psi(x)
}
\\
&&
    - {1 \over 2} \mbox{tr} \left( F^{\mu\nu}(x) F_{\mu\nu}(x) \right)
    + \int d^4y {\bar \psi}(x) K(x,y) \psi(y)
    + {\bar J}(x)\psi(x) + {\bar \psi}(x)J(x)
   \biggr\},
\nonumber
\end{eqnarray}
 where $J(x)$ and ${\bar J}(x)$ are the local sources,
 and $K(x,y)$ is the non-local source.
The normalization $Z[K=0,J=0,{\bar J}=0]=1$ is understood.
The connected Green functions like
 $\langle T \psi \psi \cdots {\bar \psi} {\bar \psi} \cdots \rangle_c$
 are obtained by differentiating $Z[K,J,{\bar J}]$
 with respect to $J$'s and ${\bar J}$'s.
The full quark propagator
 is obtained by differentiating it with respect to $-iK$.
We integrate out the quark and gluon fields
 and define the CJT effective action.

Using the derivative with respect to the sources gives
\begin{eqnarray}
\lefteqn{
 Z[K,J,{\bar J}]
 = {1 \over Z}
   \exp i \int
    \biggl\{
     - {\delta \over {i \delta J}}
        \left( i \not\!\partial - m \right)
       {\delta \over {i \delta {\bar J}}}
     - {\delta \over {i \delta J}} K {\delta \over {i \delta {\bar J}}}
    \biggr\}
}
\nonumber\\
&& \times
   \int {\cal D}\psi {\cal D}{\bar \psi} {\cal D}G
   \exp i \int
    \bigg\{
       {\bar \psi} g \not\!G \psi
     - {1 \over 2} \mbox{tr} \left( F^{\mu\nu}(x) F_{\mu\nu}(x) \right)
\nonumber\\
&& \qquad\qquad\qquad\qquad\qquad
     + {\bar J}(x)\psi(x) + {\bar \psi}(x)J(x)
    \biggr\}.
\end{eqnarray}
We introduce the full quark propagator $S$.
\begin{eqnarray}
\lefteqn{
 Z[K,J,{\bar J}]
 = {1 \over Z}
   \exp i \int
    \biggl\{
       \mbox{tr} \left[
                  \left( i \not\!\partial - m - iS^{-1} \right)
                  {\delta \over {i \delta {\bar J}}}
                  {\delta \over {i \delta J}}
                 \right]
     + \mbox{tr} \left[
                  K {\delta \over {i \delta {\bar J}}}
                  {\delta \over {i \delta J}}
                 \right]
    \biggr\}
}
\nonumber\\
&& \times \int {\cal D}\psi {\cal D}{\bar \psi} {\cal D}G
   \exp i \int
    \biggl\{
       {\bar \psi} iS^{-1} \psi
     + {\bar \psi} g \not\!G \psi
     - {1 \over 2} \mbox{tr} \left( F^{\mu\nu}(x) F_{\mu\nu}(x) \right)
\nonumber\\
&& \qquad\qquad\qquad\qquad\qquad
     + {\bar J}(x)\psi(x) + {\bar \psi}(x)J(x)
    \biggr\}.
\end{eqnarray}
At this stage,
 $S$ need not be the full propagator,
 but we will ensure it later.
The integration out on the gluon field
 gives the set of connected two-particle irreducible (2PI) vacuum diagrams
 $D[S,J,{\bar J}]$.
\begin{eqnarray}
\lefteqn{
 Z[K,J,{\bar J}]
 = {1 \over Z}
   \exp i \int
    \biggl\{
       \mbox{tr} \left[
                  \left( i \not\!\partial - m - iS^{-1} \right)
                  {\delta \over {i \delta {\bar J}}}
                  {\delta \over {i \delta J}}
                 \right]
     + \mbox{tr} \left[
                  K {\delta \over {i \delta {\bar J}}}
                  {\delta \over {i \delta J}}
                 \right]
    \biggr\}
}
\nonumber\\
&& \times \int {\cal D}\psi {\cal D}{\bar \psi}
   \exp i \int
    \biggl\{
       {\bar \psi} iS^{-1} \psi
     + {\bar J}(x)\psi(x) + {\bar \psi}(x)J(x)
    \biggr\}
   \times
   e^{D[S,J,{\bar J}]}.
\end{eqnarray}
The fact that $S$ is the full propagator
 makes the set of the vacuum diagrams to be the two-particle irreducible one.
Now we can integrate out the quark fields.
\begin{eqnarray}
 Z[K,J,{\bar J}]
 &=& {1 \over Z}
   \exp i \int
    \biggl\{
       \mbox{tr} \left[
                  \left( i \not\!\partial - m - iS^{-1} \right)
                  {\delta \over {i \delta {\bar J}}}
                  {\delta \over {i \delta J}}
                 \right]
     + \mbox{tr} \left[
                  K {\delta \over {i \delta {\bar J}}}
                  {\delta \over {i \delta J}}
                 \right]
    \biggr\}
\nonumber\\
&& \qquad\qquad\qquad\qquad\qquad
   \times
   {\sl Det} \left( - S^{-1} \right)
   e^{- \int {\bar J} S J}
   e^{D[S,J,{\bar J}]}.
\end{eqnarray}
Carrying out the functional derivatives
 and setting $J$ and ${\bar J}$ to zero gives
\begin{equation}
 Z[K] = {1 \over Z} \exp
    \biggl\{
       \ln {\sl Det} \left( - S^{-1} \right)
     -i \mbox{Tr}
        \left[ \left( i \not\!\partial - m - i S^{-1} \right) S \right]
     -i \mbox{Tr} \left[ K S \right]
     + D[S]
    \biggr\},
\end{equation}
 where $\mbox{Tr}$ denotes the trace in the functional space
 in addition to the trace in the spinor and flavor spaces.
The generating functional for the connected Green functions,
 $W[K] = \ln Z[K]$, is obtained as
\begin{equation}
 W[K] =
 \ln {\sl Det} \left( - S^{-1} \right)
  -i \mbox{Tr} \left[ \left( i \not\!\partial - m - i S^{-1} \right) S \right]
  -i \mbox{Tr} \left[ K S \right]
  + D[S],
\end{equation}
 where we neglect a constant $\ln Z$.
Now we can say that $S$ is really the full propagator, since
\begin{equation}
 - {1 \over i}
 \left. {{\delta W[K]} \over {\delta K}} \right|_{K \rightarrow 0} = S.
\end{equation}
The Legendre transformation gives the CJT effective action as
\begin{eqnarray}
 \Gamma[S]
 &=& W[K] + i \mbox{Tr} \left[ K S \right]
\nonumber\\
 &=& \ln {\sl Det} \left( - S^{-1} \right)
     -i \mbox{Tr}
        \left[ \left( i \not\!\partial - m - i S^{-1} \right) S \right]
     + D[S]
\end{eqnarray}
 with
\begin{equation}
 {{\delta \Gamma[S]} \over {\delta S}} = iK.
\label{transform-condition}
\end{equation}
In momentum space it becomes
\begin{equation}
 \Gamma[S]
  = \ln {\sl Det} S^{-1}(k)
    - \mbox{Tr} \left[ \left(\not\!k - m \right) S(k) \right]
    + D[S].
\end{equation}
The difference on the constant and over all factor $(2\pi)^4 \delta^4(0)$
 (four dimensional volume) is neglected.

Since $K=0$ in the real system,
 the condition eq.(\ref{transform-condition})
 becomes the stationary condition of the effective action.
The condition is equivalent to the Schwinger-Dyson equation for propagator.
If we take the 2PI vacuum diagram in fig.(\ref{hamburger}), namely
\begin{equation}
 D[S] = - {1 \over 2} \mbox{tr}
          \int {{d^4p} \over {(2\pi)^4 i}} {{d^4q} \over {(2\pi)^4 i}}
          g \gamma^\mu T^a S(p) g \gamma^\nu T^a S(q)
          D^{\mbox{\rm free}}_{\mu\nu}(p-q),
\end{equation}
\begin{figure}[t]
\caption{The vacuum diagram for ladder approximation.}
\label{hamburger}
\end{figure}
 the condition gives
 the Schwinger-Dyson equation in ladder approximation
 (cf. eq.(\ref{SD-ladder}))
\begin{equation}
 S^{-1}(k)
  = - \left( \not\!k - m \right)
    - \int {{d^4q} \over {(2\pi)^4 i}}
      g \gamma^\mu T^a S(q) g \gamma^\nu T^a
      D^{\mbox{\rm free}}_{\mu\nu}(k-q).
\end{equation}

The effective action formalism is convenient for the systematic analysis.
We will use this formalism in chapter 6 and 7
 to estimate the dynamics of the tumbling gauge theory.
\chapter{Technicolor model}

The standard model
 which describes the three fundamental interactions
 (electromagnetic, weak, and strong interactions)
 is consistent with almost all the experiments.
Especially,
 the agreement with the recent precision experiments in LEP is excellent.
It is believed that we finally get the true theory of
 electromagnetic and weak interactions.

The theory of the electromagnetic and weak interactions
 is based on the gauge principle.
The gauge symmetry of these interactions is
 the electroweak symmetry $SU(2)_L \times U(1)_Y$
 which describes the four massless gauge bosons.
The gauge symmetry
 is spontaneously broken to the electromagnetic gauge symmetry $U(1)_{em}$
 by the vacuum expectation value of the Higgs field.
The three of four gauge bosons become massive
 by the Higgs mechanism, and they mediate the weak interaction.
The mass explains the fact that the weak interaction is the short range one.
Residual one gauge boson remains massless,
 and it mediates the electromagnetic interaction.

It is interesting that
 the electroweak symmetry forbids the mass of the quarks and leptons,
 since it is the chiral symmetry.
Electroweak symmetry must be broken
 also to generate the mass of the quarks and leptons.
In the standard model,
 they get their masses through the Yukawa interactions like
\begin{equation}
 {\cal L}_{Yukawa} = - g_\psi {\bar \psi}_{Li} \psi_R \Phi^i + h.c.,
\end{equation}
 where $\psi$ is the quark or lepton, $\Phi$ is the Higgs field,
 and $i$ is the index of the doublet of the $SU(2)_L$ symmetry.
The vacuum expectation value of the Higgs field,
 $\langle \Phi \rangle
  = \left( \begin{array}{cc} 0 & v / \sqrt{2} \end{array} \right)^T$,
 which breaks the electroweak symmetry
 gives the mass to the fermions as $m_\psi = g_\psi v / \sqrt{2}$.
The vacuum expectation value of the Higgs field
 is not only the origin of the mass of the weak bosons,
 but also the origin of the mass of the quarks and leptons.

The standard model, however, has some theoretical problems in it.
In the next section,
 we will explain it as the motivation for technicolor theory
 in which the vacuum expectation value of the Higgs field
 is replaced by the fermion pair condensate
 $\langle {\bar T} T \rangle \neq 0$.
In section 2, the one family model of the technicolor theory
 is explained as an example.

\section{Motivation for technicolor}

One of the problems of the standard model is the fact that
 the masses of the quarks and leptons are not predicted.
The angles of the quark mixing and CP violating phase are not also predicted,
 since the origin is the same of the masses\cite{Kobayashi-Maskawa}.
There is no principle to predict the Yukawa couplings
 which are given as the free parameters.
We still do not know the origin of mass.
The existence of certain new physics should be expected
 at the high energy beyond the standard model.

There is also the problem which
 is called fine tuning problem, or naturalness problem, or hierarchy problem
 \cite{naturalness}.
The parameters in the Higgs potential must be fine tuned,
 if the standard model is looked upon as an effective theory.
Suppose that
 the standard model is valid until certain high energy scale $\Lambda$
 (GUT scale, for example.).
Then the Higgs potential
\begin{equation}
 V(\Phi) = - m^2 \Phi^{\dag} \Phi + {\lambda \over 4} (\Phi^{\dag} \Phi)^2
\end{equation}
 receive the radiative correction as
\begin{equation}
 V(\Phi)_{\mbox{one-loop}}
  = - \left( m^2 - \lambda {{\Lambda^2} \over {16\pi^2}} \right)
      \Phi^{\dag} \Phi + {\lambda \over 4} (\Phi^{\dag} \Phi)^2
\end{equation}
 up to the log correction.
The vacuum expectation value of the Higgs field $\Phi$ is
\begin{equation}
 \langle \Phi \rangle
 = {v \over {\sqrt{2}}}
    \left( \begin{array}{c} 0 \\ 1 \end{array} \right)
 = \sqrt{{2 \left(m^2 - \lambda \Lambda^2 / 16\pi \right)} \over \lambda}
    \left( \begin{array}{c} 0 \\ 1 \end{array} \right).
\end{equation}
This value must be the electroweak scale $v \simeq 250$GeV
 to give the realistic weak boson mass.
We must take the mass $m^2$ a large value,
 and fine tune to cancel the large value $\lambda \Lambda^2 / 16\pi$
 (like $1.00000000000001-1=0.00000000000001$)
 \footnote{
  This manipulation is the same of the renormalization.
  But here, $\Lambda$ is the physical ultraviolet cut off.}.
This work must be iterated order by order in loop calculation.
This unnatural situation means that
 we still do not know the mechanism of the electroweak symmetry breaking
 (Why the scale is so small?).
The ``natural'' physics beyond the standard model should exist.

Technicolor theory\cite{TC} is one of the candidate of the new physics.
It is a ``natural'' theory,
 and has a possibility to explain the mass generation.
The Higgs field as an elementary particle
 is eliminated and the new fermions (technifermions) are introduced
 with the additional strong interaction (technicolor interaction).
The technifermions
 are not the singlet of the electroweak symmetry $SU(2)_L \times U(1)_Y$
 so that the condensate $\langle {\bar T} T \rangle \neq 0$
 due to the technicolor interaction spontaneously breaks the symmetry.
The vacuum expectation value of the elementary Higgs field
 is replaced by the fermion pair condensate $\langle {\bar T} T \rangle$.
Namely, the electroweak symmetry is dynamically broken
 \footnote{
  This is very similar to the chiral symmetry breaking in QCD
  due to the quark condensate.}.
The Nambu-Goldstone bosons
 which are eaten by the weak bosons
 are the composite particles of the technifermions.
The physical Higgs particle is also the composite one.
Its width will be very broad,
 since it strongly decays to the pair of the Nambu-Goldstone bosons.

Consider the hierarchy of the scales
 between the weak scale and the scale of certain new physics.
If we renormalize the gauge coupling constants
 at the renormalization point $\mu=M$,
 then we have the running gauge couplings at one-loop level as
\begin{eqnarray}
 \alpha_{TC}^{-1}(\mu) &=& \alpha_{TC}^{-1}(M) + c_{TC} \ln({\mu \over M}),
\\
 \alpha_{new}^{-1}(\mu) &=& \alpha_{new}^{-1}(M) + c_{new} \ln({\mu \over M}),
\end{eqnarray}
 where $\alpha_{TC}$ is the gauge coupling of technicolor interaction,
 $\alpha_{new}$ is the gauge coupling constant
 by which the scale of new physics is specified.
Both coefficients $c_{TC}$ and $c_{new}$ take positive values
 so that the gauge interactions are the asymptotically free ones,
 and these magnitude is supposed to be of the order unity.
The technicolor condensate scale $M_{weak}$ (weak scale)
 and the new physics scale $M_{new}$ are roughly defined by
\begin{eqnarray}
 M_{weak} &:& \qquad \alpha_{TC}(M_{weak}) = 1,
\\
 M_{new} &:& \qquad \alpha_{TC}(M_{new}) = 1.
\end{eqnarray}
These scales are defined as the scales
 at which the non-perturbative effects of the gauge interactions
 become important.
The hierarchy $M_{new} \gg M_{weak}$
 can be obtained by choosing $\alpha_{TC}(M)$ without the fine tuning.
For example, consider the case $M_{new} / M_{weak} \simeq 10^{13}$.
Then, we can choose $\alpha_{TC}(M)$ as
\begin{equation}
 {{1 - \alpha_{new}^{-1}(M)} \over {c_{new}}}
 - {{1 - \alpha_{TC}^{-1}(M)} \over {c_{TC}}}
 = \ln {{M_{new}} \over {M_{weak}}} \simeq 32
\end{equation}
 without the fine tuning.
The point is that
 the large difference of the scales only comes through the form of logarithm.

To generate the masses of the quarks and leptons,
 we must introduce new interactions
 which act the role of the Yukawa couplings in the Standard model.
The technifermion condensate
 breaks the chiral symmetry in the technifermion sector,
 but it does not break the one in the ordinary fermion sector.
These two chiral symmetry should be connected
 to generate the ordinary fermion mass
 through the technifermion condensation.
The new interactions
 should reduce these two chiral symmetry to a single one.
The typical form of the new interaction is
\begin{equation}
 {\cal L}_{new} = {{g_L^\psi g_R^\psi} \over {M_\psi^2}}
                   \left( {\bar T}_{Li} T_R \right)
                   \left( {\bar \psi}_R \psi_L^i \right) + h.c.,
\label{effective-4F}
\end{equation}
 where $\psi$ denotes the ordinary fermion,
 and $i$ is the index of the $SU(2)_L$ doublet.
$g_L^\psi$ and $g_R^\psi$ are the coupling constants
 and $M_\psi$ is the appropriate energy scale.
This interaction is not invariant
 under the independent chiral rotation of the ordinary fermions
 or the technifermions,
 but it is invariant under the simultaneous chiral rotation of both fermions
 (Of course, it must be invariant under the electroweak gauge transformation).
The ordinary fermions get their masses
\begin{equation}
 m_\psi \simeq {{g_L^\psi g_R^\psi} \over {M_\psi^2}}
               \left| \langle {\bar T}_L T_R \rangle \right|
\end{equation}
 through the technifermion condensate $\langle {\bar T}T \rangle \neq 0$
 \footnote{The mean field approximation is used.
           We will treat it more rigorously in next chapter.}.
The variety of the masses, the quark mixing angles,
 and the CP violation in the weak interaction
 originate from the structure of the coupling $g_L^\psi g_R^\psi / M_\psi^2$.
As the origin of these interaction,
 there is an interesting idea so called extended technicolor theory\cite{ETC}.

In extended technicolor theory,
 the technicolor gauge interaction is extended
 to generate the above interactions
 as the effective four fermion interactions.
The extended gauge multiplets
 contain both ordinary fermions and technifermions.
\begin{equation}
 \overbrace{
  \begin{array}{cccccc}
  ( & \psi_1 & \psi_2 & \psi_3 &
  \underbrace{
   \begin{array}{cccc}
     & T_1 & T_2 & T_3
   \end{array}
  }_{\mbox{Technicolor}}
   & )
  \end{array}
 }^{\mbox{Extended technicolor}}
\end{equation}
When the extended gauge group
 is spontaneously broken to the technicolor gauge group
 by the effect of certain dynamics,
 the gauge bosons which cause
 the transition of the ordinary fermions to the technifermions
 (so called sideways gauge bosons) become massive.
Such gauge bosons generate the effective four fermion interaction like
\begin{equation}
 {\cal L}_{4F} =
  \left( {\bar T}_L {{g_L^\psi} \over {\sqrt{2}}} \gamma_\mu \psi_L \right)
  {1 \over {-M_\psi^2}}
  \left( {\bar \psi}_R {{g_R^\psi} \over {\sqrt{2}}} \gamma_\mu T_R \right)
 + h.c...
\end{equation}
Firtz transformation gives
 the four fermion interaction as eq.(\ref{effective-4F}).

The mass difference between the ordinary fermions is due to
 the difference of the scales $M_\psi$ and the couplings
 $g_L^\psi$ and $g_R^\psi$.
It is naturally expected that
 the mass difference between the different generations
 comes from the difference of the scales $M_\psi$,
 and the mass splitting in each generation
 comes from the difference of the couplings $g_L^\psi$ and $g_R^\psi$.
The mass splitting in a weak doublet
 comes from the difference of $g_R^\psi$,
 since $g_L^\psi$ must be common in a weak doublet
 according to the gauge invariance.
For example, $g_R^\psi$ for u-quark
 should be different from the one for d-quark: $g_R^u \neq g_R^d$.
Therefore, the extended gauge theory
 should be the chiral gauge theory with many (at least three) mass scales.

No concrete realistic model of the extended technicolor theory
 has been proposed until now,
 since the dynamics of the gauge theory,
 especially for the dynamics of the chiral gauge theory,
 is not completely understood.
The dynamics of the tumbling gauge theory is expected as the dynamics
 of the extended technicolor gauge interaction.
The theory is a class of the chiral gauge theory
 in which many mass scales are expected.
It will be discussed in detail in chapter 6.

In next section we introduce a model so called one-family model
 \cite{one-family}.
While the model is not realistic,
 it is worth studying to get the insight
 about what phenomenology generally appears in the technicolor theory.

\section{One-family model}

Consider the technicolor gauge group $SU(N_{TC})$,
 and introduce the one family technifermions
 (with right-handed techni-neutrino)
\begin{eqnarray*}
 \left( \begin{array}{c} U \\ D \end{array} \right)_L &\sim& (2, 1/6),
\\
 U_R \quad &\sim& (1,2/3),
\\
 D_R \quad &\sim& (1,-1/3),
\\
 \left( \begin{array}{c} N \\ E \end{array} \right)_L &\sim& (2, -1/2),
\\
 N_R \quad &\sim& (1,0),
\\
 E_R \quad &\sim& (1,-1),
\end{eqnarray*}
 as the fundamental representation of the gauge group.
Here,
 $(2,1/6)$ means the $SU(2)_L$ doublet with hypercharge $1/6$, and so on.
The one-family technifermion is considered to cancel the gauge anomalies
 \footnote{In view of the gauge anomaly cancellation, the one-doublet model
           in which the one weak doublet technifermion is introduced
           is more simple.
           But since the technifermions have the unusual electric charge
           $Q = \pm 1/2$ in that case, it is not cosmologically acceptable.}.
The technifermions condensate is expected as
\begin{equation}
 \langle {\bar U}U \rangle = \langle {\bar D}D \rangle
 = \langle {\bar N}N \rangle = \langle {\bar E}E \rangle \neq 0
\end{equation}
 like the quark condensate in QCD.
There is the (approximate) chiral symmetry $SU(8)_L \times SU(8)_R$
 in the technifermion sector,
 and it is spontaneously broken to $SU(8)_V$ by the condensates.
Therefore, $64$ Nambu-Goldstone bosons should exist. The contents are
\begin{eqnarray}
 \theta^{ai} &\sim& {\bar Q} i \gamma_5 \lambda^a \tau^i Q,
\nonumber\\
 \theta^a &\sim& {\bar Q} i \gamma_5 \lambda^a Q,
\nonumber\\
 T^{\alpha i} &\sim& {\bar Q}^\alpha i \gamma_5 \tau^i L,
 \qquad\qquad ({\bar T}^{\alpha i} \sim {\bar L}^\alpha i \gamma_5 \tau^i Q),
\nonumber\\
 T^\alpha &\sim& {\bar Q}^\alpha i \gamma_5 L,
 \qquad\qquad\quad
              ({\bar T}^\alpha \sim {\bar L}^\alpha i \gamma_5 Q),
\nonumber\\
 \Pi^i &\sim& {\bar Q} i \gamma_5 \tau^i Q + {\bar L} i \gamma_5 \tau^i L,
\nonumber\\
 P^i &\sim& {\bar Q} i \gamma_5 \tau^i Q - 3 {\bar L} i \gamma_5 \tau^i L,
\nonumber\\
 P^0 &\sim& {\bar Q} i \gamma_5 Q - 3 {\bar L} i \gamma_5 L,
\label{NG}
\end{eqnarray}
 where $Q = \left( \begin{array}{cc} U & D \end{array} \right)^T$
 and $L = \left( \begin{array}{cc} N & E \end{array} \right)^T$.
There are $56$ colored pseudoscalars:
 $\theta^{ai}$ and $\theta^a$ are the color octets (index $a$),
 $T^{\alpha i}$ and $T^\alpha$ are the color triplets (index $\alpha$)
 which are called leptoquarks.
$\Pi^i$ are identified as
 the would be Nambu-Goldstone bosons which are eaten by the weak gauge bosons.
There are four other color singlet Nambu-Goldstone bosons: $P^i$ and $P^0$.
$P^3$ and $P^0$ is called axions
 which are the singlets of both QCD and QED.

The current of the broken chiral symmetry is
\begin{equation}
 J^a_\mu = \left( \begin{array}{cc} {\bar Q} & {\bar N} \end{array} \right)
           \gamma_\mu \gamma_5 T^a
           \left( \begin{array}{c} Q \\ N \end{array} \right),
\end{equation}
 where $T^a$ is the generators of $SU(8)$,
 and the decay constant of the Nambu-Goldstone bosons $F_\pi$ is defined as
\begin{equation}
 \langle 0 | J^a_\mu | NG^b(k) \rangle = i k_\mu \delta^{ab} F_\pi.
\label{decay-const}
\end{equation}
The current contains the weak axial current $j^i_\mu$ as
\begin{equation}
 j^i_\mu = \sqrt{N_D} J^{a=i}_\mu,
\end{equation}
 where $N_D=4$ is the number of the weak doublet and
\begin{equation}
 T^{a=i} = {1 \over {\sqrt{N_D}}}
  \left(
  \begin{array}{cccc}
   {{\tau^i} \over 2} & & & \\
   & {{\tau^i} \over 2} & & \\
   & & {{\tau^i} \over 2} & \\
   & & & {{\tau^i} \over 2} \\
  \end{array}
  \right)
\end{equation}
 since the normalization $\mbox{tr}(T^a T^b) = \delta^{ab}/2$.
Therefore,
 the mass of the weak bosons due to the Higgs mechanism is
\begin{eqnarray}
 M_W^2 &=& \left( {{g_2} \over 2} \right)^2 N_D F_\pi^2,
\\
 M_Z^2 &=& M_W^2 / \cos^2 \theta_W.
\end{eqnarray}
This means that the larger $N_D$ results smaller $F_\pi$.

The condensate $\langle {\bar T} T \rangle$ ($T = U$, $D$, $N$, or $E$)
 is related with the decay constant.
If we assume the Euclidean mass function of the technifermion as
\begin{equation}
 \Sigma(x) = \left\{
             \begin{array}{cc}
               m & \quad x \leq m^2 \\
               0 & \quad x > m^2,
             \end{array}
             \right.
\label{mass-approx}
\end{equation}
 where $m$ is the typical scale of the dynamics, then we get
\begin{eqnarray}
 \langle {\bar T}_L T_R \rangle
  &=& - {{N_{TC} m^3} \over {8\pi^2}} \left( 1 - \ln 2 \right),
\\
 F_\pi^2 &=& {{N_{TC} m^2} \over {8\pi^2}} \left( 2 \ln 2 - 1 \right).
\end{eqnarray}
The condensate is calculated by the integration of the propagator,
 and the decay constant is calculated by using the Pagels-Stokar formula.
We find the relation
 \footnote{This is the naive relation.
           To tell the truth, the condensate should depend on
           the renormalization point, but the decay constant do not.
           We must use the technique of the operator product expansion
           for more rigorous arguments.}
\begin{eqnarray}
 \langle {\bar T}_L T_R \rangle
  &=& - 2\pi F_\pi^3 \sqrt{{2 \over {N_{TC}}}}
        {{1 - \ln 2} \over {\left( 2 \ln 2 - 1 \right)^{3/2}}}
\\
 &\simeq& - 2\pi F_\pi^3.
\end{eqnarray}
Similar relation
 $\langle {\bar T} T \rangle \simeq - 4\pi F_\pi^3$ is obtained
 from the argument of the naive dimensional analysis
 in the chiral perturbation theory\cite{Manohar-Georgi}.

To generate the mass of ordinary fermions,
 the technicolor interaction is extended.
The multiplet of the extended gauge group
 should be very complicated one to generate the realistic mass spectrum.
No one knows the realistic gauge group and representations.
Here, we naively consider the extended gauge group $SU(N_{TC}+3)$
 and the fundamental representations as follows
 to explain the general idea of the mass generation.
\begin{equation}
 \underbrace{
 \begin{array}{cccccc}
  ( & u & c & t &                  U^1 \cdots U^{N_{TC}} & )_{L,R} \\
  ( & d & s & b &                  D^1 \cdots D^{N_{TC}} & )_{L,R} \\
  ( & \nu_e & \nu_\mu & \nu_\tau & N^1 \cdots N^{N_{TC}} & )_{L,R} \\
  ( & e & \mu & \tau &             \underbrace{
                                   E^1 \cdots E^{N_{TC}}
                                   }_{\mbox{Technicolor}} & )_{L,R} \\
 \end{array}.
 }_{\mbox{Extended technicolor}}
\end{equation}
The right-handed neutrinos are introduced for simplicity.
We assume the dynamics
 which hierarchically breaks the extended gauge group
 to the technicolor gauge group.
Three hierarchical masses scales
 are assumed for each generations: $M_1$, $M_2$, and $M_3$.
At the scale $M_1$, $SU(N_{TC}+3)$ breaks to $SU(N_{TC}+2)$,
 and the first generation is decoupled from the gauge multiples.
Some gauge bosons get mass of order $M_1$ at this stage.
It is the same for second and third generations.
Then, the up-type quarks, for instance, get the following masses.
\begin{eqnarray}
 m_u &=& {{g_{ETC}^2(M_1)} \over {M_1^2}} \langle {\bar U}_L U_R \rangle
    \simeq {{g_{ETC}^2(M_1)} \over {M_1^2}} 2\pi F_\pi^3,
\\
 m_c &=& {{g_{ETC}^2(M_2)} \over {M_2^2}} \langle {\bar U}_L U_R \rangle
    \simeq {{g_{ETC}^2(M_2)} \over {M_2^2}} 2\pi F_\pi^3,
\\
 m_t &=& {{g_{ETC}^2(M_3)} \over {M_1^3}} \langle {\bar U}_L U_R \rangle
    \simeq {{g_{ETC}^2(M_3)} \over {M_3^2}} 2\pi F_\pi^3.
\end{eqnarray}
The down-type quarks have the same mass spectrum with up-type,
 since we are considering the vector-like extended gauge group.
So this naive model is not realistic.
If we consider the flavor dependent condensation like
\begin{equation}
 \langle {\bar U} U \rangle
 \neq \langle {\bar D} D \rangle
 \neq \langle {\bar N} N \rangle
 \neq \langle {\bar E} E \rangle,
\end{equation}
 we can get the mass splitting in a weak doublet.
But these splitting of the condensates
 cause the large effect on the $\rho$ parameter, in general.
This will be discussed in the next chapter in detail.

Since the chiral symmetry of the technifermion is explicitly broken
 by the extended technicolor and the standard gauge group,
 some Nambu-Goldstone bosons in (\ref{NG}) become massive
 (pseudo-Nambu-Goldstone bosons)
 \footnote{The extended technicolor interaction
            and the standard strong and electroweak interactions
            are considered perturbatively.}.
These masses can evaluated by the Dashen's formula\cite{Dashen}
 \footnote{This formula is derived
           by using the current algebra and PCAC relation.}
\begin{equation}
 m_\pi^2 =
  {1 \over {F_\pi^2}} \langle 0 | [Q_\pi, [ Q_\pi, {\cal H}]] | 0 \rangle,
\end{equation}
 where $Q_\pi$ is the charge corresponding to the Nambu-Goldstone boson $\pi$,
 and ${\cal H}$ is the Hamiltonian density.
But here, we briefly estimate it in the effective theoretical point of view
 \footnote{A detailed estimation has been given
            in ref.\cite{Dimopoulos-Raby-Kane}}.

We can consider the coupling
 between the color octet Nambu-Goldstone bosons and gluons as
\begin{equation}
 {\cal L}_{eff}
  = \mbox{tr}\left\{ (D_\mu \theta)^{\dag} (D^\mu \theta) \right\},
\end{equation}
 where
\begin{eqnarray}
 D_\mu \theta &=& \partial_\mu - i g_3 [G_\mu, \theta],
\\
 G_\mu &=& G_\mu^a {{\lambda^a} \over 2},
\\
 \theta &=& \theta^a {{\lambda^a} \over 2}.
\end{eqnarray}
Namely,
 we assume the standard gauge coupling of the elementary colored scalar field.
The mass is obtained by evaluating the diagram of fig.(\ref{NG-mass}).
\begin{figure}[t]
\caption{The diagrams for the mass of the Nambu-Goldstone bosons.}
\label{NG-mass}
\end{figure}
Since the Nambu-Goldstone bosons are the composite particles,
 we assume the form factor as
\begin{equation}
 g_3 \rightarrow g_3 \cdot {{M_V^2} \over {M_V^2 - k^2}},
\end{equation}
 where $k^2$ is the momentum of the gluon
 and $M_V$ is the mass of the vector boundstate
 composed by the technifermions which can mix with gluons.
We get the value as
\begin{equation}
 M_V = m_\rho {{F_\pi} \over {f_\pi}} \sqrt{{3 \over {N_{TC}}}}
     \simeq \sqrt{{3 \over {N_{TC}}}} \cdot 1\mbox{TeV}
\end{equation}
 by the naive scaling up from the $\rho$ meson mass in QCD.
Then, we get
\begin{eqnarray}
 m_\theta^2 &\simeq& {9 \over {4\pi}} C_2(8) \alpha_s M_V^2
\nonumber\\
            &\simeq& (460\mbox{GeV})^2 {3 \over {N_{TC}}},
\end{eqnarray}
 where $C_2(8)=N_c$ (Casimir coefficient), and we set $\alpha_s = 0.1$.
This is the same for color triplet Nambu-Goldstone bosons.
We get
\begin{eqnarray}
 m_T^2 &\simeq& {9 \over {4\pi}} C_2(3) \alpha_s M_V^2
\nonumber\\
        &\simeq& (300\mbox{GeV})^2 {3 \over {N_{TC}}}.
\end{eqnarray}
The mass of the triplet is smaller than that of the octet,
 because of the difference of the Casimir coefficient.
The mass of the charged Nambu-Goldstone bosons
 is also estimated in the same way by considering photon instead of gluon.
The result is
\begin{eqnarray}
 m_{P^\pm}^2 &\simeq& {9 \over {4\pi}} \alpha M_V^2
\nonumber\\
       &\simeq& (70\mbox{GeV})^2 {3 \over {N_{TC}}}.
\end{eqnarray}
The mass of $P^3$ and $P^0$ is expected to be very small,
 since they do not couple with the gluon and photon.
Only the explicit breaking of the chiral symmetry
 due to the weak interaction, the extended technicolor interaction,
 and the other additional interactions mediated by the massive particles
 is the small source of their mass.
The explicit breaking due to the extended technicolor interaction
 is highly model dependent.
If these masses can not become large,
 and the coupling of these bosons with the ordinary fermions
 (Yukawa coupling) is not small enough,
 this model is phenomenologically ruled out.

The Yukawa couplings
 between the pseudo-Nambu-Goldstone bosons and ordinary fermions
 are estimated by considering the diagram of fig.(\ref{Yukawa}).
\begin{figure}[t]
\caption{The diagrams for the Yukawa coupling.}
\label{Yukawa}
\end{figure}
We use the mass function of eq.(\ref{mass-approx}),
 and the formula of Bethe-Salpeter amplitude
\begin{equation}
 F_\pi P^a(k,k) = 2 i \Sigma(k^2) T^a
\end{equation}
 which has already derived in the previous chapter.
The coupling depend on the generation and flavor of the ordinary fermions as
\begin{equation}
 \left( g_Y^a \right)^{ij}_n F_\pi
  = N_{TC} {{g_L^i g_R^j} \over {16\pi^2}} \left( T^a \right)^{ij}
    {m \over {M_n^2}} \int_0^{m^2} dk^2 {{k^2} \over {m^2+k^2}},
\end{equation}
 where $i$ and $j$ are the indexes of the flavor
 and $n$ is the index of the generation.
On the other hand,
 we get the mass of ordinary fermions by evaluation the diagram
 of fig.(\ref{mass-diagram}).
\begin{figure}[t]
\caption{The diagrams for the masses of the quarks and leptons.}
\label{mass-diagram}
\end{figure}
We get
\begin{equation}
 m^{ij}_n = 2 N_{TC} {{g_L^i g_R^j} \over {16\pi^2}} {m \over {M_n^2}}
            \int_0^{m^2} dk^2 {{k^2} \over {m^2+k^2}}.
\end{equation}
Therefore he have the relation
\begin{equation}
 \left( g_Y^a \right)^{ij}_n F_\pi = 2 \left( T^a \right)^{ij} m^{ij}_n.
\end{equation}
This is nothing but the Goldberger-Treiman relation
 \footnote{In QCD, the Yukawa coupling of the nucleons with the pions,
           $g_{\pi NN}$, is given by the Goldberger-Treiman relation
           $g_{\pi NN} f_\pi = 2 g_A m_N$ \cite{Goldberger-Treiman},
           where $g_A$ is the axial vector coupling of nucleon
           and $m_N$ is the mass of nucleon.}.
This formula is similar to the mass formula in the standard model,
 $m_\psi = g_Y^\psi v / \sqrt{2}$,
 where $v$ is the vacuum expectation value of the elementary Higgs field.

There is another important couplings for the pseudo-Nambu-Goldstone bosons.
Since the axial current have the anomaly, in general,
 we have the coupling like in fig.(\ref{triangle})\cite{Ellis-et-al}.
\begin{figure}[t]
\caption{Coupling through the anomaly.}
\label{triangle}
\end{figure}
This coupling is important
 for the production and decay of the pseudo-Nambu-Goldstone bosons.
The strength of the coupling can be obtained as follows.

We have the anomalous Ward-Takahashi identity
\begin{equation}
 \partial^\alpha J^a_\alpha (x)
  = - \left( \delta {\cal L} \right)^a
    + {{g^{(b)} g^{(c)} S^{abc}} \over {16\pi^2}}
      \epsilon^{\mu\nu\rho\sigma} F^b_{\mu\nu}(x) F^c_{\rho\sigma}(x),
\end{equation}
 where $\left( \delta {\cal L} \right)^a$ means the explicit breaking
 (See chapter 2),
 $F^b_{\mu\nu}$ and $F^c_{\rho\sigma}$ are the gauge field strength,
 $g^{(b)}$ and $g^{(c)}$ are the corresponding gauge coupling constant,
 and
\begin{equation}
 S^{abc} = {1 \over 2} \mbox{tr} \left( T^a \{ T^b, T^c \} \right)
\end{equation}
 with the generators of the gauge interactions $T^b$ and $T^c$
 ($8 \times 8$ matrixes acting on the one family technifermions).
We neglect the effect of the explicit breaking
 as the first approximation of the perturbation on it.
Consider the matrix element
\begin{equation}
 \langle A^b(k_1) A^c(k_2) | \partial^\alpha J^a_\alpha (0) | 0 \rangle
 = {{g^{(b)} g^{(c)} S^{abc}} \over {16\pi^2}} \epsilon^{\mu\nu\rho\sigma}
   \langle A^b(k_1) A^c(k_2) | F^b_{\mu\nu}(0) F^c_{\rho\sigma}(0) | 0 \rangle,
\end{equation}
 where $A^b$ and $A^c$ are the gauge fields
 corresponding the the strength $F^b_{\mu\nu}$ and $F^c_{\rho\sigma}$,
 respectively.
The right-hand side becomes
\begin{equation}
 i \int d^4 e^{-ikx} [\mbox{right-hand side}]
 = (2\pi)^4 i \delta^4 (k_1+k_2-k)
   {{g^{(b)} g^{(c)} S^{abc}} \over {4\pi^2}} \epsilon_{\mu\nu\rho\sigma}
   \epsilon^*(k_1)^\mu \epsilon^*(k_2)^\nu k_1^\rho k_2^\sigma
\end{equation}
 in the lowest order of the perturbation on the gauge interactions.
To evaluate the left-hand side, we assume
\begin{equation}
 \partial^\alpha J^a_\alpha \simeq - F_\pi \Box \Pi^a,
\end{equation}
 since the Nambu-Goldstone theorem says that
 the Nambu-Goldstone boson couple with the current as eq.(\ref{decay-const}).
By assuming of this operator relation,
 the higher contribution, the axial vector resonance, for instance,
 is neglected.
Therefore,
 the following result is valid
 only for the low energy pseudo-Nambu-Goldstone bosons.
Then,
\begin{eqnarray}
 i \int d^4x e^{-ikx} [\mbox{left-hand side}]
 &\simeq& - i F_\pi \int d^4x e^{-ikx} \Box \langle A^b(k_1) A^c(k_2) |
                                   \Pi^a(x) | 0 \rangle
\nonumber\\
 &=& - F_\pi \langle A^b(k_1) A^c(k_2) | \Pi^a(k) \rangle,
\end{eqnarray}
 where we used the Lehmann-Symanzik-Zimmermann reduction formula.
Therefore, we obtain
\begin{equation}
 \langle A^b(k_1) A^c(k_2) | \Pi^a(k) \rangle
 = - (2\pi)^4 i \delta^4 (k_1+k_2-k)
   {{g^{(b)} g^{(c)} S^{abc}} \over {4\pi^2}} \epsilon_{\mu\nu\rho\sigma}
   \epsilon^*(k_1)^\mu \epsilon^*(k_2)^\nu k_1^\rho k_2^\sigma,
\end{equation}
 and the vertex of the diagram of fig.(\ref{triangle})
\begin{equation}
 \Gamma^{abc}_{\mu\nu}(k_1,k_2)
 = - {{g^{(b)} g^{(c)} S^{abc}} \over {4\pi^2}}
     \epsilon_{\mu\nu\rho\sigma} k_1^\rho k_2^\sigma.
\end{equation}
If the two gauge bosons are the same ones
 (for example, $\langle \gamma\gamma | \Pi^a \rangle$),
 the factor $2$ must be multiplied.

Finally, we briefly review
 the experimental bound on the mass of the pseudo-Nambu-Goldstone bosons.
The quarkonium can decay into the axions $P^3$ and $P^0$,
 through the anomaly induced vertex as fig.(\ref{quarkonium-decay}).
\begin{figure}[t]
\caption{Quarkonium going to $P^3 \gamma$ or $P^0 \gamma$.}
\label{quarkonium-decay}
\end{figure}
{}From $\Upsilon (1S)$ decay, we have the bound
\begin{equation}
 m_{P^0, P^3} > 9\mbox{GeV}
\end{equation}
 for the axions with decay constant $F_\pi = 125\mbox{GeV}$
 (one-family model)\cite{axion-limit}.
The colored Pseudo-Nambu-Goldstone bosons
 can be produced in hadron collider
 by the process in fig.(\ref{production-colored}).
\begin{figure}[t]
\caption{$\theta^i_a$ or $\theta_a$ production at the hadron collider.}
\label{production-colored}
\end{figure}
We have the bound
\begin{equation}
 m_\theta > 86\mbox{GeV}
\end{equation}
 for the color octet stable particles\cite{colored-limit}.
The leptoquarks, $T^{\alpha i}$ and $T^\alpha$,
 can be produced in $S$-channel at the $ep$ collider HERA.
We have the bound\cite{leptoquark-limit}
\begin{equation}
 m_T > 98 \sim 121 \mbox{GeV}.
\end{equation}
The charged color singlet Pseudo-Nambu-Goldstone bosons, $P^\pm$,
 can be produced in pair at LEP.
We have the bound
\begin{equation}
 m_{P^\pm} > 41.7 \mbox{GeV}
\end{equation}
 as the bound on the charged Higgs\cite{charged-limit}.
\chapter{Phenomenology of the Technicolor theory}

In this chapter
 we discuss the phenomenological bounds on the technicolor models.
Flavor-changing neutral current (FCNC) problem
 is discussed in the first section.
The solution of the problem by the large anomalous dimension dynamics
 is explained in the second section.
The bound from the recent precision experiments
 is discussed in the third section.
The prediction of the technicolor theory
 on the oblique correction $S$, $T$, and $U$
 will be compared with the experiments.
The bound through the non-oblique correction
 is also discussed in the fourth section.

\section{Flavor-changing neutral current problem}

In naive model of the extended technicolor theory,
 pseudo-Nambu-Goldstone bosons mediate large FCNC.
The Yukawa couplings
 between the pseudo-Nambu-Goldstone bosons and the ordinary fermions
 depend on the models of the extended technicolor interaction.
Consider the general two doublet Higgs system with Yukawa coupling
\begin{eqnarray}
 {\cal L}_{Yukawa}
  &=& \left( g^u_1 \right)^i{}_j {\bar \psi}_{Li} {\tilde \Phi}_1 u_R^j
   + \left( g^u_2 \right)^i{}_j {\bar \psi}_{Li} {\tilde \Phi}_2 u_R^j
\nonumber\\
  &+& \left( g^d_1 \right)^i{}_j {\bar \psi}_{Li} \Phi_1 d_R^j
   + \left( g^d_2 \right)^i{}_j {\bar \psi}_{Li} \Phi_2 d_R^j + h.c.
\nonumber\\
  &+& \mbox{(lepton sector)},
\end{eqnarray}
 where ${\tilde \Phi} = i \tau^2 \Phi^*$.
If we consider the one-family model,
 the fields $\Phi_1$ and $\Phi_2$ can be identified
 as the fields which contain $\Pi^i$ and $P^i$, respectively.
Assume that only $\Phi_1$ get the vacuum expectation value
\begin{equation}
 \langle \Phi_1 \rangle = {v \over {\sqrt{2}}}
  \left( \begin{array}{c} 0 \\ 1 \end{array} \right).
\end{equation}
Then, we get the quark mass matrixes
\begin{eqnarray}
 \left( M_u \right)^i{}_j &=& \left( g^u_1 \right)^i{}_j {v \over {\sqrt{2}}},
\\
 \left( M_d \right)^i{}_j &=& \left( g^d_1 \right)^i{}_j {v \over {\sqrt{2}}}.
\end{eqnarray}
These mass matrixes should be diagonalized to go to the mass eigenstate.
It can be done by the bi-unitary chiral transformation
\begin{eqnarray}
 \left( U^u_L \right)^{\dag} M_u U^u_R &=&
  \left( \begin{array}{ccc}
          m_u &    &     \\
              &m_c &     \\
              &    & m_t \\
         \end{array}
  \right),
\\
 \left( U^d_L \right)^{\dag} M_d U^d_R &=&
  \left( \begin{array}{ccc}
          m_d &    &     \\
              &m_s &     \\
              &    & m_b \\
         \end{array}
  \right).
\end{eqnarray}
Though the Yukawa couplings of $\Phi_1$ ($g^u_1$ and $g^d_1$)
 become diagonal by this transformation,
 the Yukawa couplings of $\Phi_2$ ($g^u_2$ and $g^d_2$)
 do not always simultaneously become diagonal.
Therefore, $\Phi_2$ can mediate the FCNC at tree level, in general.
For example, it causes the large mixing of $K^0$ and ${\bar K}^0$
  ($|\Delta S| = 2$ transition) as fig.(\ref{FCNC-pseudo}).
\begin{figure}[t]
\caption{Example of FCNC: $K^0$-${\bar K}^0$ mixing
         mediated by the pseudo-Nambu-Goldstone boson.
         The quarks with suffix $m$ denote the mass eigenstate.}
\label{FCNC-pseudo}
\end{figure}
There exists FCNC in the standard model,
 but it is the small one loop effect
 suppressed by the GIM mechanism\cite{GIM}.
The prediction of the standard model on the FCNC
 is consistent with the experiments.
So, this tree level contribution is dangerous.
The off diagonal element of the Yukawa couplings, $g^u_2$ and $g^d_2$,
 must be very small, or the mass of $\Phi_2$ is heavy.
We have the constraint
 so that this tree level contribution is smaller
 than the contribution of the standard model.
The stringent constraint comes from the $K^0$-${\bar K}^0$ mixing
\begin{equation}
 \left( {{\left( g_Y^d \right)_{sd}} \over {M_P}} \right)^2
  < 10^{-12} \mbox{GeV}^{-2}.
\end{equation}
It is naturally expected that
 $\left( g_Y^d \right)_{sd}$ is of the order of the Yukawa coupling
 of the strange quark $g_Y^s = \sqrt{2} m_s / v \simeq 5 \times 10^{-4}$,
 where $m_s \simeq 83\mbox{MeV}$ (at $300\mbox{GeV}$).
Then we get
\begin{equation}
 M_P > 500 \mbox{GeV}.
\end{equation}
We must naturally explain
 why the off diagonal elements of the Yukawa couplings are so small,
 if the masses of the pseudo-Nambu-Goldstone bosons
 are smaller than a few hundred GeV.
In the naive model,
 many pseudo-Nambu-Goldstone bosons are possibly lighter than a few GeV,
 which is estimated in previous chapter.

The gauge bosons of the extended technicolor interaction
 can also mediate the FCNC.
The massive gauge boson exchanges
 generate the effective four fermion interactions like fig.(\ref{FCNC-ETC}a),
 where quarks are at the eigenstate of the gauge interactions.
These four fermion interactions become the FCNC interaction,
 when we go to the mass eigenstate of the quarks.
The four fermion interaction of fig.(\ref{FCNC-ETC}a)
 results $K^0$-${\bar K}^0$ mixing as fig.(\ref{FCNC-ETC}b).
\begin{figure}[t]
\caption{Example of FCNC: $K^0$-${\bar K}^0$ mixing
         mediated by the extended technicolor gauge bosons.
         (a) The process in gauge eigenstate.
         (b) The same process in mass eigenstate.
             The Cabibbo angle $\theta_c$
             is taken as the typical mixing angle.}
\label{FCNC-ETC}
\end{figure}
We have the constraint for the mass and coupling of the gauge boson
\begin{equation}
 \left( {{g_{ETC} \sin \theta_c} \over M} \right)^2
  < 10^{-12} \mbox{GeV}^{-2}.
\end{equation}
Since $\sin^2 \theta_c \simeq 1/20$, we obtain
\begin{equation}
 \left( {{g_{ETC}} \over M} \right)^2
  < \left( 200 \mbox{TeV} \right)^{-2}.
\label{FCNC-bound}
\end{equation}
The squared coupling $g_{ETC}^2$ is the largest one
 in $g_R(M)^2$, $g_L(M)^2$, and $g_R(M) \times g_L(M)$,
 where $g_L$ and $g_R$ are the extended technicolor effective gauge couplings
 for the left-handed fermions and the right-handed fermions, respectively.

The mass of the gauge boson $M$
 likely to be equal to the mass of the sideways boson
 which is related with the strange quark mass
 \footnote{In the naive one-family model in the previous section,
           the sideways bosons of the strange quark
           is associated with the broken off diagonal generator
           in $SU(N_{TC}+2) \rightarrow SU(N_{TC}+1)$ breaking.
           The gauge boson which mediates FCNC
           is associated with the broken diagonal generator of the breaking.}.
Namely, we have
\begin{equation}
 m_s = {{g_L(M) g_R(M)} \over {M^2}} 2\pi F_\pi^3.
\end{equation}
Therefore we get
\begin{equation}
 {{g_L(M) g_R(M)} \over {M^2}}
  = {{m_s} \over {2\pi F_\pi^3}} \simeq \left( 10 \mbox{TeV} \right)^{-2},
\end{equation}
 where we use $F_\pi=125\mbox{GeV}$ by assuming the one-family model.
There are about three order excess from the bound of eq.(\ref{FCNC-bound}).

In general, in the naive extended technicolor model,
 there are large additional FCNCs
 which are mediated by the pseudo-Nambu-Goldstone bosons
 and the extended technicolor gauge bosons.
This is called FCNC problem in the extended technicolor theory\cite{FCNC}.
Of course, we can consider the complicated models to avoid FCNCs:
 complicated representation of the extended technicolor multiplet,
 complicated breaking pattern of the extended technicolor gauge group, etc.
But it is unnatural.
Technicolor theory had once died of this problem.
But it was pointed out that
 if we consider the special dynamics for the technicolor,
 the masses of the pseudo-Nambu-Goldstone bosons and
 the extended technicolor gauge bosons can be large
 without the change of the masses of the ordinary fermions and weak bosons.
The dynamics is called the large anomalous dimension dynamics\cite{large}.
In next section, we explain how the dynamics solves the problem.

\section{Technicolor with large anomalous dimension}

We know that
 the non-Abelian gauge theory has the nature of the asymptotic freedom.
The gauge interaction goes down at the high energy.
This is the reason that
 the mass function of the technifermion dumps as $\Sigma(x) \sim 1/x$
 for large Euclidean squared momentum $x=p^2$.
If the technicolor interaction does not quickly go down at the high energy,
 the dump of the mass function becomes slow as
\begin{equation}
 \Sigma(x) \sim {1 \over x} \left( {x \over {m_T^2}} \right)^{\gamma_m/2}
\end{equation}
 with the anomalous dimension $\gamma_m$,
 where $m_T$ is the typical scale of the dynamics
 (Remember the discussion in section 3.1 and eq.(\ref{asymptotic-form}).).

Two kind of explicit models with large anomalous dimension are proposed.
One is the gauge theory with slowly running coupling
 (walking coupling)\cite{walking},
 another is the system of the gauge plus strong four fermion interaction
 (gauged Nambu-Jona-Lasinio model)\cite{strong-4F}.
Walking dynamics can realize the large anomalous dimension $\gamma_m \simeq 1$.
More large anomalous dimension $\gamma_m \simeq 2$
 can be realized in the gauged Nambu-Jona-Lasinio model,
 since the four fermion interaction does not go down at high energy.

As we will explain in the following,
 if the anomalous dimension is large,
 the masses of the ordinary fermions and the pseudo-Nambu-Goldstone bosons
 are extensively enhanced,
 while the decay constant of the Nambu-Goldstone boson
 (mass of the weak boson) is unchanged.
Enhancement of the ordinary fermion mass
 allows us to consider the heavy extended technicolor gauge bosons.
This is the idea to solve the FCNC problem
 by the large anomalous dimension dynamics\cite{large}.

The mass function of the ordinary fermions can be calculated as
 (fig.(\ref{mass-ordinary}))
\begin{equation}
 \Sigma^f(q)
  = - N_{TC} \int {{d^4p} \over {(2\pi)^4i}}
           {{g_{ETC}} \over \sqrt{2}} \gamma_\mu
           {{1-\gamma_5} \over 2} S(p) {{1-\gamma_5} \over 2}
           {{g_{ETC}} \over \sqrt{2}} \gamma_\nu D^{\mu\nu}(q-p),
\end{equation}
 where $S(p)$ is the technifermion full propagator,
 $g_{ETC}$ is the sideways gauge coupling,
 and $D^{\mu\nu}(q-p)$ is the sideways massive gauge boson propagator.
\begin{figure}[t]
\caption{The diagram for the mass function of the ordinary fermion.}
\label{mass-ordinary}
\end{figure}
In section 3.1, we estimated
 the full technifermion propagator at the large momentum as
\begin{equation}
 \lim_{p \rightarrow \infty}
  {{1-\gamma_5} \over 2} S(p) {{1-\gamma_5} \over 2}
  = i \lim_{p \rightarrow \infty} U_{{\bar T} T}(p,g,\mu)
    {{1-\gamma_5} \over 2} \langle 0 | {\bar T}_R(0) T_L(0) | 0 \rangle,
\end{equation}
\begin{equation}
 \lim_{p \rightarrow \infty} U_{{\bar T} T}(p,g,\mu)
 = c {1 \over i} {3 \over {2N_{TC}}} g^2 C_2
   {1 \over {p^4}} \left( {{p^2} \over {\mu^2}} \right)^{\gamma_m / 2},
\end{equation}
 where we take the Landau gauge in the technicolor interaction
 (see eqs.(\ref{OPE-limit}) and (\ref{Wilson-coefficient}).).
We get the ordinary fermion mass
 which is defined by $m_f \equiv {1 \over 4} \mbox{tr}(\Sigma^f(q=0))$ as
\begin{eqnarray}
 m_f
  &\simeq& - {1 \over 4} N_{TC} \mbox{tr} \int {{d^4p} \over {(2\pi)^4i}}
  {{g_{ETC}} \over \sqrt{2}} \gamma_\mu
  \cdot i
  \cdot c {1 \over i} {3 \over {2N_{TC}}} g^2 C_2
   {1 \over {p^4}} \left( {{p^2} \over {m_T^2}} \right)^{\gamma_m / 2}
\nonumber\\
&& \qquad\qquad\qquad
  \times {{1-\gamma_5} \over 2} {{m_T^3} \over 2}
  {{g_{ETC}} \over \sqrt{2}} \gamma_\nu
  {{g^{\mu\nu}} \over {p^2 - M_f^2}}
\nonumber\\
  &\simeq&
   {{3c} \over {64\pi^2}} (-1)^{\gamma_m /2} g_{ETC}^2 g^2 C_2
   {{m_T^3} \over {M_f^2}}
   \int_0^{M_f^2} dx
    {1 \over x} \left( {x \over {m_T^2}} \right)^{\gamma_m / 2}
\nonumber\\
  &=& C {{g_{ETC}^2 m_T^3} \over {2M_f^2}}
      \left( {{M_f} \over {m_T}} \right)^{\gamma_m},
\end{eqnarray}
 where
 we set $\langle 0 | {\bar T}(0) T(0) | 0 \rangle = m_T^3$ and $\mu = m_T$,
 $M_f$ is the mass of the sideways gauge bosons, and
\begin{equation}
 C = {{3c} \over {32\pi}} g^2 C_2 (-1)^{\gamma_m /2}
     B(2-\gamma_m/2, 2-\gamma_m/2) {\pi \over {\sin \pi \gamma_m/2}}
\end{equation}
 which is expected to be of the order unity ($B(p,q)$ is the beta function).
The naive estimation in the previous chapter by the mean field approximation
\begin{equation}
 m_f \simeq {{g_{ETC}^2} \over {M_f^2}}
            \left| \langle {\bar T}_L T_R \rangle \right|
\end{equation}
 corresponds to the case $\gamma_m = 0$ (QCD-like dynamics).
If the running coupling of the technicolor interaction is constant
 (as an extreme limit of the walking dynamics),
 we get the asymptotic behavior of the technifermion mass function as
 $\Sigma(x) \sim 1/\sqrt{x}$, and get $\gamma_m = 1$
 (Remember the discussion in the last of the section 3.1.).
In that case, the mass of the ordinary fermion is proportional to $1/M_f$
 instead of $1/M_f^2$.
Namely,
 the mass of the ordinary fermion is enhanced by the anomalous dimension.
Therefore, heavier sideways gauge bosons are allowed.
The mass of the sideways bosons
 which is related with the strange quark mass becomes
\begin{equation}
 {{M_2} \over {g_{ETC}^2}} \simeq {{(4\pi F_\pi^3)^{2/3}} \over {2m_s}}
                           \simeq 510 \mbox{TeV}.
\end{equation}
The bound from the $K^0$-${\bar K}^0$ mixing, eq.(\ref{FCNC-bound}),
 can be satisfied if $g_{ETC}$ is of the order of unity.

The masses of the pseudo-Nambu-Goldstone bosons are enhanced by this dynamics.
The mass can be estimated
 by considering the diagram of fig.(\ref{mass-PNGB}).
\begin{figure}[t]
\caption{Diagram for the mass of pseudo-Nambu-Goldstone bosons.}
\label{mass-PNGB}
\end{figure}
The diagram denotes that
 the explicit chiral symmetry breaking by the gauge interactions
 (strong, electromagnetic, weak, extended technicolor interaction,
  or the others which are spontaneously broken)
 makes the mass of the Nambu-Goldstone boson.
Consider the case that
 the chiral symmetry is explicitly broken
 by the additional gauge interaction which is spontaneously broken,
 for example.
This is the case for the axions, $P^3$ and $P^0$, in the one-family model
 (Corresponding chiral symmetries
  are broken by the Pati-Salam gauge interaction\cite{Pati-Salam}.).
The mass can be estimated as
\begin{eqnarray}
 M_a^2
 &=& - \mbox{tr} \int {{d^4p} \over {(2\pi)^4i}} {{d^4k} \over {(2\pi)^4i}}
    P^a(p-k,p-k) {1 \over {-(\not\!p - \not\!k)}}
     g {\cal T}^\alpha \gamma_\mu {1 \over {-\not\!p}}
\nonumber\\
&&\qquad\qquad
    \times
    P^a(p,p) {1 \over {-\not\!p}}
     g {\cal T}^\beta \gamma_\nu
     {1 \over {-(\not\!p - \not\!k)}}
    {{\delta_{\alpha\beta} g^{\mu\nu}} \over {-M^2}}.
\end{eqnarray}
The index $a$ is not summed over.
Since the Bethe-Salpeter amplitude at the zero momentum transfer
 can be written by the mass function of the technifermion as
\begin{equation}
 P^a(p,p) = 2i \gamma_5 T^a {{\Sigma(p)} \over {F_\pi}},
\end{equation}
 we obtain
\begin{equation}
 M_a^2 \simeq
 C \mbox{tr}\left( [T^a, {\cal T}^\alpha] [T^a, {\cal T}^\alpha] \right)
 g^2 {{m_T^4} \over {F_\pi^2}} {{m_T^2} \over {M^2}}
 \left( {{M^2} \over {m_T^2}} \right)^{\gamma_m}
\end{equation}
 by using
\begin{equation}
 \Sigma(x)
  \simeq {{m_T^3} \over x} \left( {x \over {m_T^2}} \right)^{\gamma_m/2}
\end{equation}
 in Euclidean space,
 where $C$ is the dimensionless constant
 which is expected to be of the order of unity,
 and the trace of the generators should not be zero
 if the considering gauge interaction breaks the chiral symmetry
 \footnote{The contribution which is proportional to
           $\mbox{tr}(T^a T^a {\cal T}^\alpha {\cal T}^\alpha)$
           comes from the other diagrams,
           since $\mbox{tr}([T^a, {\cal T}^\alpha] [T^a, {\cal T}^\alpha])
                  = 2\mbox{tr}(T^a {\cal T}^\alpha T^a {\cal T}^\alpha)
                  - 2\mbox{tr}(T^a T^a {\cal T}^\alpha {\cal T}^\alpha)$.}.
If the anomalous dimension is small,
 the squared mass is proportional to the small factor $m_T^2 / M^2$.
But if $\gamma_m=1$, the suppression by $M$ is disappeared.
Suppose that $M=350\mbox{TeV}$
 \footnote{The experimental lower bound
            on the mass of the Pati-Salam gauge boson\cite{Pati-Salam-bound}.
           Pati-Salam gauge interaction in the technifermion sector
            breaks the chiral symmetries which are related with the axions.}.
Then, the mass of the pseudo-Nambu-Goldstone boson is
\begin{equation}
 M^2 \simeq {1 \over {F_\pi^2}}
   \cdot {{\left( 4\pi F_\pi^3 \right)^2} \over {(350 \mbox{TeV})^2}}
   \simeq \left( 560 \mbox{MeV} \right)^2
\end{equation}
 for $\gamma_m=0$, but
\begin{equation}
 M^2 \simeq {1 \over {F_\pi^2}} \cdot \left( 4\pi F_\pi^3 \right)^{4/3}
     \simeq \left( 680 \mbox{GeV} \right)^2
\end{equation}
 for $\gamma_m=1$,
 where we set $m_T^3 = 4\pi F_\pi^3$ and $F_\pi = 125 \mbox{GeV}$
 (one family model).
The pseudo-Nambu-Goldstone bosons become heavy,
 and the FCNCs mediated by them are suppressed.

The order of the decay constant of the Nambu-Goldstone boson
 is not affected whether the anomalous dimension is large or not.
We can estimate the decay constant
 by using the Pagels-Stokar formula (eq.(\ref{PS-formula}))
\begin{equation}
 F_\pi^2 = {{N_{TC}} \over {4\pi^2}} \int dx
       {{x \Sigma(x)} \over {\left( \Sigma^2(x) + x \right)^2}}
       \left\{
        \Sigma(x) - {1 \over 2} x {{d \Sigma(x)} \over {dx}}
       \right\}.
\end{equation}
Consider the mass function
\begin{equation}
 \Sigma(x) = \left\{
             \begin{array}{cc}
               m_T & \quad x \leq m_T^2 \\
               {{m_T^3} \over x}
                \left( {x \over {m_T^2}} \right)^{\gamma_m / 2}
               & \quad x > m_T^2,
             \end{array}
             \right.
\end{equation}
Then we obtain
\begin{equation}
 F_\pi^2 = {{N_{TC}} \over {8\pi^2}} m_T^2
           \left\{
            1 + {1 \over 2} \cdot {{3-\gamma_m/2} \over {1-\gamma_m/2}}
           \right\}
\end{equation}
As far as the anomalous dimension $\gamma_m$ is not extremely near $2$,
 the variation of the anomalous dimension
 does not change the order of the decay constant.
The reason is that
 the integral of the Pagels-Stokar formula is infrared dominant.
The decay constant is the quantity
 which is determined mainly by the low energy behavior of the dynamics.

In this way, the FCNC problem can be solved
 by considering the large anomalous dimension dynamics
 for the technicolor interaction.
The walking gauge coupling is obtained
 by assuming large number of technifermions (large $N_F$)
 so that the $\beta$-function becomes very small.
But it should be noted that
 the number of the pseudo-Nambu-Goldstone bosons
 increases together with the number of the technifermions.
The large number of pseudo-Nambu-Goldstone bosons
 are dangerous for FCNC, even if they are heavy.
And it should be noted that
 the value of the decay constant decreases
 as the increasing of the number of the technifermion weak doublets.
We know that
 the small value of the decay constant
 means the light pseudo-Nambu-Goldstone bosons.

In the gauged Nambu-Jona-Lasinio model,
 the four fermion interaction must be the effective interaction
 which is mediated by some massive particles,
 since the four fermion interaction is not renormalizable
 \footnote{Recently, it is shown that
           the system of the gauge plus four fermion interaction
           can be renormalizable in ladder approximation
           \cite{4F-renormalization}.}.
Someone consider that
 the origin of the strong four fermion interaction
 is the extended technicolor interaction.
But it is not clear that
 such strong four fermion interaction
 can be realized as the low energy effective interaction of the gauge theory
 \footnote{We do not consider the gauge symmetry breaking
            by the vacuum expectation value of the elementary scalar fields.}.
This is the main topics of this thesis, namely,
 whether the effective strong four fermion interaction is possible or not
 in the tumbling gauge theory
 which is a candidate of the dynamics of the extended technicolor theory.
We will discuss this topics in and after the next chapter.

\section{Technicolor and precision experiments I}

The electroweak theory is precisely tested on the Z-pole at LEP.
The experiments with extremely high precision
 makes possible to test the one-loop radiative correction.
Since the new physics beyond the standard model
 affects on the low energy phenomena through the radiative correction,
 we can obtain some constraint on the new physics.
Recently,
 the radiative correction on the electroweak interaction
 due to the new physics is parametrized by three parameters,
 $S$, $T$, and $U$\cite{Peskin-Takeuchi}.
These three parameters describe so called oblique correction
 due to the new physics which is characterized by a certain high energy scale
 in comparison with the electroweak scale.

The radiative correction on the self-energies of the electroweak gauge bosons
 is called oblique correction (fig.(\ref{oblique})).
\begin{figure}[t]
\caption{Oblique correction.}
\label{oblique}
\end{figure}
In the standard model the non-oblique correction
 is suppressed by the factor $m_f / M_Z$
 relative to the oblique correction,
 where $m_f$ is the mass of the quarks and leptons.
Consider the radiative correction on the four fermion process
 due to the Higgs field (fig.(\ref{correction-Higgs})).
\begin{figure}[t]
\caption{Corrections due to the Higgs.
        (a) Oblique correction.
        (b) Vertex correction.
        (c) Box diagram.}
\label{correction-Higgs}
\end{figure}
The order of the oblique correction of fig.(\ref{correction-Higgs}a)
 to the amplitude is $g_2^2 (g_2 / \sqrt{2} \cos \theta)^2$,
 where $\theta$ is the Weinberg angle.
The factor $(g_2 / \sqrt{2} \cos \theta)^2$ comes from the Z-Higgs coupling.
The order of the vertex correction of fig.(\ref{correction-Higgs}b) is
 $g_2^2 (g_2 / \sqrt{2} \cos \theta)^2 (m_f / M_Z)^2$.
The factor $m_f / M_Z$ comes from the Yukawa coupling
 $g_Y^f = \sqrt{2} m_f / v$ and $M_Z = g_2 v / 2 \cos \theta$,
 where $v/\sqrt{2}$ is the vacuum expectation value of the Higgs field.
The contribution of the box diagram fig.(\ref{correction-Higgs}c)
 is also suppressed by the same factor.
The order of the contribution is
 $(g_2 / \sqrt{2} \cos \theta)^4 (m_f / M_Z)^4$.
Essentially,
 as far as we consider the light quarks and light leptons
 as the incoming and the outgoing particle,
 the non-oblique correction can be neglected.
But the heavy top quark can generate the large non-oblique correction,
 even if it does not appear as the incoming and outgoing particle
 (Now we know that the top quark is heavier than the Z-boson.).
For example,
 consider the contributions of the diagrams of fig.(\ref{non-oblique-top}).
\begin{figure}[t]
\caption{Non-oblique correction due to the top quark.}
\label{non-oblique-top}
\end{figure}
These non-oblique correction should be considered.

If we consider the new physics
 which is characterized by the very high energy scale $M_{new} \gg M_Z$,
 the oblique correction is parametrized only by three parameters.
There are four kind of vacuum polarization of gauge bosons
 in the electroweak theory:
 $\Pi^{\mu\nu}_{11}(q) = \Pi^{\mu\nu}_{22}(q)$, $\Pi^{\mu\nu}_{33}(q)$,
 $\Pi^{\mu\nu}_{YY}(q)$, $\Pi^{\mu\nu}_{3Y}(q)$,
 where indexes $1$, $2$, and $3$ mean the $SU(2)_L$ gauge bosons,
 and index $Y$ means the $U(1)_Y$ gauge boson.
The $U(1)_{em}$ symmetry ensures the relation
 $\Pi^{\mu\nu}_{11}(q) = \Pi^{\mu\nu}_{22}(q)$.
The new physics contribution
 is defined by subtracting the contribution of the standard model
\begin{equation}
 \Pi_{new}^{\mu\nu}(q)
 = \Pi_{full}^{\mu\nu}(q) - \Pi_{SM}^{\mu\nu}(q;m_t,M_H).
\end{equation}
Since the masses of the top quark and Higgs are unknown,
 we must fix the masses as an reference point of the subtraction.
The true value of these masses may be different from the reference value.
The difference is treated as the contribution of the new physics.
We simply write $\Pi_{new}^{\mu\nu}(q)$ as $\Pi^{\mu\nu}(q)$ from now on.

If the scale of the new physics $M_{new}$ is sufficiently large,
 four $\Pi^{\mu\nu}(q)$ can be expended as
\begin{eqnarray}
 \Pi^{\mu\nu}(q)
 &=& \left( g^{\mu\nu} - {{q^\mu q^\nu} \over {q^2}} \right) \Pi(q^2)
\nonumber\\
 &=& \left( g^{\mu\nu} - {{q^\mu q^\nu} \over {q^2}} \right)
     \left( \Pi(0) + q^2 \Pi'(0) + O(q^2/M_{new}^2) \right).
\end{eqnarray}
Up to the order $q^2/M_{new}^2$,
 the correction to the Lagrangian becomes as follows.
\begin{eqnarray}
 {\cal L}_{gauge}
 = &-& {1 \over 4}
        \left( 1-g_2^2 \Pi'_{11}(0) \right) W_1^{\mu\nu} W_{1\mu\nu}
     + {1 \over 2}
        g_2^2 \left( {{v^2} \over 4} + \Pi_{11}(0) \right) W_1^\mu W_{1\mu}
\nonumber\\
 &+& \mbox{($W_2 \rightarrow W_1$)}
\nonumber\\
 &-& {1 \over 4}
      \left( 1-g_2^2 \Pi'_{33}(0) \right) W_3^{\mu\nu} W_{3\mu\nu}
\nonumber\\
 &-& {1 \over 4}
      \left( 1-g_1^2 \Pi'_{YY}(0) \right) B^{\mu\nu} B_{\mu\nu}
\nonumber\\
 &+& {1 \over 2}
      {{4\pi\alpha} \over {sc}} \Pi'_{3Y}(0) W_3^{\mu\nu} B_{\mu\nu}
\nonumber\\
 &+& {1 \over 2}
      \left( \begin{array}{cc} W_{3\mu} & B_\mu \end{array} \right)
      \left( {{v^2} \over 4} + \Pi_{33}(0) \right)
      \left( \begin{array}{cc} g_2^2     & - g_1 g_2 \\
                               - g_1 g_2 & g_1^2
             \end{array} \right)
      \left( \begin{array}{c} W_3^\mu \\ B^\mu \end{array} \right),
\end{eqnarray}
 where $\alpha$ is the fine structure constant,
 and $s$ and $c$ are the sine and cosine of the Weinberg angle, respectively.
Here, we used the relations
\begin{eqnarray}
 \Pi_{3Y}(0) + \Pi_{33}(0) &=& 0,
\nonumber\\
 \Pi_{3Y}(0) + \Pi_{YY}(0) &=& 0,
\end{eqnarray}
 which come from the $U(1)_{em}$ symmetry.
There are six parameters ($\Pi$s and $\Pi'$s) in the Lagrangian.
Three of the six parameters
 can be absorbed into the gauge couplings and vacuum expectation value as
\begin{eqnarray}
 \Pi'_{11}(0) &\rightarrow& \mbox{$W$ fields} \rightarrow g_2,
\nonumber\\
 \Pi'_{YY}(0) &\rightarrow& \mbox{$B$ field} \rightarrow g_1,
\nonumber\\
 \Pi_{11}(0) &\rightarrow& v.
\end{eqnarray}
Therefore, these are not the physical quantities.
After that, we obtain
\begin{eqnarray}
 {\cal L}_{gauge}
 = &-& {1 \over 4}
        \left( W_1^{\mu\nu} W_{1\mu\nu} + W_2^{\mu\nu} W_{2\mu\nu} \right)
\nonumber\\
  &+& {1 \over 2} g_2^2 {{v^2} \over 4}
       \left( W_1^\mu W_{1\mu} + W_2^\mu W_{2\mu} \right)
\nonumber\\
 &-& {1 \over 4}
      \left( 1+{{\alpha} \over {4s^2}} U \right) W_3^{\mu\nu} W_{3\mu\nu}
\nonumber\\
 &-& {1 \over 4} B^{\mu\nu} B_{\mu\nu}
\nonumber\\
 &-& {1 \over 4}
      {{\alpha} \over {2sc}} S W_3^{\mu\nu} B_{\mu\nu}
\nonumber\\
 &+& {1 \over 2}
      \left( \begin{array}{cc} W_{3\mu} & B_\mu \end{array} \right)
      {{v^2} \over 4} \left( 1 - \alpha T \right)
      \left( \begin{array}{cc} g_2^2     & - g_1 g_2 \\
                               - g_1 g_2 & g_1^2
             \end{array} \right)
      \left( \begin{array}{c} W_3^\mu \\ B^\mu \end{array} \right),
\end{eqnarray}
 where
\begin{eqnarray}
 S &\equiv& - 16\pi \Pi'_{3Y}(0),
\nonumber\\
 T &\equiv& {{4\pi} \over {s^2 c^2 M_Z^2}}
             \left[ \Pi_{11}(0) - \Pi_{33}(0) \right],
\nonumber\\
 U &\equiv& 16\pi \left[ \Pi'_{11}(0) - \Pi'_{33}(0) \right].
\end{eqnarray}
$S$ denotes the $B$-$W_3$ kinetic mixing,
 $T$ denotes the breaking of the custodial symmetry\cite{custodial}
 (The deviation of the $\rho$ parameter from unity),
 and $U$ also denotes the breaking of the custodial symmetry.
$U$ is the next order of the $q^2 / M_{new}^2$ expansion
 in comparison with $T$.
These three parameters, $S$, $T$, and $U$, are the physical parameters
 which values should be restricted by the experiments.

The physical quantities can be written with the new physics contribution.
For example, the total width of the Z-boson is obtained as
\begin{equation}
 \Gamma_Z = \Gamma_Z^{SM} + a S + b T,
\end{equation}
 where $\Gamma_Z^{SM}$ is the prediction of the standard model and
\begin{equation}
 a=-{{\alpha^2 M_Z} \over {12 s^2 c^2 (c^2-s^2)}}
    \sum_f \left( I_{3f} - s^2 Q_f \right) Q_f N_f,
\end{equation}
\begin{equation}
 b={{\alpha^2 M_Z} \over {6 s^2 c^2}}
    \sum_f \left\{ \left( I_{3f} - s^2 Q_f \right)^2 N_f
                   + {{2 s^2 c^2} \over {c^2 - s^2}}
                     \left( I_{3f} - s^2 Q_f \right) Q_f N_f \right\},
\end{equation}
 where $I_{3f}$, $Q_f$, and $N_f$
 denote weak isospin, electric charge, and effective number of color
 with respect to the fermion of flavor $f$, respectively
 \cite{Peskin-Takeuchi}.
The measured total width with small error
 restricts the values of $S$ and $T$ to a band in $S$-$T$ plane.
The $U$ parameter appears only in $M_W/M_Z$,
 if we consider only the physical quantities
 which have already precisely measured.
Therefore, $U$ parameter is not important at present.

Technicolor theory
 predicts the existence of the pseudo-Nambu-Goldstone bosons
 and the vector boundstates (techni-$\rho$) at high energy.
These particle contribute to the $S$, $T$ (and $U$) parameters.
Now, we briefly estimate the prediction of the technicolor.

Calculation of the one-loop diagram of fig.(\ref{one-loop}a) gives
\begin{equation}
 S = {{N_{TC}} \over {6\pi}}
     \sum_{doublets} \left( 1 - Y_L \ln {{m_N^2} \over {m_E^2}} \right),
\end{equation}
 where $m_N$ and $m_E$ are the masses of the technifermions
 of weak isospin $1/2$ and $-1/2$, respectively,
 and $Y_L$ is the hypercharge of the weak doublet.
\begin{figure}[t]
\caption{One-loop diagrams for the oblique corrections.
         (a) $S$ parameter.
         (b) $T$ (and $U$) parameter.}
\label{one-loop}
\end{figure}
The summention is taken over all the weak doubles.
The contribution of the one doublet is $S \simeq 0.05 N_{TC}$,
 if the masses are degenerate in the doublet.
The non-perturbative effect of the technicolor interaction
 is not included in this estimation.
But we know the tendency that
 $S$ is proportional to the number of the techni-doublet and $N_{TC}$.
Namely, large technicolor sector results large value of $S$.
The difference of the mass in the weak doublet
 can not be the large source of the change of the value of $S$.
The dependence is not the power but only logarithmic,
 since $S$ denotes the kinetic mixing of $W_3$ and $B$.

If the dynamics of the technicolor is postulated as the QCD-like one,
 we can estimate the value of $S$ including the non-perturbative effect
 by scaling up in energy.
The boundstates due to QCD, $\pi$, $\rho$ and $a_1$ mason and so on,
 contribute to the vacuum polarization $\Pi_{3Y}^{\mu\nu}$.
The contribution of the techni-$\rho$ and techni-$a_1$ and so on is obtained
 by scaling up the experimental data in the low energy to the high energy.
The value
\begin{equation}
 S \simeq 0.1 N_{TC} N_D
\end{equation}
 is obtained\cite{Peskin-Takeuchi}, where $N_D$ is the number of the doublet.
This value is about two times larger
 than the value from the above naive estimation.
In the technicolor theory with large anomalous dimension,
 it is expected that $0.05 < S/N_{TC}N_D < 0.1$.
Namely, $S/N_{TC}N_D$ will take the value
 between the values of the above two estimation,
 since the constant technifermion mass in naive estimation
 is expected to be the case $\gamma_m=2$
 (For numerical estimation, see ref.\cite{S-large-gamma}.).

The naive estimation of $T$ parameter
 by the calculation of the one-loop diagram of fig.(\ref{one-loop}b) gives
\begin{equation}
 T = {{N_{TC}} \over {16\pi s^2 c^2 M_Z^2}}
     \sum_{doublets} \left[ m_N^2 + m_E^2
                          - {{2 m_N^2 m_E^2} \over {m_N^2 - m_E^2}}
                            \ln {{m_N^2} \over {m_E^2}}
                    \right].
\end{equation}
This is the same formula for
 $\delta \rho = \rho - 1$ ($\delta \rho^{new} = \alpha T$).
The difference of the masses in a doublet cause the large effect,
 since the dependence is in power.
If we consider the technicolor model
 in which the custodial symmetry is not broken in the technifermion sector
 (equal values of the condensates in a doublet),
 $T$ parameter can be very small.
Only the effect of the heavy extended technicolor bosons
 makes the deviation of $T$ from zero.

Now, we compare the above estimation with the bound from the experiments.
Especially, the experimental restriction on the value of $S$ is important,
 since $S$ reflects the ``size'' of the technicolor sector.
It should also be noted that
 the experiments can directly restrict the technicolor sector itself.
This is not the test of the extended technicolor theory,
 but the technicolor theory.
The bound from FCNC
 which is discussed in the previous section is the model dependent one,
 since FCNC strongly depends on the model-dependent
 extended technicolor interaction.

The bound from the experiment is displayed in fig.(\ref{S-T})
 with the prediction of the technicolor theory.
\begin{figure}[t]
\caption{Bound form the experiments
          and the prediction of the technicolor theory.
         Technicolor theory predicts the value
          of the right-hand region from the vertical line
          (prediction of the $SU(2)_{TC}$ one-family model).}
\label{S-T}
\end{figure}
The reference values of the masses of the top quark and Higgs are taken as
 $m_t=150\mbox{GeV}$ and $M_H=1\mbox{TeV}$.
We use the following experiments:
 total width of Z-boson, R-ratio at the Z-pole,
 b quark forward-backward asymmetry, $\tau$ polarization asymmetry,
 the left and right couplings in the deep inelastic neutrino scattering,
 and atomic parity violation of ${}^{133}_{53}Cs$\cite{experiments}.
We see that the technicolor theory predicts too large $S$,
 and it is disfavored by the experiments.
The technicolor theory
 with large technicolor sector (large $N_{TC}$, $N_D$)
 is strongly disfavored by the experiments.

At present, there are no definite solution of this difficulty.
The mass difference in the doublets of the technifermions
 can give the negative contribution to $S$.
But if we set the mass difference
 so that the $S$ parameter becomes small enough,
 the value of $T$ parameter becomes too large
 to be consistent with the experiments.
It is difficult to simultaneously suppress $S$ and $T$ parameters.
It is pointed out that
 the Majorana technifermions can give the negative contribution
 to the $S$ parameter\cite{Terning}
 without drastical changing of the $T$ parameter.
But the contribution is very small
 ($\delta S \simeq -10^{-2}$ and $\delta T \simeq -10^{-3}$
   at every set with two Majorana technifermions.),
 and the many Majorana technifermion are needed
 to be consistent with the experiment.

We can consider the special dynamics to avoid this difficulty.
The large anomalous dimension dynamics
 and the dynamics of the chiral gauge theory are the candidates.
But it is the difficult work to find the good dynamics,
 since we do not have the enough technique
 to treat the non-perturbative effect.

The cancellation between the oblique corrections
 by assuming the contribution
 from the other additional new physics can be expected.
Extended technicolor interaction,
 the additional broken gauge interactions inspired by GUTs, and so on
 can be the additional new physics.
But the model will become very complicated and unnatural one.

For the completeness,
 the non-oblique correction should be considered
 in addition to the oblique correction.
It is pointed out that
 the non-oblique correction due to the extended technicolor interaction
 is large,
 when the top quark is heavy\cite{non-ob}.
The non-oblique correction may cancel out the oblique correction.
In next section
 we estimate the effect of the non-oblique correction
 in the extended technicolor theory.

\section{Technicolor and precision experiments II}

It is pointed out that
 the non-oblique correction to the $Zb{\bar b}$ vertex
 is generally large in the extended technicolor theory\cite{non-ob}.
In this section,
 we explicitly calculate the non-oblique correction
 in a naive model of the extended technicolor theory.
We show that
 the non-oblique correction is too large to be consistent with the experiments.

Many extended technicolor gauge bosons become massive,
 in the process of the breaking of the extended technicolor gauge symmetry.
Some of them called sideways
 cause the transition of the ordinary fermions to the technifermions,
 some of them called horizontal
 connect the ordinary fermions themselves,
 and the others called ``diagonal'' diagonally interact
 with both the ordinary fermions and technifermions.
The sideways bosons
 must exist in the realistic model
 to generate the quark and lepton masses,
 while the existence of the horizontal and the ``diagonal'' bosons
 is rather model-dependent.
We consider the effect of the sideways and the ``diagonal'' from now on.
The lightest sideways and ``diagonal'' gauge bosons
 are associated with the top quark.
They make the largest contributions to the radiative corrections.

Consider four fundamental representations
 of the extended technicolor gauge group $SU(N_{TC}+1)$
 containing the top and the bottom quark
 and the technifermions $U$ and $D$.
\begin{equation}
 \left(\begin{array}{cc}
  \quad
  \left(\begin{array}{c} U^1 \\ \vdots \\ U^{N_{TC}} \\ t
        \end{array}\right)_L
  \quad
  \left(\begin{array}{c} D^1 \\ \vdots \\ D^{N_{TC}} \\ b
        \end{array}\right)_L
  \quad
 \end{array}\right),
 \qquad
 \left(\begin{array}{c} U^1 \\ \vdots \\ U^{N_{TC}} \\ t
       \end{array}\right)_R,
 \qquad
 \left(\begin{array}{c} D^1 \\ \vdots \\ D^{N_{TC}} \\ b
       \end{array}\right)_R.
\end{equation}
Two left-handed representations form an $SU(2)_L$ doublet
 and two right-handed representations are $SU(2)_L$ singlets.
(We assume that the extended technicolor gauge group commute
 with the weak interaction gauge group.)
After the breaking of extended technicolor gauge group
 down to the technicolor gauge group $SU(N_{TC})$,
 the massive sideways and ``diagonal'' gauge bosons are generated.

In this model the mass of top quark
 is equal to the mass of bottom quark,
 because of the common mass and coupling of the sideways
 and $\langle \bar{U} U \rangle = \langle \bar{D} D \rangle$.
In the realistic model, however,
 the right-handed top quark and the right-handed bottom quark
 must be contained in the different representations
 of the extended technicolor gauge group
 to realize the different sideways couplings.
Therefore,
 the extended technicolor gauge theory must be a chiral gauge theory.
However,
 realistic representations of extended technicolor gauge group
 are not known yet.
Instead of considering explicit extended technicolor model,
 we here simply assume different extended technicolor couplings
 for top and bottom right-handed fundamental representations,
 while keeping the technicolor interaction vector-like.
We also assume that
 $\langle \bar{U} U \rangle = \langle \bar{D} D \rangle$.
This is a good toy model for the isospin breaking,
 although the extended technicolor gauge symmetry is destroyed.
More explicitly,
 we set the sideways coupling $\xi_t g_t$ for the left-handed quarks,
 $g_t / \xi_t$ for the right-handed top quark,
 and $g_t / \xi_b$ for the right-handed bottom quark,
 where $g_t$ is given by
\begin{equation}
 m_t \simeq {{g_t^2} \over {M_X^2}} 4 \pi F_\pi^3.
\end{equation}
The scale $M_X$ is the mass of sideways
 and the relation $\langle \bar{U} U \rangle \simeq 4 \pi F_\pi^3$
 (naive dimensional analysis) is used,
 where $F_\pi$ is the decay constant of the Nambu-Goldstone bosons
 of the technicolor chiral symmetry breaking.
Large top quark mass
 indicates large value of $g_t$ or small value of $M_X$.
We assume that the sideways effect can be treated perturbatively
 in loop calculation, namely, ${{(\xi_t g_t)^2} \over {4\pi}} < 1$
 and ${{(g_t / \xi_t)^2} \over {4\pi}} < 1$.
This relations restrict the value of $\xi_t$.
For the realistic bottom quark mass,
 $\xi_b$ is restricted by
 ${1 \over {\xi_b}} \leq {1 \over {\xi_t}}{{m_b} \over {m_t}}$.

The ``diagonal'' couplings
 are fixed through the relation to the sideways couplings.
For technifermion, we obtain the ``diagonal'' coupling
 by multiplying the factor
 $-{1 \over {N_{TC}}} \sqrt{{N_{TC}} \over {N_{TC} + 1}}$
 to their sideways couplings.
For quark, we obtain it by multiplying the factor
 $\sqrt{{N_{TC}} \over {N_{TC} + 1}}$
 to their sideways couplings.
These factors
 come from the normalization and traceless property
 of the diagonal generator.
The ``diagonal'' interaction is also chiral.

The sideways bosons
 yield potentially a large non-oblique correction
 to the $Z b \bar{b}$ vertex.
By using the approach of the effective Lagrangian,
 the correction to the left-handed and the right-handed couplings
 of the bottom quark are derived as (Chivukula et al. in ref.\cite{non-ob})
\begin{eqnarray}
 \delta g_L^b &=& {{\xi_t^2} \over 4}
                  {{m_t} \over {4 \pi F_\pi}} {e \over {cs}},
\nonumber\\
 \delta g_R^b &=&-{1 \over {4\xi_b^2}}
                  {{m_t} \over {4 \pi F_\pi}} {e \over {cs}},
\label{sideways}
\end{eqnarray}
 where $c$ and $s$
 are the cosine and sine of the Weinberg angle, respectively.
The suppression factor $m_t / 4\pi F_\pi$
 is not small for large $m_t$.
The diagram corresponding to this correction
 is shown in fig.(\ref{corrections}).
\begin{figure}[t]
\caption{The diagrams of the non-oblique correction to the $Zb\bar{b}$ vertex:
        (a) Sideways contribution.
        (b) ``Diagonal'' contribution.}
\label{corrections}
\end{figure}

The ``diagonal'' boson ``$X$'' also yields the non-oblique correction
 through the mixing with $Z$ boson\cite{Holdom}.
The mixing is parametrized by three parameters $x$, $y$, and $w$ as
\begin{eqnarray}
 {\cal L}_{AZX}
  = & - & {1 \over 4}
     \left(
       \begin{array}{ccc}
         X_{\mu\nu} & Z_{\mu\nu} & A_{\mu\nu}
       \end{array}
     \right)
     \left(
       \begin{array}{ccc}
         1 \quad    & y \quad    & w          \\
         y \quad    & 1 \quad    & 0          \\
         w \quad    & 0 \quad    & 1
       \end{array}
     \right)
     \left(
       \begin{array}{c}
         X^{\mu\nu} \\
         Z^{\mu\nu} \\
         A^{\mu\nu}
       \end{array}
     \right)
\nonumber\\
   & + & {1 \over 2}
     \left(
       \begin{array}{ccc}
         X_\mu      & Z_\mu      & A_\mu
       \end{array}
     \right)
     \left(
       \begin{array}{ccc}
         M_X^2      & x M_Z^2    & 0          \\
         x M_Z^2    & M_Z^2      & 0          \\
         0          & 0          & 0
       \end{array}
     \right)
     \left(
       \begin{array}{c}
         X^\mu \\
         Z^\mu \\
         A^\mu
       \end{array}
     \right),
\end{eqnarray}
 where we set the ``diagonal'' mass equal to the sideways mass.
Within the leading order of $x$, $y$,
 and $w$ in the four-fermion amplitude,
 the non-oblique correction to the $Zb\bar{b}$ vertex is obtained as
\begin{eqnarray}
 \delta g_L^b &=& \xi_t g_t
                  \sqrt{{N_{TC}} \over {N_{TC}+1}}
                  {{M_Z^2} \over {M_X^2-M_Z^2}} (y-x),
\nonumber\\
 \delta g_R^b &=& {{g_t} \over {\xi_b}}
                  \sqrt{{N_{TC}} \over {N_{TC}+1}}
                  {{M_Z^2} \over {M_X^2-M_Z^2}} (y-x).
\label{diagonal}
\end{eqnarray}

We get $x$, $y$, and $w$
 by calculating one-loop diagrams of fig.(\ref{loops})
 with constant fermion mass.
\begin{figure}
\caption{The one-loop diagrams for calculating $X$-$W_3$ and $X$-$B$ mixing.}
\label{loops}
\end{figure}
The results are
\begin{equation}
 x = N_C \sqrt{{N_{TC}} \over {N_{TC} + 1}}
     {{g_t e} \over {(4\pi)^2}} {1 \over {cs}}
     \left[
           (\xi_t - {1 \over {\xi_t}} ) {{m_U^2-m_t^2} \over {M_Z^2}}
          -(\xi_t - {1 \over {\xi_b}} ) {{m_D^2-m_b^2} \over {M_Z^2}}
     \right],
\end{equation}
\begin{eqnarray}
 y = &-& N_C \sqrt{{N_{TC}} \over {N_{TC} + 1}}
     \xi_t
     {{g_t e} \over {(4\pi)^2}} {1 \over {cs}} {1 \over 3}
     \left[
            \ln {{m_t^2} \over {m_b^2}} - \ln {{m_U^2} \over {m_D^2}}
     \right]
\nonumber\\
     &-& N_C \sqrt{{N_{TC}} \over {N_{TC} + 1}}
     {{g_t e} \over {(4\pi)^2}} {s \over c} {2 \over 3}
     \left[
           {2 \over 3} (\xi_t + {1 \over {\xi_t}} )
            \ln {{m_U^2} \over {m_t^2}}
          -{1 \over 3} (\xi_t + {1 \over {\xi_b}} )
            \ln {{m_D^2} \over {m_b^2}}
     \right],
\end{eqnarray}
\begin{equation}
 w = N_C \sqrt{{N_{TC}} \over {N_{TC} + 1}}
     {{g_t e} \over {(4\pi)^2}} {2 \over 3}
     \left[
           {2 \over 3} (\xi_t + {1 \over {\xi_t}} )
            \ln {{m_U^2} \over {m_t^2}}
          -{1 \over 3} (\xi_t + {1 \over {\xi_b}} )
            \ln {{m_D^2} \over {m_b^2}}
     \right].
\end{equation}
Because the diagonal generator is traceless,
 the kinetic mixing parameters $y$ and $w$ are naturally finite.
The mass mixing parameter $x$ should be naturally finite
 if we use the dynamical fermion mass having momentum dependence.
To include the effect of dynamical mass,
 we set the momentum cutoff to the fermion mass
 in the individual loops, and get a finite $x$.
The diagram corresponding to the correction of eq.(\ref{diagonal})
 is shown in fig.(\ref{corrections}).

The effect of technicolor resonances and Pseudo-Nambu-Goldstone bosons
 is not included in this one-loop calculation of $x$, $y$, and $w$.
It can be expected that these effects
 do not change the sign and the order of magnitude of the correction
 unless we consider a special model.
Actually
 in the estimation of the oblique correction due to the technicolor dynamics,
 the $S$ parameter estimated by the QCD scale up
 is just twice of the one estimated by the one loop calculation,
 and the loop effect of pseudo-Nambu-Goldstone bosons is small
 without many light pseudo-Nambu-Goldstone bosons
 \cite{Peskin-Takeuchi}.
Although these effects are important
 when we rigorously compare the theory with experiments,
 it can be expected that these effects are not important
 to show the importance of the non-oblique correction
 due to the extended technicolor gauge bosons.
Therefore,
 we do not consider these effects in the estimation of $x$, $y$, and $w$.

The $\xi_t$ dependence of the both contributions
 to the left-handed coupling of the bottom quark are shown in fig.(\ref{gl}).
\begin{figure}[t]
\caption{The $\xi_t$ dependence of $\delta g_L^b$.
         The contributions of ``diagonal'' and Sideways boson
          are shown separately.
         Total contribution of extended technicolor bosons is also shown.
         The region of $\xi_t$
          in which the perturbative loop calculation is valid
         is $0.7 < \xi_t < 1.4$.}
\label{gl}
\end{figure}
We set $M_X=1\ $TeV, $F_\pi=125\ $GeV (one family model),
 $m_U=m_D=(4\pi F_\pi^3)^{1/3}$,
 $N_{TC}=4$, $m_t=150\ $GeV, and $M_H=1\ $TeV.
The region of $\xi_t$ in which the perturbative loop calculation is reliable
 is $0.7 < \xi_t < 1.4$.
The $\xi_t$ dependence of the sideways contribution is quadratic and strong.
As for the ``diagonal'' contribution, it is approximately quadratic,
 but the dependence is very weak
 and the contribution is almost constant
 within the possible region of $\xi_t$.
Both contributions are positive and do not cancel each other.
The ``diagonal'' contribution
 is $30$\% of the sideways contribution when $\xi_t=1$.
We can approximately write the ``diagonal'' contribution when $\xi_t=1$ as
\begin{equation}
 \delta g^b_L = {1 \over {(4\pi)^{4/3}}}
                {{N_C N_{TC}} \over {N_{TC}+1}}
                {{m_t} \over {4\pi F_\pi}} {e \over {cs}}
\end{equation}
 which should be compared with eq.(\ref{sideways}) with $\xi_t=1$.

Both the sideways and the ``diagonal'' contributions
 to the right-handed $Zb\bar{b}$ coupling,
 eqs.(\ref{sideways}) and (\ref{diagonal}) are suppressed
 at least by the power of ${1 \over {\xi_b \xi_t}}$.
Namely,
 these contributions are suppressed
 by the power of ${{m_b} \over {m_t}}{1 \over {\xi_t^2}}$
 compared with the contributions to the left-handed coupling.
Therefore,
 the contributions to the right-handed coupling are not important
 unless the predictions are highly sensitive
 to the change of right-handed coupling.
We ignore it from now on.

Both sideways and ``diagonal'' bosons
 also contribute to the oblique correction, $S$, $T$ and $U$.
However, since the contribution is the two-loop effect,
 it is small in comparison with the contribution
 due to the technicolor dynamics.
For example,
 the ``diagonal'' contribution to the $S$ parameter is $S=0.022$
 for $M_X=1\ $TeV, $F_\pi=125\ $GeV, $N_{TC}=4$, $\xi_t=1$,
 $m_t=150\ $GeV, and $M_H=1\ $TeV,
 which is small in comparison with the contribution $S > 0.4$
 due to the technicolor dynamics.
\begin{figure}[t]
\caption{Contours in $S$-$T$ plane ($90\%$ C.L.).
         A contour with doted line is the result of
          two parameter ($S$ and $T$) fitting.
         Another contour is a cross section (at $\delta g_L^b = -0.0045$)
          of the ellipsoid which is the result of three parameter fitting.}
\label{best}
\end{figure}

The non-oblique corrections eqs.(\ref{sideways}) and (\ref{diagonal})
 change the predictions.
For example, the total $Z$ width can be written as
\begin{equation}
 \Gamma_Z = \Gamma_Z^{SM} + \delta \Gamma_Z + a S + b T,
\end{equation}
 where $\Gamma_Z^{SM}$ is the one-loop prediction of the standard model and
\begin{equation}
 \delta \Gamma_Z
  = {{\partial \Gamma_Z^{SM}} \over {\partial g_L^b}} \delta g_L^b.
\end{equation}
The coefficients $a$ and $b$ are written by the standard model parameters as
\begin{equation}
 a=-{{\alpha^2 M_Z} \over {12 s^2 c^2 (c^2-s^2)}}
    \sum_f \left( I_{3f} - s^2 Q_f \right) Q_f N_f,
\end{equation}
\begin{equation}
 b={{\alpha^2 M_Z} \over {6 s^2 c^2}}
    \sum_f \left\{ \left( I_{3f} - s^2 Q_f \right)^2 N_f
                   + {{2 s^2 c^2} \over {c^2 - s^2}}
                     \left( I_{3f} - s^2 Q_f \right) Q_f N_f \right\},
\end{equation}
 where $I_{3f}$, $Q_f$, and $N_f$
 denote weak isospin, electric charge, and effective number of color
 with respect to the fermion of flavor $f$, respectively
 \cite{Peskin-Takeuchi}.
The predictions for the R ratio on $Z$ pole
 and forward-backward asymmetry of $b$ quark
 are also changed in the same way as $\Gamma_Z$.
We analyze the experiments
 by using three free parameters $S$, $T$, $\delta g_L^b$
 and get an ellipsoid in $S$-$T$-$\delta g_L^b$ space.
The region inside the ellipsoid is favored by the experiments.
We do not consider the experiments
 which are parametrized by $U$ parameter, for simplicity.
\begin{figure}[t]
\caption{Cross sections of the ellipsoid of $90\%$ C.L.
          at $\delta g_L^b = 0$, $0.003$, $0.006$.}
\label{S-T-2}
\end{figure}

The cross section of the ellipsoid of $90\%$ C.L.,
 which contains a best-fit point, is shown in fig.(\ref{best}).
The following experiments are considered;
 total $Z$ width, $R$ ratio on $Z$ pole,
 forward-backward asymmetry of $b$ and $\mu$,
 polarization asymmetry of $\tau$,
 deep inelastic neutrino scattering ($g_L$ and $g_R$),
 and atomic parity violation $Q_W(^{133}_{\ 53}C_S)$
 \cite{experiments}.
The most favorable values
 are $\delta g_L^b=-0.0045$, $S=-0.73$, and $T=-0.11$.
The favorable region in $S$-$T$ plane
 becomes larger than the one from two parameter analysis.
Fig.(\ref{S-T-2}) shows the favorable regions in $S$-$T$ plane ($90\%$ C.L.)
 when $\delta g_L^b$  is fixed in several values.
We find that
 the new physics contribution to the left-handed coupling of b quark
 must be smaller than $0.007$ in $90\%$ C.L.
This value is much smaller than the extended technicolor contributions
 which is shown in fig.(\ref{gl}).

Fig.(\ref{S-G}) shows
 the cross sections of ellipsoid at several values of $T$.
\begin{figure}[t]
\caption{Cross sections of the ellipsoid of $90\%$ C.L.
          at $T = -0.11$ (best fit), $0.5$, $1.0$.
         Horizontal line means the lower bound
          of the orediction of the extended technicolor thoery.}
\label{S-G}
\end{figure}
We also see that
 the extended technicolor contributions
 are at the outside of the ellipsoid of $90\%$ C.L.

We showed that
 the non-oblique corrections to $Zb{\bar b}$ vertex
 which are naturally expected to exist
 in the realistic extended technicolor model
 are too large to be consistent with the experiments.
The reason is that
 only the predictions to the total Z width, the R ratio on Z pole,
 and the forward-backward asymmetry of the b quark
 which relate with the $Zb{\bar b}$ vertex are modified.
There must be some mechanisms
 which reduce or cancel this non-oblique correction
 in order that the extended technicolor model
 is consistent with the experiments.

In the phenomenological test of the extended technicolor models
 we must consider both non-oblique and oblique corrections.
The physics
 which simultaneously reduce both oblique and non-oblique corrections
 is expected in model building of the extended technicolor theory.
\chapter{Tumbling gauge theory}

Tumbling gauge theory\cite{tumbling} contains the many hierarchical scales.
No elementary scalar fields
 are needed to realize the hierarchical gauge symmetry breaking.
Many effective four fermion interactions with the hierarchical scales
 are generated in the process of ``tumbling''.
Therefore,
 the theory has been considered
 as the candidate of the dynamics of the extended technicolor theory.
Tumbling gauge theory is the special kind of the chiral gauge theory.
It is also a good property
 to be the dynamics of the extended technicolor theory,
 since the extended technicolor theory
 should be the chiral gauge theory for the realistic fermion masses,
 as we have already explained in chapter 4.
In this chapter,
 we formulate the tumbling gauge theory
 by using the Cornwall-Jackiw-Tomboulis (CJT) effective action\cite{CJT},
 which have already introduced in section 3.3.

The chiral gauge theory
 is the gauge theory with the fermions
 \footnote{Here, we do not consider the elementary scalar fields.}
 which representation is not real as a whole.
As an example of the chiral gauge theory,
 consider the one generation of the standard model.
\begin{eqnarray*}
 \left( \begin{array}{c} u \\ d \end{array} \right)_L &\sim& (3, 2, 1/6),
\\
 \left( u_R \right)^c \quad &\sim& (3^*, 1, -2/3),
\\
 \left( d_R \right)^c \quad &\sim& (3^*, 1, 1/3),
\\
 \left( \begin{array}{c} \nu_e \\ e \end{array} \right)_L &\sim& (1, 2, -1/2),
\\
 \left( e_R \right)^c \quad &\sim& (1, 1, 1),
\end{eqnarray*}
 where $(3,2,1/6)$, for instance,
 means the $SU(3)_c$ triplet, the $SU(2)_L$ doublet,
 and $Y/2 = 1/6$ ($Y/2$ is the charge of $U(1)_Y$).
All the fermion fields are written as the left-handed chiral fields.
If we take the complex conjugate of this representation, we get
\begin{eqnarray*}
 \left( \begin{array}{c} {\tilde u} \\
                         {\tilde d}
        \end{array} \right)_L &\sim& (3^*, 2, -1/6),
\\
 \left( {\tilde u}_R \right)^c \quad &\sim& (3, 1, 2/3),
\\
 \left( {\tilde d}_R \right)^c \quad &\sim& (3, 1, -1/3),
\\
 \left( \begin{array}{c} {\tilde \nu}_e \\
                         {\tilde e}
        \end{array} \right)_L &\sim& (1, 2, 1/2),
\\
 \left( {\tilde e}_R \right)^c \quad &\sim& (1, 1, -1).
\end{eqnarray*}
(Remember that all the representation of the $SU(2)$ group is real.)
If we consider only the $SU(3)$ gauge group,
 this representation is equivalent to the original one as a whole
There are same number of $3$ and $3^*$.
We call this representation real or vector-like one with respect to $SU(3)_c$.
But, the conjugated representation
 are not equivalent to the original one
 when we consider the $SU(2)_L \times (1)_Y$ gauge group.
We call such representation chiral representation.
The standard model is a chiral gauge theory.

In the chiral gauge theory,
 the cancellation of the gauge anomaly is non-trivial.
It must be canceled so that the theory is renormalizable.
We must impose the condition
\begin{equation}
 \mbox{tr} \left( T^a \left\{ T^b, T^c \right\} \right) = 0
\end{equation}
 for any combination of the generators of the gauge symmetry,
 $T^a$, $T^b$, and  $T^c$,
 where the generators are represented as the large matrixes
 which act on the space of the whole fermion contents.
It is famous that
 the non-trivial cancellation is accomplished in the standard model.

The tumbling gauge theory
 can be formulated by using the CJT effective action.
Let us consider the non-Abelian gauge group $G$ and left-handed fermions
 in the irreducible representation $r_1$, $r_2$, $\cdots$.
We consider the chiral representation
 and impose the cancellation of the gauge anomaly.
And we consider the situation that the fermion contents
 are compatible with asymptotic freedom.
In order to know the remaining symmetry at the low energy
 and the symmetry breaking scales,
 we must solve the dynamics of this theory.
We consider the CJT effective action
 which includes only one gauge boson exchange vacuum diagram,
 and introduce the most attractive channel (MAC) hypothesis\cite{tumbling}.

For simplicity, we assume that $G$ is a simple group.
It is convenient to represent all the left-handed fermions
 as a Majorana fermion field
\begin{equation}
 \Psi \equiv
 \left(
 \begin{array}{c}
   \psi_L (r_1) \\
   \psi_L (r_2) \\
   \vdots
 \end{array}
 \right)+\left(
 \begin{array}{c}
   {\psi_L}(r_1)^c  \\
   {\psi_L}(r_2)^c  \\
   \vdots
 \end{array}
 \right).
\end{equation}
Note that the generators of $G$ acting on this field
 are defined as
\begin{equation}
 {\cal T}^a \equiv T^a({1-\gamma_5 \over 2})
 -T^{a \ast}({1+\gamma_5 \over 2}),
\end{equation}
 where $T^a$ denotes the generator
 which acts on the space of the whole fermion contents.
Then the effective action can be written as
\begin{equation}
 \Gamma[S] = {1 \over 2} \ln{\sl Det} (S^{-1})
             - {1 \over 2} {\rm Tr} \left( \not\!p S \right)
             + D[S],
\end{equation}
 where $S$ denotes the full propagator of the Majorana field $\Psi$
 and $D[S]$ is the set of
 all two-particle irreducible (2PI) vacuum diagrams
 (fig.(\ref{diagrams})).
\begin{figure}
\caption{Two particle irreducible vacuum diagrams in effective action.}
\label{diagrams}
\end{figure}
The full propagator $S$ contains the mass function
$\Sigma$ as
\begin{equation}
 S^{-1}(p) = \Sigma(p) - \not\!p.
\end{equation}
When we consider the one gauge boson exchange diagram
 in fig.(\ref{diagrams}) only,
 we get the Schwinger-Dyson equation (gap equation)
 in ladder approximation
\begin{equation}
 \Sigma(p) = - g^2 \int {{d^4k} \over {(2\pi)^4i}}
             \gamma_\nu {\cal T}^a S(k)
             \gamma_\mu {\cal T}^a D^{\mu\nu}(k-p)
\end{equation}
 as a stationary condition of this effective action (See section 3.3),
 where $g$ is the gauge coupling constant and $D_{\mu\nu}$
 is the free gauge boson propagator.
We take Landau gauge
 so that the ladder approximation is consistent
 with the Ward-Takahashi identity (See section 3.1.).
In Landau gauge,
 $\Sigma(p)$ does not contain the $\gamma$-matrix,
 and it becomes the function of $p^2$.
The mass function $\Sigma(p^2)$ can be decomposed as
\begin{equation}
 \Sigma(p^2) = \Sigma(p^2)_{r_1 \times r_2 \rightarrow r_c^{12}}
        \oplus \Sigma(p^2)_{r_1 \times r_3 \rightarrow r_c^{13}}
        \oplus \cdots
\label{decomposition-mass}
\end{equation}
 corresponding to the channel of the pair condensation
 $r_1 \times r_2 \rightarrow r_c^{12}$,
 $r_1 \times r_3 \rightarrow r_c^{13}$, $\cdots$,
 where $r_c$'s denote the representation of the condensates.

It is enough to consider the linearized gap equation
 to estimate the scale of the condensate.
By linearizing the equation with respect to the mass function,
 each condensation channels can be treated separately
 \footnote{We get the independent equations for each channels.
           There is not the mixing of the channels.},
 since we can decompose the mass function as eq.(\ref{decomposition-mass}).
We concentrate on the channel $r_1 \times r_2 \rightarrow r_c$ from now on.
We get the linearized gap equation of the channel
\begin{equation}
 \Sigma_A(-p^2) =
   \lambda(r_1, r_2, r_c)
   \int dk^2 {{\Sigma_A(-k^2)} \over {k^2 + m^2}}
   \big\{ \theta(k^2-p^2)
         + {{k^2} \over {p^2}} \theta(p^2-k^2) \big\},
\end{equation}
\begin{equation}
 \lambda(r_1, r_2, r_c) \equiv
  {3 \over {4\pi}} \cdot {{g^2} \over {4\pi}} \cdot {1 \over 2}
  \big\{ C_2(r_1) + C_2(r_2) - C_2(r_c) \big\},
\end{equation}
 where $A$ is the index of the representation $r_c$,
 $C_2(r)$ is the coefficient of the second Casimir operator
 of representation $r$,
 and $m = \Sigma_A(m^2)$.
The integration is performing in Euclidean space
 and the angular integration has already done.
The effective coupling of the channel
 $r_1 \times r_2 \rightarrow r_c$
 is defined by $\lambda(r_1, r_2, r_c)$.
When the gauge group $G$ is not a simple group but a direct product group,
 the effective coupling becomes the sum of the effective couplings
 with respect to each gauge interactions.
We call the channel with the largest effective coupling
 most attractive channel (MAC).
The above gap equation has a non-trivial solution
 when $\lambda$ exceeds the critical value $\lambda_c=1/4$,
 as we estimated in section 3.1.

We can obtain hierarchical condensation scales in the MAC hypothesis.
The quantities in the (exact) effective action,
 the gauge coupling $g$, for instance,
 are renormalized quantities with scale dependence.
Here we approximately replace the gauge coupling constant $g$
 by the running coupling $g(\mu)$ of the
 one-loop perturbative calculation,
 where $\mu$ is a reference of the energy scale.
As the running coupling becomes large in the low energy region,
 the effective coupling of the MAC reaches ``firstly''
 the critical value $\lambda_c = 1/4$ at a certain scale $M$
 defined by
\begin{equation}
 \lambda(M) = {3 \over {4\pi}} \cdot {{g^2(M)} \over {4\pi}} \cdot {1 \over 2}
              \big\{ C_2(r_1) + C_2(r_2) - C_2(r_c) \big\}
            = {1 \over 4}.
\end{equation}
Then the condensate of this channel is likely to emerge
 or become non-negligible in comparison
 with the reference of the energy scale $M$.
We here assume the MAC hypothesis that
 the condensate is formed in not all the attractive channels
 but only in the MAC.
Since the MAC condensation is gauge non-singlet in general,
 the symmetry breaking takes place not only on the global symmetry
 but also on the gauge symmetry itself.
Some gauge bosons and fermions
 get their masses of the order of $M$ by this symmetry breaking.
The physics below the scale $M$ can be described by an effective theory
 containing gauge bosons of reduced gauge symmetry and massless fermions.

In the original tumbling scenario,
 the particles which get masses of order $M$
 are neglected and it is naively assumed
 that they decouple from the low energy dynamics.
In the effective theory
 the scale $M$ has a meaning of the ultraviolet momentum cutoff.
The values of the running gauge couplings
 in the theory above and below the scale $M$
 must coincide with each other at the scale $M$
 because of the continuity of the theory
 and the uniqueness of the fundamental scale of the theory.
We can apply the above effective action argument to the effective theory
 and find the MAC and the next condensation scale $M'$.
The effective theory below the scale $M'$
 has a further reduced gauge symmetry.
In this way,
 the gauge symmetry can be broken one after another by itself
 and the scale hierarchy $M$, $M'$, $\cdots$ can be generated (``tumbling'').
The sequence of the gauge symmetry breaking
 continues until the representation of the massless fermions
 becomes real as a whole,
 provided that the asymptotic freedom of the gauge coupling holds.
The sequence of the condensation scales
 (not always taking with the gauge symmetry breaking) terminates
 when all the fermions with attractive gauge interaction become massive.

As an example,
 consider the gauge group $SU(5)$
 and the chiral representations $5^*$ and $10$
 which can be written as
\begin{equation}
 \Psi_{5^*}
  = \left( \begin{array}{c}
            \psi_{5^*}^1 \\
            \psi_{5^*}^2 \\
            \psi_{5^*}^3 \\
            \psi_{5^*}^4 \\
            \psi_{5^*}^5
           \end{array}
    \right),
\qquad\qquad
 \Psi_{10}
 = \left( \begin{array}{ccccc}
             0         &   \psi^{12} &
             \psi^{13} &   \psi^{14} &   \psi^{15} \\
           - \psi^{12} &   0              &
             \psi^{23} &   \psi^{24} &   \psi^{25} \\
           - \psi^{13} & - \psi^{23} &
             0         &   \psi^{34} &   \psi^{35} \\
           - \psi^{14} & - \psi^{24} &
           - \psi^{34} &   0         &   \psi^{45} \\
           - \psi^{15} & - \psi^{25} &
           - \psi^{35} & - \psi^{45} &   0         \\
          \end{array}
   \right).
\end{equation}
These fermions are the left-handed fermions.
This representation
 is the same as the one generation in the $SU(5)$ GUT\cite{Georgi-Glashow}.
The gauge anomaly is canceled out,
 and the theory is asymptotically free.
The following channel of the pair condensation can be considered.
\begin{eqnarray*}
 10 \times 10 &=& 5^* + 45 + 50,
\\
 5^* \times 5^* &=& 10^* + 15^*,
\\
 5^* \times 10 &=& 5 + 45^*.
\end{eqnarray*}
Since there is no singlet in the right-hand side
 (representation of the condensate),
 the pair condensates should break the gauge symmetry.
The coefficients of the effective couplings of each channels
 $\left\{ C_2(r_1) + C_2(r_2) - C_2(r_c) \right\}$ are as follows.
\begin{eqnarray*}
 10 \times 10 &\rightarrow&
  \left\{ \begin{array}{ccr}
           5^* & \quad & {{24} \over 5}   \\
           45  & \quad & {4 \over 5}      \\
           50  & \quad & - {6 \over 5},
          \end{array}
  \right.
\\
 5^* \times 5^* &\rightarrow&
  \left\{ \begin{array}{ccr}
           10^* & \quad & {6 \over 5}    \\
           15^* & \quad & - {4 \over 5},
          \end{array}
  \right.
\\
 5^* \times 10 &\rightarrow&
  \left\{ \begin{array}{ccr}
           5    & \quad & {{18} \over 5} \\
           45^* & \quad & - {2 \over 5}.
          \end{array}
  \right.
\end{eqnarray*}
The MAC is the channel $10 \times 10 \rightarrow 5^*$.
The condensation of the representation $5^*$
 breaks the $SU(5)$ gauge symmetry to the $SU(4)$ gauge symmetry.

The fermion contents after the breaking is
\begin{equation}
 \Psi_{4^*}
  = \left( \begin{array}{c}
            \psi_{5^*}^1 \\
            \psi_{5^*}^2 \\
            \psi_{5^*}^3 \\
            \psi_{5^*}^4 \\
           \end{array}
    \right),
\qquad\qquad
 \Psi_{1} = \psi_{5^*}^5,
\end{equation}
\begin{equation}
 \Psi_6
 = \left( \begin{array}{cccc}
           0           &   \psi^{12} &
             \psi^{13} &   \psi^{14} \\
           - \psi^{12} &   0              &
             \psi^{23} &   \psi^{24} \\
           - \psi^{13} & - \psi^{23} &
             0              &   \psi^{34} \\
           - \psi^{14} & - \psi^{24} &
           - \psi^{34} &   0
          \end{array}
   \right),
\qquad\qquad
 \Psi_4
  = \left( \begin{array}{c}
            \psi^{15} \\
            \psi^{25} \\
            \psi^{35} \\
            \psi^{45}
           \end{array}
    \right) .
\end{equation}
The condensed fermion is $\Psi_6$,
 and it has the Majorana mass of the order of the $SU(5)$ breaking scale.
Other fermions remain massless at this stage.

The $SU(4)$ gauge coupling becomes large at low energy,
 and the subsequent pair condensation occurs.
The MAC is the singlet channel $4 \times 4^* \rightarrow 1$,
 and the $SU(4)$ gauge symmetry is not broken.
The two fermions $4$ and $4^*$ form a Dirac massive fermion.
All the fermions except for the singlet $\Psi_1$ become massive.
No subsequent condensation occurs.

The above estimation is not the rigorous one,
 since there are some assumptions:
 the ladder approximation, the MAC hypothesis,
 and neglecting the massive particles in the effective theory.
Moreover,
 we did not specify the regularization scheme and the renormalization.
In fact,
 the explicit momentum cut off regularization in the above argument
 is dangerous,
 since it breaks the chiral gauge symmetry
 \footnote{We have not known the regularization scheme
           which preserve the chiral gauge symmetry.
           But, recently a method
            which use the infinite number of the Pauli-Villars fields
            is proposed\cite{Frolov-Slavnov}.
           We hope that the regularization method
            proceeds our understanding on the chiral gauge theory.
           Especially, we hope the lattice calculation.}.
More rigorous treatment is needed to apply this dynamics to the physics.
In this thesis,
 we believe that the result is at least qualitatively good,
 while it may not be quantitatively good.

It is interesting
 whether the gauge group of the standard model
 can be included or not in the low energy effective theory
 of the tumbling gauge theories.
It is systematically estimated
 for the tumbling gauge theories with simple gauge group $SU(N)$
 \cite{Sualp-Kaptanoglu}.
Assuming the simple group
 means to consider the grand unified theory with the extended technicolor.
The restrictions of the asymptotic freedom
 and the gauge anomaly cancellation
 reduces the number of the models which should be considered.
The result is the discouraging one.
There are no candidate
 which includes the gauge group of the standard model
 in the low energy effective theory.
If we replace the MAC hypothesis into more rigorous one,
 we may have another pattern of the tumbling
 and may get the candidates.

In the next chapter,
 we consider the tumbling including the massive particles
 which get their masses in the process of tumbling.
After the second stage of the tumbling,
 there exist massive fermions and the effective four fermion interactions
 which are mediated by the massive gauge bosons.
\chapter{Tumbling and technicolor}

As we explained in chapter 5,
 the technicolor dynamics should be accompanied
 with the large anomalous dimension
 to avoid the flavor-changing neutral current problem.
Two explicit dynamical models with large anomalous dimension are proposed.
One is the gauge theory with very slowly running coupling
 (walking dynamics)\cite{walking},
 and another is the gauge theory with strong four fermion interaction
 (gauged Nambu-Jona-Lasinio model)\cite{strong-4F}.

It is also explained that
 the dynamics of the tumbling gauge theory is a candidate of the dynamics
 of the extended technicolor theory.
The effective four fermion interactions with hierarchical scales
 can be translated to the hierarchy of the masses of the ordinary fermions.
Moreover,
 if the four fermion interactions between the technifermions can be strong,
 the large anomalous dimension is realized.
In the following,
 we estimate the strength of the effective four fermion interactions
 in the tumbling gauge theory,
 and get the simple formula
 to give the values of the coupling constants of the interactions.
We also discuss of the decoupling of the massive gauge bosons
 which get their masses in the process of tumbling
 from the low energy dynamics.
The decoupling of the massive fermions
 which get their masses in the process of tumbling is also discussed.
The decoupling of the massive particles
 has already been proved in the perturbative point of view\cite{decouple}.
But it is still non-trivial in the non-perturbative point of view.

\section{The massive gauge boson effect}

In the process of tumbling,
 gauge bosons associated with the broken generators become massive.
In the ordinary scenario, these massive gauge bosons
 are assumed to decouple from the low energy dynamics,
 and they are only perturbatively considered
 as the weak effective four fermion interactions in the low energy region.
However, if the four fermion interaction
 can be strong, their non-perturbative effect
 must be considered.

Consider a certain stage of tumbling such that
 the gauge group $G$ breaks to $H$ at the scale $M$.
For simplicity, we regard the groups $G$ and $H$ as the simple ones.
Extension to the direct product groups
 is straightforward and we will discuss it later.
The gauge bosons
 with respect to the generators of $G/H$ become massive
 in this process.
The representation of the fermions are denoted
 by $r^G_1$, $r^G_2$, $\cdots$ in the theory above the scale $M$,
 and by $r^H_1$, $r^H_2$, $\cdots$ in the theory below the scale $M$.

The effective action below the scale $M$
 is given by decomposing the effective action at the scale $M$.
Above the scale $M$, the effective action
 with respect to the channel $r^G_1 \times r^G_2 \rightarrow r^G_c$ is
\begin{equation}
 \Gamma[\Sigma]
 = H[\Sigma^{\ast A} \Sigma_A]
 + {{g^2_G} \over 2}
   \big\{ C_2(r_1^G) + C_2(r_2^G) - C_2(r_c^G) \big\}
   F[\Sigma^{\ast A} \Sigma_A],
\end{equation}
 where the functional
\begin{equation}
 H[\Sigma^{\ast A} \Sigma_A]
 \equiv -2 \int {{d^4p} \over {(2\pi)^4}} {1 \over {p^2-m^2}}
           \Sigma^{\ast A}(p^2) \Sigma_A(p^2)
\end{equation}
 comes from the fermion integration, the functional
\begin{eqnarray}
\lefteqn{ F[\Sigma^{\ast A} \Sigma_A] \equiv }
\nonumber\\
&&
     - 2 i \int {{d^4p} \over {(2\pi)^4i}} {{d^4k} \over {(2\pi)^4i}}
                {1 \over {(k+p)^2-m^2}} {1 \over {p^2-m^2}}
        \cdot D^\mu_{\ \mu}(k) \Sigma^{\ast A}((k+p)^2) \Sigma_A(p^2)
\end{eqnarray}
 comes from one gauge boson exchange vacuum diagram,
 $g_G$ is the gauge coupling constant in the theory
 above the scale $M$,
 and $A$ is the group index of the condensate
 of representation $r^G_c$
 composed of $r^G_1$ and $r^G_2$
 \footnote{Since we take the Landau gauge,
            the mass function is the function of $p^2$ instead of $p$.}.
This effective action has already been linearized
 with respect to $\Sigma^{\ast A} \Sigma_A$,
 therefore we get the linearized gap equation
 in ladder approximation directly
 by differentiating with respect to $\Sigma^{\ast A}$.
We are assuming that the condensation in MAC emerges at the scale $M$
 and gauge group $G$ is broken to $H$,
 according to the formulation which is explained in the previous chapter.

The mass function $\Sigma_A(p^2)$ is decomposed
 into the irreducible components of the group $H$,
 $r_{c1}^H$, $r_{c2}^H$, $\dots$ $r_{cn}^H$, as
\begin{equation}
 \Sigma_A(p^2) =      \Sigma_{i_1}^{(r_{c1}^H)}(p^2)
               \oplus \Sigma_{i_2}^{(r_{c2}^H)}(p^2)
               \oplus \dots
               \oplus \Sigma_{i_n}^{(r_{cn}^H)}(p^2),
\end{equation}
 where $i_k$ is the group index of the representation $r_{ck}^H$.
The fermion representation can also be decomposed as
\begin{eqnarray}
 r_1^G &=& r_{11}^H \oplus r_{12}^H \oplus \dots \oplus r_{1n_1}^H,
\\
 r_2^G &=& r_{21}^H \oplus r_{22}^H \oplus \dots \oplus r_{2n_2}^H,
\\
 &\vdots&.
\nonumber
\end{eqnarray}
The decomposed mass functions
 correspond to all the possible condensates composed of
 $r^H_{11}$, $r^H_{12}$, $\cdots$, $r^H_{1n_1}$
 and $r^H_{21}$, $r^H_{22}$, $\cdots$, $r^H_{2n_2}$.
But the correspondence is not one to one.
For example,
 if both pairs,
 $r_{11}^H \times r_{21}^H$ and $r_{12}^H \times r_{22}^H$,
 contain the same representation $r^H_{c1}$,
 the mass function $\Sigma_{i_1}^{(r_{c1}^H)}$
 is the sum of two independent mass functions
 of the two channels with certain weight.
This means the mixing of the two channels of
 $r_{11}^H \times r_{21}^H \rightarrow r^H_{c1}$
 and $r_{12}^H \times r_{22}^H \rightarrow r^H_{c1}$
\footnote{The mixing is caused by the exchange of the massive gauge bosons
           corresponding to the broken off-diagonal generators
           which cause the transition $r_{11}^H \longrightarrow r_{12}^H$
           and $r_{22}^H \longrightarrow r_{21}^H$, for example.
          The other contributions are diagonal.}.
Except for these mixings, the correspondence is one to one.
We ignore the mixings for simplicity from now on.

Accordingly, the effective action is decomposed as
\begin{eqnarray}
 \Gamma[\Sigma]
 &=& H[\Sigma^{\ast i_1 (r_{c1}^H)} \Sigma_{i_1}^{(r_{c1}^H)}]
  + {{g_G^2} \over 2}
    \big\{ C_2(r_1^G) + C_2(r_2^G) - C_2(r_c^G) \big\}
    F[\Sigma^{\ast i_1 (r_{c1}^H)} \Sigma_{i_1}^{(r_{c1}^H)}]
\nonumber\\
 &+& \left( r_{c1}^H \longrightarrow r_{c2}^H, r_{c3}^H, \cdots \right).
\end{eqnarray}
There are no cross terms
 of the decomposed mass functions of the different representations.
We concentrate on
 a channel $r^H_{11} \times r^H_{21} \rightarrow r^H_{c1}$
 for a while.
Because the functional $F$
 linearly contains a gauge boson propagator,
 it can be decomposed into two parts concerning $H$ and $G/H$,
 respectively as
\begin{eqnarray}
\lefteqn{
 {{g^2} \over 2} \big\{C_2(r_1^G) + C_2(r_2^G) - C_2(r_c^G) \big\}
  F[\Sigma^{\ast i_1 (r_{c1}^H)} \Sigma_{i_1}^{(r_{c1}^H)}]
}
\nonumber\\
&&
 = a_H F[\Sigma^{\ast i_1 (r_{c1}^H)} \Sigma_{i_1}^{(r_{c1}^H)}]
 + a_{G/H} F[\Sigma^{\ast i_1 (r_{c1}^H)} \Sigma_{i_1}^{(r_{c1}^H)}],
\end{eqnarray}
 where $a_H$ and $a_{G/H}$ are constants satisfying
\begin{equation}
 a_H + a_{G/H}
 = {{g_G^2} \over 2}
   \big\{ C_2(r_1^G) + C_2(r_2^G) - C_2(r_c^G) \big\}.
\end{equation}
The contribution of the unbroken gauge interaction is known as
\begin{equation}
 a_H
  = {{g_H^2} \over 2}
    \big\{ C_2(r_{11}^H)+C_2(r_{21}^H)-C_2(r_{c1}^H) \big\},
\end{equation}
 where $g_H$ is the gauge coupling in the theory below the scale $M$.
As a matching condition we have $g_H(M)=g_G(M)$.
The above decomposition has to be done at the scale $M$,
 then we obtain the value of $a_{G/H}$ as
\begin{equation}
 a_{G/H}
 = {{g_G^2(M)} \over 2}
    \big\{ C_2(r_1^G)+C_2(r_2^G)-C_2(r_c^G) \big\}
 - {{g_H^2(M)} \over 2}
    \big\{ C_2(r_{11}^H)+C_2(r_{21}^H)-C_2(r_{c1}^H) \big\}.
\label{eq:rel}
\end{equation}
We can get the effective action below the scale $M$
 by giving masses of the order of $M$ to the gauge bosons $G/H$
 and the fermions which would become massive in this process.

We can relate $a_{G/H}$
 to the coupling constant of the effective four fermion interaction.
The gap equation with respect to the channel
 $r^H_{11} \times r^H_{21} \rightarrow r^H_{c1}$
 is given by differentiating the effective action
 with respect to $\Sigma^{(r^H_{c1}) \ast}$ as
\begin{eqnarray}
 \Sigma^{(r_{c1}^H)}_{i_1}(-q^2)
 &=& \lambda^{(r_{c1}^H)} \int^{M^2} dp^2
             {{\Sigma^{(r_{c1}^H)}_{i_1}(-p^2)} \over {p^2 + m^2}}
             \big\{ \theta(q^2-p^2) {{p^2} \over {q^2}}
                               + \theta(p^2-q^2) \big\}
\nonumber\\
 &+& {{g^{(r_{c1}^H)}_{4F}} \over {M^2}} \int^{M^2} dp^2
                      \Sigma^{(r_{c1}^H)}_{i_1}(-p^2),
\label{gap-equation}
\end{eqnarray}
 where we approximate the massive gauge boson propagator
 as $g_{\mu\nu} / M^2$
 and the integration is performed in Euclidean space.
Here we ignore the massive fermions
 which get their masses of the order of $M$
 at the breaking of $G \rightarrow H$.
The effect of the massive fermions will be discussed in the
 next section.
This is the linearized gap equation
 of the gauged Nambu-Jona-Lasinio model in the ladder approximation
 and we defined the effective couplings of gauge
 and four fermion interaction as
\begin{equation}
 \lambda^{(r_{c1}^H)} = {3 \over {4\pi}} \cdot {{a_H} \over {4\pi}}
\label{eq:def1}
\end{equation}
 and
\begin{equation}
 g^{(r_{c1}^H)}_{4F}
  = {3 \over {4\pi}} \cdot {{a_{G/H}} \over {4\pi}},
\label{eq:def2}
\end{equation}
 respectively.
This four fermion coupling
 measures the size of the contribution of the massive gauge boson
 to the dynamics of this channel.

The solution of this gap equation and phase diagram
 (fig.(\ref{phase-diagram})) are given in ref.\cite{phase}.
\begin{figure}
\caption{Phase diagram of gauged Nambu-Jona-Lasinio model
          in ladder and fixed coupling approximation.
         The region inside the critical line
          is symmetric phase and outside is broken phase.
         The anomalous dimension of the mass function is given by
          $\gamma_m = 1 + \left( 1 - \lambda / \lambda_c \right)^{1/2}$.
         It increases from $\gamma_m=1$ to $\gamma_m=2$
          along the critical line.}
\label{phase-diagram}
\end{figure}
The critical line can be obtained as follows.
The eq.(\ref{gap-equation}) with $m=0$
 (expanding the Schwinger-Dyson equation around the zero of the mass function)
 can be transformed to the differential equation and boundary conditions
\begin{equation}
 \left( x \Sigma(x) \right)''
  + \lambda {{\Sigma(x)} \over x} = 0,
\label{differential-eq-4F}
\end{equation}
\begin{equation}
 \lim_{x \rightarrow \mu^2} x^2 \Sigma'(x) = 0,
\label{IRBC-4F}
\end{equation}
\begin{equation}
 \lim_{x \rightarrow M^2}
    \left[ \left( x \Sigma(x) \right)' + {{g_{4F}} \over \lambda} x \Sigma(x)'
    \right] = 0,
\label{UVBC-4F}
\end{equation}
 where $x$ is the squared Euclidean momentum: $x=q^2$
 (We write $\Sigma(-q^2)$ as $\Sigma(x)$.),
 and $\mu$ is the scale which will be set to zero in the following.
We neglect the group indexes for simplicity.
The effect of the four fermion interaction
 comes in only through the ultraviolet boundary condition
 (see eqs.(\ref{differential-eq}), (\ref{IRBC-bif}), and (\ref{UVBC-bif}).).
Although the bifurcation theory\cite{Atkinson}
 can not be used to obtain the critical line,
 because we have two parameters $g_{4F}$ and $\lambda$,
 it can be obtained as a condition to exist the non-trivial solution.
It is enough to consider the above linearized equation,
 if we consider only the small solution.
While both couplings are small, there is only a trivial solution.
And when the couplings become large and satisfy a certain relation,
 the small non-trivial solution will be {\it bifurcated}
 from the trivial solution.
The relation is nothing but the critical line.

The general solution of the differential equation is
\begin{equation}
 \Sigma(x) = c_1 x^{-(1-\gamma)/2} + c_2 x^{-(1+\gamma)/2},
\end{equation}
 where $c_1$ and $c_2$ are the constants, and $\gamma=\sqrt{1-4\lambda}$.
The boundary conditions give
\begin{equation}
 c_1 {{1-\gamma} \over 2} \mu^{1+\gamma}
 + c_2 {{1+\gamma} \over 2} \mu^{1-\gamma} = 0,
\end{equation}
\begin{equation}
 c_1 \left(
       1 - \left( 1 + {{g_{4F}} \over \lambda} \right) {{1-\gamma} \over 2}
     \right) M^{-1+\gamma}
 +
 c_2 \left(
       1 - \left( 1 + {{g_{4F}} \over \lambda} \right) {{1+\gamma} \over 2}
     \right) M^{-1-\gamma}
 = 0.
\end{equation}
By combining these conditions, we obtain
\begin{equation}
 c_2 \left[
      1 - {{1-\gamma} \over {1+\gamma}} \cdot
       {
        {1-\left( 1 + {{g_{4F}} \over \lambda} \right) {{1+\gamma} \over 2}}
       \over
        {1-\left( 1 + {{g_{4F}} \over \lambda} \right) {{1-\gamma} \over 2}}
       }
       \left( {\mu \over M} \right)^{2\gamma}
      \right] = 0.
\end{equation}
The non-trivial solution exists if the inside of the square bracket vanishes.
In the naive limit $\mu/M \rightarrow 0$,
 the condition $\gamma=0$, namely $\lambda=\lambda_c=1/4$,
 should be satisfied for any $g_{4f}$ and $\lambda<1/4$.
But, we have another solution for $g_{4F} \geq 1/4$.
The condition for non-trivial solution can be written as
\begin{equation}
 {{1-\gamma^2} \over {1+\gamma^2}} \cdot
 {
  {1+\gamma-{1 \over 2} \left( 1- \gamma^2 + 4 g_{4F} \right)}
 \over
  {1-\gamma-{1 \over 2} \left( 1- \gamma^2 + 4 g_{4F} \right)}
 }
 = \left( {\mu \over M} \right)^{2\gamma}.
\end{equation}
Taking the limit $\mu/M \rightarrow 0$ gives the another condition
\begin{equation}
 \lambda
  = {1 \over 4} \left\{ 1 - \left( \sqrt{4g_{4F}} - 1 \right)^2 \right\}.
\end{equation}
We have obtained the critical line
 which is displayed in fig.(\ref{phase-diagram}).

The value of the critical gauge coupling $\lambda_c^\prime$
 depends on the value of the effective four fermion coupling as
\begin{equation}
 \left\{
  \begin{array}{ll}
    \lambda_c^\prime
     = {1 \over 4} \left\{ 1-\left( \sqrt{4 g_{4F}}-1 \right)^2 \right\} &
    \qquad
    g_{4F} > {1 \over 4} \\
   \noalign{\vskip 0.2cm}
    \lambda_c^\prime = {1 \over 4} &
    \qquad
    g_{4F} \le {1 \over 4}. \\
  \end{array}
 \right.
\end{equation}
When the four fermion coupling is larger than $1/4$,
 the critical gauge coupling is reduced.
The condensation scale of this channel is defined
 by using the phase diagram.
We replace the gauge coupling constant by the running one
 given by the one-loop perturbative calculation.
We, however, ignore the running of the massive gauge boson-fermion coupling,
 because the coupling does not become large.
The reason is that the negative contribution
 to the $\beta$-function of the coupling of the broken gauge interaction
 can be neglected,
 since the gauge boson loop in the self-energy of the massive gauge bosons
 should contain at least one massive gauge boson.
The condensation scale can be defined as the scale
 at which the gauge coupling takes critical value.
The larger effective four fermion coupling constant means
 the smaller condensation scale and the larger anomalous dimension
 as $\gamma_m = 1+\sqrt{1-4\lambda_c^\prime}=\sqrt{4g_{4F}}$ for
 $g_{4F} \ge 1/4$ \cite{Miransky-Yamawaki}.
We have a criterion for the attractive effective four fermion interaction
 to be significant to the dynamics:
 the coupling must be larger than $1/4$.

Now we evaluate the massive gauge boson effect
 on the dynamics of the channel with attractive gauge interaction
 and answer the question
 whether the large anomalous dimension dynamics
 due to the strong four fermion interaction can be realized
 in the channel.
We can get a relation of the effective couplings at the scale $M$
\begin{equation}
 g^{(r_{c1}^H)}_{4F}
  = \lambda(r_1^G, r_2^G, r_c^G)
  - \lambda(r_{11}^H, r_{21}^H, r_{c1}^H)
\end{equation}
 from eqs.~(\ref{eq:rel}), (\ref{eq:def1}) and (\ref{eq:def2}),
 and now
\begin{equation}
 \lambda(r_{11}^H, r_{21}^H, r_{c1}^H) \ge 0
\end{equation}
 since the gauge interaction is considered to be attractive.
Furthermore,
 the effective coupling $\lambda(r_1^G, r_2^G, r_c^G)$
 at the scale $M$
 is less than the critical coupling $\lambda_c = 1/4$
 because of the MAC hypothesis.
Therefore,
 the four fermion coupling $g^{(r_{c1}^H)}_{4F}$
 must be less than $1/4$ and we conclude that
 the attractive force by the massive gauge boson exchange is too small
 to significantly affect the dynamics
 in the channel with attractive gauge interaction.
This means that the realistic technicolor dynamics
 with the large anomalous dimension
 due to the strong four fermion interaction
 cannot be realized in the low energy effective theory
 of tumbling gauge theories,
 provided that the mixing channels are ignored.
We can also conclude that the MAC is not change,
 even if we consider the massive gauge boson effect.

This conclusion is also true
 when we consider a direct product gauge group.
We consider that the gauge group
 $G \equiv G_1 \otimes G_2 \otimes \dots \otimes G_n$
 is broken to the gauge group
 $H \equiv H_1 \otimes H_2 \otimes \dots \otimes H_m$
 at the scale $M$.
For example, we consider a channel
\begin{equation}
 R^G\{i\} \times R^G\{j\} \rightarrow R^G_c\{k\},
\end{equation}
 where
\begin{equation}
 R^G\{i\} \equiv
 \big( r_{i_1}^{G_1}, r_{i_2}^{G_2}, \dots , r_{i_n}^{G_n} \big)
\end{equation}
 denote a representation.
The representations of the group $G$
 can be decomposed into the representations of the group $H$ as
\begin{eqnarray}
 R^G\{i\} &=& \mathop\bigoplus_{\{p\}} R^H\{p\},
\\
 R^G\{j\} &=& \mathop\bigoplus_{\{q\}} R^H\{q\},
\\
 R^G_c\{k\} &=& \mathop\bigoplus_{\{l\}} R^H_c\{l\},
\end{eqnarray}
 where the summations are taken over
 the appropriate sets of the irreducible representations of $H$,
 $\{p\}$, $\{q\}$ and $\{l\}$.
If the channel mixing are ignored again,
 the decomposed representations of the mass functions in the effective action
 have the one-to-one correspondence to the
 channels in terms of the representations of $H$.
We also obtain similar relations to determine the
 effective four fermion couplings.
For example, for the channel
\begin{equation}
 R^H\{p\} \times R^H\{q\} \rightarrow R^H_c\{l\},
\end{equation}
 we get the relation
\begin{equation}
 g^{(R^H\{l\})}_{4F} = \lambda(R^G\{i\}, R^G\{j\}, R^G_c\{k\})
                     - \lambda(R^H\{p\}, R^H\{q\}, R^H_c\{l\}),
\end{equation}
 where the effective couplings are defined by
\begin{eqnarray}
 \lambda(R^G\{i\}, R^G\{j\}, R^G_c\{k\})
 &=& \sum_{\alpha=1}^n
     \lambda(r^{G_\alpha}_{i_\alpha}, r^{G_\alpha}_{j_\alpha},
             r^{G_\alpha}_{ck_\alpha}),
\\
 \lambda(R^H\{p\}, R^H\{q\}, R^H_c\{l\})
 &=& \sum_{\beta=1}^m
     \lambda(r^{H_\beta}_{p_\beta}, r^{H_\beta}_{q_\beta},
             r^{H_\beta}_{cl_\beta}).
\end{eqnarray}
Therefore,
 the four fermion coupling must be less than $1/4$
 in the channel with attractive gauge interaction
 and the conclusion does not change.
The attractive force of the massive gauge boson is too small
 to significantly affect the dynamics
 of the channel with attractive gauge interaction.

\section{The massive fermion effect}

Several fermions get their masses in the process of tumbling.
In the original tumbling scenario it is naively assumed that
 the massive fermions do not cause significant effect
 on the lower energy dynamics.
In this section we will argue
 the decoupling of the fermions which get Majorana masses
 in the process of tumbling
 \footnote{Some fermions get Dirac masses (mixing Majorana masses)
            in the process of tumbling.
           These masses are due to the mixing of the condensation.
           We do not discuss the mixing, here.}.

We examine the linearized ladder gap equation
 for the channel with the massive fermions,
\begin{eqnarray}
\lefteqn{
 \Sigma_A (-q^2) =
}
\nonumber\\
&&
 + \lambda \int^{M^2} dk^2 {{k^2} \over {(m_1^2+k^2)(m_2^2+k^2)}}
             \bigg\{
               \theta(k^2-q^2) + {{k^2} \over {q^2}} \theta(q^2-k^2)
             \bigg\} \Sigma_A(-k^2)
\nonumber\\
&&
 - \lambda \int^{M^2} dk^2
              {{m_1 m_2} \over {(m_1^2+k^2)(m_2^2+k^2)}}
             \bigg\{
               \theta(k^2-q^2) + {{k^2} \over {q^2}} \theta(q^2-k^2)
             \bigg\} \big( \Sigma_A(-k^2) \big)^{\ast},
\end{eqnarray}
 where $M$ is the scale of the previous tumbling,
 $m_1$ and $m_2$ are the Majorana masses,
 $\lambda$ is the gauge coupling
 and $A$ is the index of the representation of condensate
 (fig.(\ref{gapeq})).
\begin{figure}
\caption{Gap equation for fermions with Majorana mass.}
\label{gapeq}
\end{figure}
Here we neglect the massive gauge boson effect,
 because we have found in the previous section that
 such effect is small for $\lambda \ge 0$.
This equation can be translated into a differential equation
 with boundary conditions,
\begin{equation}
 {{d^2} \over {dx^2}} \big( x \Sigma(x))
 + \lambda {{x \Sigma(x) - m_1 m_2 \Sigma(x)}
             \over {(m_1^2+x)(m_2^2+x)}}
 = 0,
\end{equation}
\begin{equation}
 \mathop{\rm lim}_{x \rightarrow 0} x^2 {{d \Sigma} \over {dx}} = 0,
\end{equation}
\begin{equation}
 \mathop{\rm lim}_{x \rightarrow M^2}
 {d \over {dx}} \big( x \Sigma(x) \big) = 0,
\end{equation}
 where $x=q^2$ and index $A$ is omitted.
We assume that the mass function is real.

First, we consider the case $m_1 = m \simeq M$ and $m_2 = 0$.
In this case
 the differential equation
 can be reduced to the Bessel differential equation
\begin{equation}
 {{d^2 S} \over {d z^2}} + {1 \over z} {{dS} \over {dz}}
  + \left( 1 - {1 \over {z^2}} \right) S = 0,
\end{equation}
 where $S(z)=z\Sigma(x)$ and $z=2\sqrt{\lambda x / m^2}$.
We can see that the condition
\begin{equation}
 J_0(2 \sqrt{\lambda M^2 / m^2}) = 0
\end{equation}
 must be satisfied for the non-trivial solution to exist,
 where $J_0$ is the zero-th order Bessel function.
The critical value of the coupling is defined
by the lowest zero point of $J_0$ as
\begin{equation}
 \lambda_c \simeq 1.45 {{m^2} \over {M^2}} \simeq 1.45,
\end{equation}
 and this is very large compared
 with the value of massless fermion case $1/4$.

Next, in the case $m_1 = m_2 \simeq M$,
 the equation can be reduced
 to the modified Bessel differential equation
\begin{equation}
 {{d^2 S} \over {d z^2}} + {1 \over z} {{dS} \over {dz}}
  - \left( 1 + {1 \over {z^2}} \right) S = 0.
\end{equation}
We can see that
 there is no non-trivial solution which satisfies the boundary conditions.

Therefore,
 the channels containing the fermions with Majorana masses
 are unlikely to be the MAC.
The fermions which get Majorana masses in the process of tumbling
 may decouple from the low energy dynamics.
We have non-perturbative shown it.
\chapter{Conclusion}

The idea of the technicolor theory and the extended tehcnicolor theory
 is very beautiful.
It can be expected that
 the theory solves some problems of the standard model.
The hierarchical gauge symmetry breaking
 is expected in the extended technicolor theory
 to explain the hierarchy of the mass of the quarks and leptons.
Tumbling gauge theory
 is a candidate of the dynamics of such hierarchical gauge symmetry breaking.

But some new problems peculiar to the technicolor theory is emerged.
One of the problem is the flavor-changing neutral current problem.
Extended technicolor theory results
 too large mixing of $K^0$ and ${\bar K}^0$
 to be consistent with the experiment, for example.
The problem is solved
 by considering the special technicolor dynamics
 with the large anomalous dimension.
Therefore, it is expected that
 the large anomalous dimension is realized in the tumbling gauge theory.

We formulated the tumbling gauge theory
 using the Cornwall-Jackiw-Tomboulis effective action
 in the one gauge boson exchange approximation (ladder approximation).
Non-perturbative effect of the massive particles
 which get their masses in the process of tumbling
 was estimated by using the formalism.
We got the effective action
 at the low energy region of the tumbling gauge theory
 and obtained a formula to give the coupling constants
 of the effective four fermion interactions
 which is mediated by the massive gauge bosons.
Upper bound on the coupling constants
 was obtained in the channels with attractive unbroken gauge interactions.
The effective four fermion interactions are too weak
 to cause the significant effect on the dynamics.
The large anomalous dimension due to the strong four fermion interaction
 with the attractive gauge interaction (gauged Nambu-Jona-Lasinio type system)
 cannot be realized in the tumbling gauge theories.
This result is important
 for the model building of the extended technicolor theory
 with tumbling dynamics.
The possibility of walking dynamics
 (gauge theory with very slowly running coupling)
 in the tumbling gauge theory is not excluded.

We also estimated non-perturbative effect of the massive fermions
 (Majorana type) in the same framework.
It was shown
 in one gauge boson exchange approximation that
 they decouple from the low energy dynamics
 \footnote{It must be noted
            that all the mixing effects were ignored in our argument.}.
The channels containing massive fermions
 are hard to be MAC (most attractive channel).
It can be expected that
 all the massive particles {\it non-perturbatively} decouple
 from the low energy dynamics.
In the perturbative sense,
 the decoupling of the heavy particles from the low energy physics
 is well known as the decoupling theorem.

The recent phenomenological issues of the technicolor theory
 were also discussed.
We discussed the relation
 between the technicolor theory and the recent precision experiments.
The radiative correction
 due to the technicolor or extended technicolor dynamics
 is classified into two parts: oblique correction and non-oblique correction.
The experimental constraint on the oblique correction ($S$ parameter)
 directory restricts the size of the technicolor sector.
The number of the technicolor degrees of freedom
 and the number of the techni-flavor are restricted.
The technicolor theory is disfavored by the experiments.
Extended technicolor theory
 generally generates the large non-oblique correction.
It was explained that the non-oblique correction on $Zb{\bar b}$ vertex
 due to the sideways and ``diagonal'' bosons are large,
 when the mass of the top quark is large.
We showed that
 the correction on the vertex is too large
 to be consistent with the experiments.
The extended technicolor theory is disfavored by the experiments.

All the results of this thesis
 are disadvantageous for the technicolor theory.
If the nature selects
 the dynamical electroweak symmetry breaking
 and the dynamical generation of the ordinary fermion masses
 without the elementary scalar fields,
 there will be some new mechanisms.
The present technique
 to analyze the non-perturbative dynamics of the gauge theory is poor.
Therefore, the breakthrough in the technique
 can make possible to find the new mechanisms.

\begin{flushleft}
 \large\bf
 Acknowledgments
\end{flushleft}
I am grateful to Professor Koichi Yamawaki
 for helpful encouragement and many discussions.
I would like to thank
 Professor A.Ichiro Sanda, Yoshio Kikukawa, Masaharu Tanabashi,
 and Masayasu Harada for many discussions and suggestions.
I also thank Sho Tsujimaru for careful reading of this thesis.
I thank all the members of the elementary particle physics laboratory.

\end{document}